\documentclass[11pt]{article}
\usepackage{epsfig,dsfont}
\usepackage{geometry}
\usepackage{amsmath}
\usepackage{amssymb}
\usepackage[latin1]{inputenc}
\usepackage{url}
\usepackage{bm}
\usepackage[normalem]{ulem}
\usepackage{color}
\usepackage{hyperref}

\newcommand{\be}{\begin{equation}}
\newcommand{\ee}{\end{equation}}
\newcommand{\SL}{\text{SL}}
\newcommand{\PSL}{\text{PSL}}

\usepackage{cite}

\begin{document}
 
\vspace*{1mm}

\begin{center}

{\LARGE
Self-dual U(1) lattice field theory with a $\theta$-term} \\
\vskip2mm
\vskip10mm
Mariia Anosova$\,^a$, Christof Gattringer$\,^b$ and Tin Sulejmanpasic$\,^c$ 
\vskip8mm
$\;^a$ Universit\"at Graz, Institut f\"ur Physik\footnote{Member of NAWI Graz.}, Universit\"atsplatz 5, 8010 Graz, Austria
\vskip1mm
$^b$ FWF - Austrian Science Fund\footnote{On leave of absence from Universit\"at Graz.}, Sensengasse 1, 1090 Vienna, Austria
\vskip1mm
$\;^c$  Department of Mathematical Sciences, Durham University, DH1 3LE Durham, United Kingdom
\end{center}
\vskip15mm

\begin{abstract}
We study U(1) gauge theories with a modified Villain action. Such theories can naturally be coupled to electric and magnetic matter, and display exact 
electric-magnetic duality. In their simplest formulation without a $\theta$-term, such theories are ultra-local. We extend the 
discussion to U(1) gauge theories with 
$\theta$-terms, such that $\theta$ periodicity is exact for a free theory, and show that imposing electric-magnetic duality results in a local, 
but not ultra-local lattice action, which is reminiscent of the L\"uscher construction of axial-symmetry preserving fermions in 4d. We discuss the coupling to electric and magnetic matter as well as to dyons. For dyonic matter
the electric-magnetic duality and shifts of the $\theta$-angle by $2\pi$ together generate an SL$(2,\mathbb Z)$ duality group of  transformations, 
just like in the continuum. We finally illustrate how the SL$(2,\mathbb Z)$ duality may be used to explore theories at finite $\theta$ without a sign problem.
\end{abstract}

\vskip15mm


\section{Introduction}

Abelian U(1) gauge theories are at the heart of electrodynamics, which, together with gravity, explains most of the world we experience. 
In modern high-energy physics, they are typically viewed as boring cousins of their non-abelian counterparts. The reason for this is 
that in four space-time dimensions they are infrared free and UV incomplete. In solid-state physics, however, U(1) gauge theories 
pop up as effective descriptions of many condensed matter systems with non-trivial and strongly coupled dynamics. Moreover they 
show up as effective descriptions of non-abelian gauge theories and were long hoped to shed light on quark confinement and mass 
gap generation in those theories. While such abelian mechanisms have shown to be relevant in both supersymmetric 
\cite{Seiberg:1994rs,Seiberg:1994aj}, and non-supersymmetric settings 
\cite{Polyakov:1976fu,Unsal:2007vu,Unsal:2008ch,Poppitz:2012sw}, a lot of justified skepticism remains about the potential of abelian 
gauge theories to give insights into the mysteries of 4d non-abelian gauge theories.

Both, in condensed matter applications as well as in the abelianized non-abelian gauge theories, monopoles play a crucial role. Monopoles 
have the ability to render the  theory confining, and can therefore completely change the IR  dynamics of the theory. The  details  of 
monopoles (e.g., their charge or mass) dictate whether or not  they are important for the IR physics. Yet, on the lattice monopoles 
typically come as artifact rather than a feature, and are not under direct control. Two of us proposed a formalism of U(1) lattice gauge 
theories which allows electric and magnetic matter to be coupled simultaneously \cite{Sulejmanpasic:2019ytl}, and provides  complete 
control over  the matter content of  the  theory -- both electric  and magnetic. Moreover the formalism allows a duality transformation, 
mapping magnetic and electric content to each other, which is a well known feature of continuum theories. Such models were used for 
avoiding the complex action problem associated with the $\theta$-term in 2d abelian gauge theories 
\cite{Gattringer:2018dlw,Goschl:2018uma,Anosova:2019quw,Sulejmanpasic:2020lyq,Sulejmanpasic:2020ubo}, and for formulating 
interacting exactly self-dual gauge theories  \cite{Anosova:2021akr,inprogress}. These modified Villain theories also found applications in fracton physics 
\cite{Gorantla:2021svj} and construction of theories with non-invertible symmetries \cite{Nguyen:2021yld,Choi:2021kmx}.

Furthermore 4d Villain theories allow for $\theta$-terms. Indeed $\theta$-terms in Villain-like models which have the correct continuum limit 
have been introduced long ago \cite{Cardy:1981qy,Cardy:1981fd}, but generically they do not preserve exact $\theta$-periodicity and self-duality, when 
applicable\footnote{The $\theta$-periodicity is lost in the presence of monopoles due to the Witten effect, since a monopole turns into a dyon which carries 
electric charge as well as a magnetic one. However, if the entire spectrum of dyons exists, whose masses correctly interchange under a $2\pi$ shift, the $
\theta$-periodicity can be valid even in the presence of magnetically charged matter. This is in fact what happens in abelianized non-abelian theories in 4d.}. 
In \cite{Sulejmanpasic:2019ytl} a class of $\theta$-terms which preserve the exact $2\pi$ periodicity of the $\theta$-angle in the pure 
gauge theory were introduced, but self-dual transformations in the presence of the $\theta$-term were not discussed in 
detail. Here we find that insisting on exact $\theta$-periodicity and on ultra-locality
ruins self-duality. More precisely, if one performs a duality transformation 
with one of the $\theta$-terms proposed in \cite{Sulejmanpasic:2019ytl}, a dual theory has a local, but not ultra-local\footnote{A lattice 
action is called \emph{ultra-local} when it couples only fields at a finite number 
of lattice distances. In a 
\emph{local} lattice theory fields at arbitrary distances may couple, but the pre-factors decrease (at least) exponentially with the lattice separation.} action, so that the theory is not self-dual for any choice of parameters.

On the other hand, one can restore self-duality by defining a local, but not ultra-local lattice action, which transforms covariantly under 
the duality transformation. We therefore reformulate the original U(1) lattice gauge theory in a local but non-ultra-local way, such that 
self-duality becomes exact. We recover the $\SL(2,\mathbb Z)$ duality, well known in continuum U(1) gauge theories 
\cite{Witten:2003ya}\footnote{It is worth noting that \cite{Cardy:1981fd} is the first proposal of the SL$(2,\mathbb Z)$ structure for a lattice gauge theory, 
which, however, is only approximate.}. Exploiting the $\SL(2,\mathbb Z)$ structure, we discuss interacting 
theories free from the sign problem.
The situation is reminiscent of exactly massless Dirac fermions on the lattice. Exactly massless Dirac fermions have axial 
symmetries in addition to vector symmetries, which have mixed 't Hooft anomalies and are subject to the Nielsen-Ninomiya theorem 
\cite{Nielsen:1980rz,Nielsen:1981xu,Nielsen:1981hk}. This was widely interpreted as as the inability to preserve axial symmetries on the 
lattice. However, L\"uscher showed \cite{Luscher:1998pqa},using Neuberger's solution 
\cite{Neuberger:1997fp} of the Ginsparg-Wilson relation \cite{Ginsparg:1981bj}, that a lattice theory possesing an exact axial symmetry exists, at the expense of abandoning ultra-locality.

In fact also abelian gauge theories  have `t Hooft anomalies 
\cite{Kravec:2013pua,Gaiotto:2014kfa,Komargodski:2017dmc,Komargodski:2017smk} (see also \cite{Honda:2020txe}), 
so a straightforward implementation of symmetries is bound to face an obstructions, much like in the case of fermionic 
chiral symmetry. What is perhaps surprising is that the lattice action can still be made ultra-local when $\theta$ angle is vanishing 
\cite{Sulejmanpasic:2019ytl}, as we will review. Nevertheless we are forced to abandon ultra-locality when insisting on the $\theta$-angle periodicity\footnote{The $2\pi$ $\theta$-angle shifts are only an invariance of a theory without dynamical monopoles, because $2\pi$ shifts cause all monopoles to become dyons. Alternatively the theory can posses an entire tower of dynamical dyons of unit magnetic charge and all electric charges, in which case the $2\pi$ periodicity will be restored. This is precisely what happens in the abelianized regimes of the $SU(N)$-gauge theory. } and imposing 
self-duality in free abelian gauge theory. As is well known, these transformations form an $\SL(2,\mathbb Z)$ group.

The paper is organized as follows: In the next section we review the approach to 
lattice discretization from \cite{Sulejmanpasic:2019ytl} that gives rise to the Villain form with an additional closedness 
constraint for the 2-form Villain variables. This closedness constraint removes monopoles and we show that pure U(1) lattice gauge 
theory becomes self-dual. In Section 3 we add electrically and magnetically charged matter and demonstrate that with our formulation 
self-duality can be extended to full QED with either bosonic or fermionic matter. For the case of bosonic matter we show that 
with a worldline formulation all sign problems are overcome (for fermionic matter the Grassmann nature of the matter fields may lead to 
a remaining sign problem). Based on the closedness constraint, in Section 4 we define a topological charge in a consistent way and 
discuss its properties including the Witten effect. Using this topological charge we include a $\theta$-term and discuss the 
generalized duality transformation that takes into account this term. To obtain self-duality with the $\theta$-term we 
generalize the action to a non-ultra-local 
form which leads to a fully self-dual lattice discretization of QED with a $\theta$-term. We discuss the resulting self-duality relations for 
several observables as well as the full $\SL(2,\mathds{Z})$ duality. Finally we employ this duality to discuss theories with non-zero topological 
angle $\theta$ that can be simulated without complex action 
problem. 

Some more technical parts are collected in several appendices: Appendix \ref{app:diff_forms} summarizes our notation and some results for 
differential forms on the lattice. In Appendix \ref{app:B} we discuss properties of our definition of the topological charge.
In Appendix \ref{app:C} we discuss properties of the action kernels used in our paper and in Appendix \ref{app:D} we provide a short derivation of a
generalized Poisson resummation formula. Appendix \ref{app:worldline} shows how self dual QED can be mapped to a worldline representation that 
avoids the complex action problem and thus makes the theory accessible to Monte Carlo simulations. Finally, in Appendix \ref{app:gen_top_charge}
we discuss a further generalization of the topological charge that implements all lattice symmetries.


\section{Generalized Villain action for U(1) gauge fields and self-duality}

In this section we briefly summarize the discretization strategy outlined in \cite{Sulejmanpasic:2019ytl}, which leads to 
a Villain form of the Boltzmann factor \cite{Villain:1974ir}. We implement additional closedness constraints for the Villain variables that 
suppress the artificial monopoles that plague the standard lattice discretization of U(1) gauge fields. We show that the construction 
leads to a self-dual discretization of the photon field which will be the starting point for more general self-dual theories 
in the subsequent sections. 


\subsection{Generalized Villain discretization with closedness constraints}

When coupling U(1) gauge fields to matter in a gauge invariant way one uses the U(1)-valued compact link variables
$U_{x,\mu} \in$ U(1). The coupling between gauge and matter fields is then implemented via nearest neighbor terms 
such as $\phi_x^{\, *} \, U_{x,\mu} \, \phi_{x + \hat{\mu}}$. Under a U(1) gauge transformation the matter fields transform 
as $\phi_x \rightarrow e^{i\lambda_x} \phi_x$, such that the transformation $U_{x,\mu} \rightarrow 
e^{i\lambda_x}  U_{x,\mu} \,e^{-\, i\lambda_{x+\hat{\mu}}}$ gives rise to gauge invariance of 
$\phi_x^{\, *} \, U_{x,\mu} \, \phi_{x + \hat{\mu}}$. 

We now parameterize the link variables in the form\footnote{To be specific, we consider a 4-d hypercubic lattice with 
lattice extents $N_\mu, \mu = 1,2,3,4$ and a  total number of sites $V \equiv N_1 N_2 N_3 N_4$. 
The lattice constant is set to $a = 1$ and all fields obey periodic boundary conditions.}
\begin{equation}
U_{x,\mu} \; \equiv \; e^{ i A^e_{x,\mu} }  \; ,
\label{links}
\end{equation}
with the  gauge fields $A^e_{x,\mu} \in [-\pi, \pi)$. We added the superscript $^e$ to mark the gauge fields 
as \emph{electric gauge fields}, a notation that will turn out to be useful when studying electric-magnetic duality.  

We may define the field strength tensor as the naive discretization of the continuum form 
$\partial_\mu A^e_\nu - \partial_\nu A^e_\mu$ and obtain
\begin{equation}
A^e_{x + \hat{\mu},\nu} - A^e_{x,\nu} - A^e_{x + \hat{\nu},\mu} + A^e_{x,\mu} \; \equiv \; (d A^e)_{x,\mu \nu} \; ,
\end{equation}
where in the last step we have defined the exterior lattice derivative $d$, here acting on the link variables. We will partly 
use the language of differential forms on the lattice and in Appendix \ref{app:diff_forms} briefly summarize our conventions and some 
of the results we use (for a more extensive presentation see the appendix of \cite{Sulejmanpasic:2019ytl}  or
the mathematical standard literature such as \cite{wallace}).
 
The compact link variables
(\ref{links}) are obviously invariant under the shifts 
\begin{equation}
A^e_{x,\mu} \; \rightarrow \; A^e_{x,\mu} \; + \; 2 \pi \, k_{x,\mu} \;  , \; \; k_{x,\mu} \; \in \; \mathds{Z} \; .
\label{shifts}
\end{equation}
However, the exterior derivative $(d A^e)_{x,\mu \nu}$ is not invariant under shifts, but instead transforms as 
\begin{equation}
(d A^e)_{x,\mu \nu} \; \rightarrow \; (d A^e)_{x,\mu \nu} \; + \; 2 \pi \; (d k)_{x,\mu \nu} \; ,
\label{shiftdA}
\end{equation}
i.e., $(d A^e)_{x,\mu \nu}$ is shifted by multiples of $2 \pi$. One way to establish invariance under the shift symmetry 
is to construct the action from the periodic function $\cos((d A^e)_{x,\mu \nu})$, which gives rise to the 
Wilson gauge action. Another option is to define the field strength as
\begin{equation}
F^e_{x,\mu \nu} \; \equiv \; (d A^e)_{x,\mu \nu} \; + \; 2 \pi \, n_{x,\mu \nu} \; = \; (d A^e \; + \; 2 \pi \, n)_{x,\mu \nu} \; ,
\label{Fdef}
\end{equation}
where $n_{x,\mu \nu} \in \mathds{Z}$ is a 2-form (i.e., plaquette based) variable which subsequently is summed over 
and obviously eats up a possible term $2 \pi \, (d k)_{x,\mu \nu}$ generated by the shifts \eqref{shifts}. In other words the Villain variables $n_{x,\mu\nu}$ can be viewed as gauge field of the shift symmetry \cite{Sulejmanpasic:2019ytl}. In its simplest 
form this construction gives rise to a gauge field Boltzmann factor
\begin{equation}
B_\beta[A^e] \, \equiv \, \prod_{x \in \Lambda} \prod_{\mu < \nu} \; \sum_{n_{x,\mu \nu} \in \mathds{Z}} \!\!\! e^{ \, - \frac{\beta}{2}
F^e_{x,\mu \nu} F^e_{x,\mu \nu}} \; = \; 
\prod_{x \in \Lambda} \prod_{\mu < \nu} \; \sum_{n_{x,\mu \nu} \in \mathds{Z}} \!\!\! e^{ \, - \frac{\beta}{2}
(d A^e \; + \; 2 \pi \, n)_{x,\mu \nu} (d A^e \; + \; 2 \pi \, n)_{x,\mu \nu} } \; ,
\label{boltzmann1}
\end{equation}
which is known as the \emph{Villain discretization} \cite{Villain:1974ir}. The first product runs over our 4-dimensional lattice 
$\Lambda$ with periodic boundary conditions. We will refer to the variables $n_{x,\mu \nu}$ as \emph{Villain variables}. As usual
$\beta$ is the inverse gauge coupling $\beta = 1/e^2$, where $e$ is the electric charge.

The Boltzmann factor (\ref{boltzmann1}) is not only invariant under the shifts (\ref{shifts}), (\ref{shiftdA}), but also under
the transformation $A^e_{x,\mu} \rightarrow A^e_{x,\mu} - \lambda_{x + \hat{\mu}} + \lambda_x$ which, up to a possible
re-projection into the interval $[-\pi, \pi)$, corresponds to the gauge transformation of the  gauge fields 
$A^e_{x,\mu}$. 

A key insight is to note that the Villain variables $n_{x,\mu \nu}$ may be constrained further: They were 
introduced to implement the invariance of the Boltzmann factor $B_\beta[A^e]$ under the shifts (\ref{shifts}), (\ref{shiftdA}) 
which leads to a shift 
of the exterior derivative $(dA^e)_{x,\mu \nu}$ by $2 \pi \, (d k)_{x,\mu \nu}$. Note that this shift obeys 
$(d(dk))_{x,\mu \nu \rho} = 0$ due to the nil-potency of the exterior derivative operator $d$ , i.e., $d^2 = 0$ (compare Appendix \ref{app:diff_forms}). 
Thus we may constrain also the Villain variables $n_{x,\mu \nu}$ to obey
\begin{equation}
(d n)_ {x,\mu \nu \rho} \; = \; 0 \; \; \; \; \forall x \; , \; \; \mu < \nu < \rho \; .
\label{Villainconstraint}
\end{equation}
This constraint implies that for all 3-cubes $(x,\mu < \nu < \rho)$ of the lattice the oriented sum over the Villain variables
on the surface of the cube vanishes (Appendix \ref{app:diff_forms}). Using again the language of differential forms, 
the Villain variables are restricted to be a \emph{closed} integer-valued 2-form. We point out that this constraint implies 
the absence of monopoles as we will discuss in more detail below. 

Taking into account this closedness constraint we may formulate the partition sum of pure U(1) gauge theory in the form 
\begin{equation}
Z(\beta) \; \equiv \; \int D[A^e] \; \sum_{\{ n \}} \; e^{ \, -\frac{\beta}{2} \sum_{x,\mu < \nu} ( F^e_{x,\mu \nu} )^2 }
\; \prod_{x} \prod_{\mu < \nu < \rho}  \delta\big( (dn)_{x,\mu \nu \rho}\big) \; .
\label{Zdef1}
\end{equation}
Here we have defined the measure for the  gauge fields and the sum over all configurations of the 
Villain variables as
\begin{equation}
\int D[A^e] \; \equiv \; \prod_{x} \prod_{\mu} \int_{-\pi}^\pi \! \! \frac{d A^e_{x,\mu}}{2 \pi} \; \; \; , \; \; \; \; 
\sum_{\{ n \}}  \; \equiv \; \prod_{x} \prod_{\mu < \nu} \sum_{n_{x,\mu \nu} \in \mathds{Z}} \; .
\label{dAsumn}
\end{equation}
The product in (\ref{Zdef1}) runs over all 3-cubes $(x,\mu < \nu < \rho)$ and implements the closedness constraint 
(\ref{Villainconstraint}) using Kronecker deltas which we here denote with $\delta(j) \equiv \delta_{j,0}$.

It is useful to write the constraints in (\ref{Zdef1}) by using the Fourier representation of the Kronecker deltas,
such that
\begin{eqnarray}
\prod_{x} \!  \prod_{\mu < \nu < \rho}  \! \! \delta\Big( \! (dn)_{x,\mu \nu \rho}\!\Big) &  = & 
\prod_{x} \! \prod_{\mu < \nu < \rho}  \int_{-\pi}^\pi \! \frac{dA^m_{x,\mu \nu \rho}}{2\pi} \, 
e^{-i A^m_{x,\mu \nu \rho} (dn)_{x,\mu \nu \rho}} \nonumber 
\\
& = &
\int \!\! D[A^m] \, e^{-i \sum_x \! \sum_{\mu < \nu < \rho} A^m_{x,\mu \nu \rho} (dn)_{x,\mu \nu \rho}}  ,
\label{constraintintegral}
\end{eqnarray}
where we have introduced auxiliary fields $A^m_{x,\mu\nu\rho} \in \mathds{R}$ assigned to the cubes $(x,\mu \nu \rho)$ 
of the lattice which, when integrated over, generate the closedness constraint (\ref{Villainconstraint}) for the Villain variables. 
The notation $A^m$ is chosen 
to reflect the fact that in the electric-magnetic duality transformations we discuss below, 
the auxiliary field $A^m$ will take over the role of 
the vector potential, while in the dual form the vector field $A^e$ will generate the dual 
closedness constraints. Due to this role in the duality
transformation we will also use the nomenclature \emph{magnetic gauge field} for $A^m$, 
while $A^e$ is referred to as \emph{electric gauge field} 
(see above).

For notational convenience, 
in the last step of (\ref{constraintintegral}) we have introduced the integral over all configurations of the magnetic gauge 
fields $A^m_{x,\mu\nu\rho}$,
\begin{equation}
\int \!\! D[A^m] \; \equiv \, 
\prod_{x} \! \prod_{\mu < \nu < \rho}  \int_{-\pi}^\pi \! \frac{dA^m_{x,\mu \nu \rho}}{2\pi}\;  .
\end{equation}
We thus may write the partition sum of our discretization of the U(1) gauge field in the form,
\begin{equation}
Z(\beta) \; = \; \int \!\! D[A^e] \int \!\! D[A^m] \; B_\beta[A^e,A^m] \; ,
\end{equation}
where we have introduced the Boltzmann factor 
\begin{equation}
B_\beta[A^e,A^m]  \; \equiv \; \sum_{\{ n \}} \; e^{ \, -\frac{\beta}{2} \sum_{x} \! \sum_{\mu < \nu} ( F^e_{x,\mu \nu} )^2 } 
\; e^{\,- i \sum_x \! \sum_{\mu < \nu < \rho} A^m_{x,\mu \nu \rho} (dn)_{x,\mu \nu \rho}}  ,
\label{boltzmann_both}
\end{equation}
that depends on both, the electric gauge field $A^e$ that in the Boltzmann factor $B_\beta[A^e,A^m]$ describes the photon dynamics, 
as well as on the magnetic gauge field $A^m$  that here generates the constraints. As announced, in the duality transformation we 
discuss in the next subsection these two fields will interchange their role.  


\subsection{Proof of self-duality}

The first step towards establishing self-duality is to rewrite the exponent of the second term in the Boltzmann factor (\ref{boltzmann_both}),
\begin{eqnarray}
\sum_x \! \sum_{\mu < \nu < \rho} A^m_{x,\mu \nu \rho} (dn)_{x,\mu \nu \rho} & = & 
\frac{1}{2\pi} \sum_x \! \sum_{\mu < \nu < \rho} A^m_{x,\mu \nu \rho} \big( d (dA^e + 2\pi n) \big)_{x,\mu \nu \rho} 
\nonumber \\
& = & 
- \frac{1}{2\pi} \sum_x \! \sum_{\mu < \nu} (\partial A^m)_{x,\mu \nu}  (dA^e + 2\pi n)_{x,\mu \nu} \; , 
\label{constraintsexponent}
\end{eqnarray}
where in the first step we used $d^{\,2} = 0$ (see Appendix \ref{app:diff_forms}) to insert the term $dA^e$. Subsequently we applied 
the partial integration formula (\ref{AppA:partint}). The Boltzmann factor \eqref{boltzmann_both} thus can be written 
in the form of a product over all plaquettes (we here inserted $F^e_{x,\mu \nu} = (d A^e \, + \, 2 \pi \, n)_{x,\mu \nu}$),
\begin{equation}
B_\beta[A^e,A^m]  \; = \; \prod_{x} \prod_{\mu <\nu} \sum_{n_{x,\mu\nu} \in \mathds{Z}} 
e^{ \, -\frac{\beta}{2} \big( (d A^e \, + \, 2 \pi \, n)_{x,\mu \nu} \big)^2 } 
\; e^{\, \frac{i}{2\pi}  (dA^e \, + \, 2\pi n)_{x,\mu \nu}  (\partial A^m)_{x,\mu \nu} }  .
\label{boltzmann_both_2}
\end{equation}
Each factor in this product is $2\pi$-periodic in the corresponding variable 
$dA^e$, where the periodicity is generated by the sum over the 
Villain variable on the plaquette. Thus we may use Poisson resummation (See Appendix \ref{app:D} for the proof 
of a more general result that collapses to the usual Poisson resummation when setting $N = 1$.),
\begin{equation}
\sum_{n \in \mathds{Z}} e^{ \, -\frac{\beta}{2} (d A^e \, + \, 2 \pi \, n)^2 } 
\; e^{\,  \frac{i}{2\pi}  (dA^e \, + \, 2\pi n) (\partial A^m)} \; = \; 
\frac{1}{\sqrt{ 2 \pi \beta}} \sum_{p \in \mathds{Z}} e^{ \, -\frac{1}{2} \frac{1}{4 \pi^2 \beta} (\partial A^m \, + \, 2 \pi \, p)^2 } 
\; e^{\,- i \, p  \, dA^e } \; ,
\end{equation}
where we have omitted the plaquette indices for notational convenience. Using this expression for all factors in the Boltzmann weight
(\ref{boltzmann_both_2}), we find
\begin{equation}
B_\beta[A^e,A^m]  \; = \; \left(\!\frac{1}{2\pi \beta}\!\right)^{\!3V} \sum_{\{ p \}} \; e^{ \, -\frac{\widetilde{\beta}}{2} 
\sum_{x,\mu < \nu} \big( (\partial A^m \, + \, 2 \pi \, p)_{x,\mu \nu} \big)^2 } 
\; e^{\,- \, i \sum_{x,\mu < \nu} (dA^e)_{x,\mu \nu} \, p_{x,\mu \nu}}  ,
\label{boltzmann_both_3}
\end{equation}
where we have defined the dual gauge coupling
\begin{equation}
\widetilde{\beta} \; \equiv \; \frac{1}{4 \pi^2 \beta} \; ,
\end{equation}
and denote the sum over all configurations of the newly introduced plaquette occupation numbers $p_{x,\mu \nu} \in \mathds{Z}$ as
\begin{equation}
\sum_{\{ p \}} \; \equiv \; \prod_{x} \prod_{\mu <\nu} \sum_{p_{x,\mu\nu} \in \mathds{Z}} \; .
\end{equation} 
The next step is to use again the partial integration formula (\ref{AppA:partint}) from Appendix \ref{app:diff_forms} to rewrite the second exponent
in (\ref{boltzmann_both_3}) such that the Boltzmann factor reads
\begin{equation}
B_\beta[A^e,A^m]  \; = \; \left(\!\frac{1}{2\pi \beta}\!\right)^{\!3V} \sum_{\{ p \}} \; e^{ \, -\frac{\widetilde{\beta}}{2} 
\sum_{x,\mu < \nu} \big( (\partial A^m \, + \, 2 \pi \, p)_{x,\mu \nu} \big)^2 } 
\; e^{\, -i \sum_{x, \mu} A^e_{x,\mu} \, (\partial \, p)_{x,\mu}}  .
\label{boltzmann_both_4}
\end{equation}

We remark that we will use the form (\ref{boltzmann_both_4}) of the Boltzmann factor to show that the self-dual formulation of 
scalar electrodynamics that we construct in the next section is free of any complex action problem. The self-dual formulation of
pure gauge theory is already free of a complex action problem when the form (\ref{Zdef1}) of the partition sum is used.

To complete the proof of self-duality of pure U(1) gauge theory, we now switch to the dual lattice. 
Using (\ref{AppA:dualforms}) we identify the $r$ forms on the original lattice with $4-r$ forms on the dual lattice
(when forms are considered on the dual lattice they are marked with ``$\widetilde{\;\;\;}$'' - compare Appendix \ref{app:diff_forms}),
\begin{equation}
A^e_{x,\mu}  = \! \!\! \sum_{\nu<\rho<\sigma} \!\! \epsilon_{\mu\nu\rho\sigma} \, 
\widetilde{A}^e_{\tilde{x} - \hat{\nu} - \hat{\rho} - \hat{\sigma},\nu\rho\sigma} \, , \; 
A^m_{x,\mu\nu\rho} =  \sum_{\sigma} \! \epsilon_{\mu\nu\rho\sigma} \, \widetilde{A}^m_{\tilde{x} - \hat{\sigma},\sigma}
\, , \; 
p_{x,\mu\nu} = \! \sum_{\rho<\sigma} \! \epsilon_{\mu\nu\rho\sigma} \, 
\widetilde{p}_{\tilde{x} - \hat{\rho} - \hat{\sigma},\rho\sigma} \, .
\label{convertdual1}
\end{equation}
The $d$ and $\partial$ operators interchange their role when switching to the dual lattice 
(see Equation (\ref{AppA:diffconvert}) in Appendix~\ref{app:diff_forms}), such that 
\begin{equation}
(\partial \, A^m)_{x,\mu\nu} \; = \, \sum_{\rho<\sigma} \! \epsilon_{\mu\nu\rho\sigma} \, 
(d \widetilde{A}^m)_{\tilde{x} - \hat{\rho} - \hat{\sigma},\rho\sigma} \; , \; 
(\partial \,  p)_{x,\mu} \; = \sum_{\nu<\rho<\sigma} \!\! \epsilon_{\mu\nu\rho\sigma} \, 
(d \, \widetilde{p}\,)_{\tilde{x} - \hat{\nu} - \hat{\rho} - \hat{\sigma},\nu\rho\sigma} \; .
\label{convertdual2}
\end{equation}
Using (\ref{convertdual1}) and (\ref{convertdual2}) in (\ref{boltzmann_both_4}) we find the following dual form of the 
Boltzmann factor,
\begin{eqnarray}
B_\beta[A^e,A^m]  & =  &  \left(\!\frac{1}{2\pi \beta}\!\right)^{\!\!3V} \!\! \sum_{\{ \widetilde{p} \}} e^{ \, -\frac{\widetilde{\beta}}{2} 
\sum_{\tilde{x},\mu < \nu} \! \big( \! (d \widetilde{A}^m \, + \, 2 \pi \, \widetilde{p}\,)_{\tilde{x},\mu \nu} \big)^2 } 
\, e^{\,- i \sum_{\tilde{x}, \mu < \nu < \rho} \widetilde{A}^e_{\tilde{x},\mu\nu\rho} \, (d \, \widetilde{p}\,)_{\tilde{x},\mu\nu\rho}} 
\nonumber  \\
& =  &  \left(\!\frac{1}{2\pi \beta}\!\right)^{\!\!3V} \!\! \sum_{\{ \widetilde{p} \}} e^{ \, -\frac{\widetilde{\beta}}{2} 
\sum_{\tilde{x},\mu < \nu} \! \big(  \widetilde{F}^m_{\tilde{x},\mu \nu} \big)^2 } 
e^{\,- i \sum_{\tilde{x}, \mu < \nu < \rho} \widetilde{A}^e_{\tilde{x},\mu\nu\rho} \, (d \, \widetilde{p}\,)_{\tilde{x},\mu\nu\rho}} , 
\label{boltzmann_both_5}
\end{eqnarray}
where we defined 
\begin{equation}
\widetilde{F}^m_{\tilde{x},\mu \nu} \; \equiv \; (d \widetilde{A}^m \, + \, 2 \pi \, \widetilde{p}\,)_{\tilde{x},\mu \nu} \; ,
\end{equation}
and the sum over all configurations of the dual plaquette occupation numbers,
\begin{equation}
\sum_{\{ \widetilde{p} \, \}} \; \; \equiv \;\; \prod_{\tilde{x}} \prod_{\mu <\nu} \; \sum_{\widetilde{p}_{\tilde{x},\mu\nu} \in \mathds{Z}} \;
= \;\; \sum_{\{ p \}}  \; ,
\end{equation} 
where the identity on the right hand side is an obvious consequence of the last equation in (\ref{convertdual1}). 
Thus, up to an overall factor the Boltzmann weights \eqref{boltzmann_both} and \eqref{boltzmann_both_5} have the same form. 

Comparing 
(\ref{boltzmann_both_5}) with (\ref{boltzmann_both}) we can summarize the duality relation for the Boltzmann factor as
\begin{equation}
B_\beta[A^e,A^m]  \; =  \; \left(\!\frac{1}{2\pi \beta}\!\right)^{\!\!3V}  \widetilde{B}_{\widetilde{\beta}}\big[\widetilde{A}^m,\widetilde{A}^e\big] 
\quad \mbox{with} \quad  \widetilde{\beta} \; = \; \frac{1}{4 \pi^2  \beta} \; .
\label{boltzmann_both_duality}
\end{equation}
Equation (\ref{boltzmann_both_duality}) 
constitutes the self-duality relation for the generalized Boltzmann factor $B_\beta[A^e,A^m]$, where in the lhs.\ form of the
Boltzmann factor the electric gauge field
$A^e$ describes the dynamics and the magnetic gauge field $A^m$  generates the constraints. Under the duality transformation the
gauge coupling $\beta$ is replaced by the dual coupling $\widetilde{\beta}$, the Boltzmann factor picks up the overall factor 
$(2\pi \beta)^{-3V}$, all fields are replaced by their dual living on the dual lattice 
(thus the notation $\widetilde{B}_{\widetilde{\beta}}[\widetilde{A}^m,\widetilde{A}^e]$), and finally, 
the electric and the magnetic field interchange their role. 
This means that in the dual form on the rhs.\ of  (\ref{boltzmann_both_duality}) the dual magnetic field $\widetilde{A}^m$ 
describes the dynamics, while the dual electric field  $\widetilde{A}^e$ now generates the constraints. 

To establish full duality of the partition sum $Z(\beta)$ we replace the integral measures $\int \! D[A^e]$ and 
$\int \! D[A^m]$ by the corresponding dual integral measures defined as,
\begin{eqnarray}
&& \int \!\! D\big[\widetilde{A}^e\big] \; \equiv \;
\prod_{\tilde{x}} \! \prod_{\mu < \nu < \rho}  \int_{-\pi}^\pi \! \frac{d\widetilde{A}^e_{\tilde{x},\mu \nu \rho}}{2\pi}\;  = \;  \int \!\! D[A^e]
\; , 
\nonumber \\
&&
\int \!\! D\big[\widetilde{A}^m\big] \; \equiv \; 
\prod_{\tilde{x}}  \prod_{\mu}  \int_{-\pi}^\pi \! \frac{d\widetilde{A}^m_{\tilde{x},\mu}}{2\pi} \;  = \;  \int \!\! D[A^m] \;  ,
\label{dualmeasures}
\end{eqnarray}
where again (\ref{convertdual1}) ensures that the dual integration measures are equal to the ones on the original lattice.  Thus we find
\begin{eqnarray}
Z(\beta) & = & \int \!\! D[A^e] \int \!\! D[A^m] \; B_\beta[A^e,A^m] \nonumber \\
& = &  \left(\!\frac{1}{2\pi \beta}\!\right)^{\!\!3V} \!\! 
\int \!\! D[\widetilde{A}^e] \int \!\! D[\widetilde{A}^m] \; \widetilde{B}_\beta\big[\widetilde{A}^m,\widetilde{A}^e\big] \; = 
 \left(\!\frac{1}{2\pi \beta}\!\right)^{\!\!3V} \!\! Z(\widetilde{\beta}) \; , 
\end{eqnarray}
and identify the final form of the self-duality relation for the partition sum
\begin{equation}
Z(\beta) \; = \;  \left(\!\frac{1}{2\pi \beta}\!\right)^{\!\!3V} \!\! Z(\widetilde{\beta}) \qquad \mbox{with} \qquad 
\widetilde{\beta} \; = \; \frac{1}{4 \pi^2 \beta} \; .
\label{duality_Z}
\end{equation} 
Before we discuss properties and consequences of the self-duality relation (\ref{duality_Z}) we note that iterating the duality
relation gives the identity map, i.e.,
\begin{equation}
Z(\beta) \; = \; \left(\!\frac{1}{2\pi \beta}\!\right)^{\!\!3V} \!\! Z\big(\widetilde{\beta}\big) 
\; = \;
\left(\!\frac{1}{2\pi \beta}\!\right)^{\!\!3V} \left(\frac{1}{2\pi \widetilde{\beta}}\!\right)^{\!\!3V} \!\! Z\Big(\,\widetilde{\!\widetilde{\beta}}\,\Big) \; = \; 
Z(\beta) \; ,
\label{dualityiter}
\end{equation}
where in the last step we used the obvious properties $\beta \widetilde{\beta} = 1/4\pi^2$ and $\widetilde{\!\widetilde{\beta}} = \beta$.
Equation (\ref{dualityiter}) constitutes an important consistency check for the self-duality relation we constructed.

The self-duality relation (\ref{duality_Z}) obviously maps the weak- and strong-coupling regions of the partition sum $Z(\beta)$ 
onto each other. Suitable derivatives of $\ln Z(\beta)$ thus will relate observables in the strong and the weak coupling region. 
To give an example, we consider the expectation value of the square of the field strength, which is proportional to the first derivative 
of  $\ln Z(\beta)$:
\begin{eqnarray}
\langle F^2 \rangle_\beta & \equiv & - \, \frac{1}{3V} \frac{\partial}{\partial \beta} \ln Z(\beta) \; = \;  - \, \frac{1}{3V} \frac{\partial}{\partial \beta}
\ln \left( (2\pi \beta)^{-3V} Z\big (\widetilde{\beta}) \right) 
\nonumber \\
& = & 
\frac{1}{\beta} \, - \, \frac{1}{3V} \left( \frac{\partial}{\partial \widetilde{ \beta}}  \ln Z(\widetilde{ \beta}) \right) 
\frac{ d \widetilde{\beta}}{d \beta}
\; = \; \frac{1}{\beta} \; - \; \langle F^2 \rangle_{\widetilde{\beta}} \; \frac{1}{4 \pi^2 \beta^2} \; . 
\end{eqnarray}
Multiplying the equation with $\beta$ we can summarize the duality relation for $\langle F^2 \rangle$ in a sum rule that connects
the  weak and strong coupling results for $\langle F^2 \rangle$,
\begin{equation}
\beta \, \langle F^2 \rangle_\beta \; + \; \widetilde{\beta} \, \langle F^2 \rangle_{\widetilde{\beta}} \;  = \; 1\; .
\label{sumruleF2}
\end{equation}
In a similar way one may relate the weak and strong coupling results of higher derivatives of $\ln Z(\beta)$, i.e., the susceptibility of the 
action density and higher moments. Self-duality relations for correlators of $F^2$ can be obtained by introducing an $x$-dependence of 
$\beta$ (which leaves the duality transformation unchanged) and by performing local derivatives that generate the correlators.


\section{Self-dual lattice QED}

In this section we generalize our self-dual discretization of pure gauge theory to self-dual lattice QED, where we explicitly discuss the coupling 
of electrically and magnetically charged scalar fields. More particularly we first discuss in detail the case where we couple separate species of 
matter fields, one that is electrically charged and a second field with magnetic charges. Subsequently we briefly address the possibility of coupling 
dyonic matter in our formulation. 

\subsection{Coupling separate species for electric and magnetic matter}

We here first present the construction for coupling separate species of electric and magnetic matter fields, then establish
self-duality and finally discuss some of its consequences. 
We remark that in Appendix \ref{app:worldline} we show that one may switch to a worldline formulation that for the case of bosonic matter solves 
the complex action problem introduced by the closedness constraints, such that the worldline form can be used for numerical simulations.

We couple electrically charged matter to the electric gauge field $A^e$ and magnetically charged matter to the magnetic gauge field
$A^m$. Note that we have defined the magnetic gauge field $A^m$ on the cubes of the original lattice, such that the corresponding 
dual form $\widetilde{A}^m$ corresponds to fields $\widetilde{A}^m_{\tilde{x},\mu}$ that live on the links $(\tilde{x},\mu)$ of the 
dual lattice (compare \eqref{convertdual1}).  We now use this dual form to couple the magnetically charged matter fields. 

The partition function with gauge fields coupling to electric and magnetic matter fields reads
\begin{equation}
Z(\beta, M^e\!,\lambda^e\!,q^e\!,\,M^m\!,\lambda^m\!,q^m) \equiv \int \!\! D[A^e] \! \int \!\! D[A^m] \; B_\beta[A^e,A^m] \;
Z_{M^e\!,\,\lambda^e\!,\,q^e} [A^e] \; \widetilde{Z}_{M^m\!,\,\lambda^m\!,\,q^m} \big[\widetilde{A}^m\big] \, ,
\label{Zfull}
\end{equation}
where $B_\beta[A^e,A^m]$ is the gauge field Boltzmann factor in the form (\ref{boltzmann_both}).  We have introduced 
the partition function $Z_{M^e\!,\,\lambda^e\!,\,q^e} [A^e]$ for electrically charged matter $\phi^e$ that couples to the electric 
background gauge field $A^e$ as
\begin{eqnarray}
\hspace*{-5mm} Z_{M^e\!,\,\lambda^e\!,\,q^e} [A^e] & \!\! \equiv  \!\! &  
\int \!\! D[\phi^e] \, e^{-S_{M^e\!,\,\lambda^e\!,\,q^e}[\phi^e,A^e]} \; , \; \; 
\int \!\! D[\phi^e] \; \equiv \; \prod_x \int_{\mathds{C}} \frac{d \phi_x^e}{2 \pi} \; ,
\label{Zelectric}
\\
\hspace*{-5mm} S_{M^e\!,\,\lambda^e\!,\,q^e}[\phi^e,A^e] &  \!\! \equiv  \!\! & 
\sum_x \left[ M^e | \phi_x^e|^2 + \lambda^e | \phi_x^e|^4 - \sum_\mu 
\Big[ \phi_x^{e\, *} \,  e^{\, i \, q^e  A_{x,\mu}^e} \, {\phi_{x+\hat{\mu}}^e} + c.c.\ \Big] \! \right] \; ,
\label{Selectric}
\end{eqnarray}
where we imparted the electric matter with a charge $q^e\in \mathbb Z$.
As already announced, we couple bosonic matter, which for the electric field is a complex scalar $\phi_x^e \in \mathds{C}$ that is 
assigned to the sites $x$ of the original lattice. The path integral measure $\int \!\! D[\phi^e]$ is the usual product measure and the 
action $S_{M^e\!,\,\lambda^e\!,\,q^e}[\phi^e,A^e]$ is the free action plus a quartic term with coupling $\lambda^e$. 
The mass parameter 
$M^e$ is related to the tree level mass $m^e$ via $M^e = 8 + (m^e)^2$. 

The magnetically charged scalar field $\widetilde{\phi}_{\tilde{x}}^m \in \mathds{C}$ 
lives on the sites $\tilde{x}$ of the dual lattice and couples to 
the magnetic gauge field $\widetilde{A}_{\tilde{x},\mu}^m$ on the links of the dual lattice. 
The corresponding partition sum has the same form 
as the partition sum (\ref{Zelectric}) for the electric matter but for the magnetic matter is defined entirely on the dual lattice,
\begin{eqnarray}
\hspace*{-5mm} \widetilde{Z}_{M^m\!,\,\lambda^m\!,\,q^m} \big[\widetilde{A}^m\big] &  \!\! \equiv  \!\! &  
\int \!\! D\big[\widetilde{\phi}^m\big] \, 
e^{-\widetilde{S}_{M^m\!,\,\lambda^m\!,\,q^m} \big[\widetilde{\phi}^m,\widetilde{A}^m\big]} \; , \; \; 
\int \!\! D \big[\widetilde{\phi}^m\big] \; \equiv \; \prod_{\tilde{x}} \int_{\mathds{C}} \frac{d \widetilde{\phi}_{\tilde{x}}^m}{2 \pi} \; ,
\label{Zmagnetic}
\\
\hspace*{-5mm}  \widetilde{S}_{M^m\!,\,\lambda^m\!,\,q^m} \big[\widetilde{\phi}^m,\widetilde{A}^m\big] &  \!\!  \equiv   \!\! & 
\sum_{\tilde{x}} \left[ M^m | \widetilde{\phi}_{\tilde{x}}^m|^2 + \lambda^m | \widetilde{\phi}_{\tilde{x}}^m|^4 - \sum_\mu 
\Big[  \widetilde{\phi}_{\tilde{x}}^{m \, *} \,  e^{ \, i \, q^m \widetilde{A}_{\tilde{x},\mu}^m}  \, \widetilde{\phi}_{\tilde{x}+\hat{\mu}}^m + c.c.\ \Big] \! \right] .
\label{Smagnetic}
\end{eqnarray}
Similarly to the electric matter, we chose to give the magnetic matter a charge $q^m\in\mathbb Z$.
As before we allow for a quartic self-interaction term with corresponding coupling $\lambda^m$, and the relation of the 
mass parameter $M^m$ to the tree level mass $m^m$ is again given by $M^m = 8 + (m^m)^2$. 
  
It is obvious that the construction can easily be generalized to coupling fermionic electric and magnetic matter fields, by simply 
replacing the partition sums $Z_{M^e\!,\,\lambda^e\!,\,q^e} [A^e]$ and 
$\widetilde{Z}_{M^m\!,\,\lambda^m\!,\,q^m} \big[\widetilde{A}^m\big]$ by the corresponding 
fermion determinants in the background fields $A^e$ and $\widetilde{A}^m$ where for the latter case the corresponding discretized 
lattice Dirac operator lives entirely on the dual lattice.

The proof of self-duality is straightforward and is essentially a corollary of the self-duality relation (\ref{boltzmann_both_duality}) 
for the gauge field Boltzmann factor $B_\beta[A^e,A^m]$. Using (\ref{boltzmann_both_duality}) in the partition sum (\ref{Zfull}) we find
(here $C = (1/2\pi \beta)^{3V}$)
\begin{eqnarray}
Z(\beta, M^e\!,\lambda^e\!,q^e\!,\,M^m\!,\lambda^m\!,q^m) \!\!\!&\!\!\!\! = \!\!\!\!\!\!&   C \!\!
\int \!\! D[A^e] \!\! \int \!\! D[{A}^m]  B_{\widetilde{\beta}}[\widetilde{A}^m\!,\widetilde{A}^e] 
Z_{M^e\!,\,\lambda^e\!,\,q^e\!} [A^e]  \widetilde{Z}_{M^m\!,\,\lambda^m\!,\,q^m\!}  \big[\!\widetilde{A}^m \big] 
\nonumber  \\
&\!\!\!\! = \!\!\!\!\!\!&   C \!\!
\int \!\! D[\widetilde{A}^m] \!\! \int \!\! D[\widetilde{A}^e]  B_{\widetilde{\beta}}[\widetilde{A}^m\!,\widetilde{A}^e] 
\widetilde{Z}_{M^m\!,\,\lambda^m\!,\,q^m\!}  \big[\!\widetilde{A}^m \big] 
\widetilde{\!\widetilde{Z}}_{M^e\!,\, \lambda^e\!, -q^e}\! \Big[\widetilde{\widetilde{A}^e\!} \Big] , 
\nonumber  \\
&& \hspace{-30mm}
\label{Zfull_aux}
\end{eqnarray}
where in the second step we have used (\ref{dualmeasures}) to replace the path integral measures by their dual 
counterparts, as well as the identity 
\begin{equation}
Z_{M^e\!,\,\lambda^e\!,\,q^e\!} [A^e]  \; = \; 
\widetilde{\!\widetilde{Z}}_{M^e\!,\,\lambda^e\!,\,-q^e\!} \Big[\widetilde{\widetilde{A}^e\!} \, \Big] \; .
\end{equation}
It is important to note that here the iterated duality relation gives rise to a flip of the sign of the electric charge $q^e$: 
For vector fields Eq.~\eqref{AppA:dualforms} from Appendix~\ref{app:diff_forms} implies
\begin{equation}
A^e_{x,\mu} \; =\; 
\sum_{\nu < \rho < \sigma} \epsilon_{\mu\nu\rho\sigma} \tilde A^e_{\tilde x- \hat\nu - \hat \rho - \hat \sigma,\nu\rho \sigma} 
\; , \; \;
\tilde A^e_{\tilde x,\mu\nu\rho} \; =\; - \, \sum_\sigma \epsilon_{\mu\nu\rho\sigma} A_{x+\hat s-\hat \sigma,\sigma} \; ,
\end{equation}
where $\hat s=\hat 1+\hat 2+\hat 3+\hat4$. As a consequence we find $\widetilde{\widetilde{A}^e\!}_{x,\mu} =  - A^e_{x,\mu}$, 
such that the link variables $e^{\, i q^e A_{x,\mu}}$ in the electric mater action \eqref{Selectric} become complex conjugate under 
iterated duality, which in turn is equivalent to changing the sign of the electric charge $q^e$. 

Comparing the second line of (\ref{Zfull_aux}) with the definition (\ref{Zfull}) of the full partition sum we can identify
the self-duality relation for our discretization of QED
\begin{eqnarray}
\hspace*{-5mm} 
&& \beta^{\frac{3V}{2}} Z(\beta, M^e\!,\lambda^e\!,q^e\!,\,M^m\!,\lambda^m\!,q^m) \; = \;  \widetilde{\beta}^{\frac{3V}{2}} \,
Z\big(\widetilde{\beta}, \widetilde{M}^e\!,\widetilde{q}^{\,e}\!, 
\widetilde{\lambda}^e\!,\widetilde{M}^m\!,\widetilde{\lambda}^m\!,\widetilde{q}^{\,m}\big) \qquad \mbox{with} 
\label{Zfullduality} \\
\hspace*{-5mm}  &&  
\widetilde{\beta} \, = \, \frac{1}{4\pi^2 \beta} \, , \;
\widetilde{M}^e \, = \, M^m  \, , \;  
\widetilde{\lambda}^e  \, = \, \lambda^m \, , \; 
\widetilde{q}^{\, e}  \, = \, q^m \, , \; 
\widetilde{M}^m  \, = \, M^e \, , \; 
\widetilde{\lambda}^m  \, = \, \lambda^e \, , \;  
\widetilde{q}^{\; m}  \, = \, - \, q^e \; .
\nonumber
\end{eqnarray}
Note that we have split the prefactor $C = (1/2\pi \beta)^{3V}$ of the partition sums in \eqref{Zfull_aux}
by dividing it into equal powers of $\beta$ and $\widetilde{\beta}$ to fully display the symmetry.
As for the case of pure gauge theory, the duality transformation generates an overall factor, and the gauge coupling 
$\beta$ is replaced by the dual coupling $\widetilde{\beta} = 1/(4\pi^2 \beta)$, thus interchanging weak and strong coupling. 
In addition the parameters $M^e, \lambda^e$ of electric matter are interchanged with the parameters 
$M^m, \lambda^m$ of magnetic matter\footnote{When coupling fermionic matter only the bare fermion masses appear as 
parameters.}. For the charge vector $\bm q = (q^e,q^m)^t$ (the superscript $^t$ denotes transposition) we find the following relation between 
the dual and the original charges,
\be
\label{eq:S_matrix}
\widetilde{\bm q} \; = \; S \, {\bm q} \quad \mbox{with} \quad
S \; =\; \begin{pmatrix}
0 & 1\\
-1 &0
\end{pmatrix} \; .
\ee
The self-duality symmetry described by \eqref{Zfullduality} and \eqref{eq:S_matrix}
will be referred to as ${\cal S}$. However, we remark already now that 
in the next section we will extend the symmetry ${\cal S}$ to also include the $\theta$-angle shifts by $2\pi$, the so called $\cal T$ transformation. 

Again we can generate self-duality relations for observables by evaluating derivatives 
of $\ln Z$ with respect to the couplings and when the 
couplings are considered to be space-time dependent suitable derivatives give rise to self-duality relations
for correlation functions. We discuss two examples 
for such self-duality relations, the first being the generalization of the sum rule (\ref{sumruleF2}),
\begin{equation}
\beta \, \langle (F^e)^2 \rangle_{\beta, M^e\!,\,\lambda^e\!,\,q^e\!,\,M^m\!,\,\lambda^m\!,\,q^m} \; + \; \widetilde{\beta} \, 
\langle (F^e)^2 \rangle_{\widetilde{\beta}, \widetilde{M}^e\!,\,\widetilde{q}^{\,e}\!,\,\widetilde{\lambda}^e\!,
\,\widetilde{M}^m\!,\,\widetilde{\lambda}^m\!,\,\widetilde{q}^{\;m}} \;  = \; 1\; .
\label{sumruleF2_qed}
\end{equation}
The second term in the sum rule is evaluated with the dual parameters, i.e., with gauge coupling $\widetilde{\beta}$ and interchanged 
electric and magnetic coupling parameters. 

Derivatives with respect to $M^e$ and $M^m$ generate field expectation values for the electric and magnetic matter fields. Exploring 
the duality relation (\ref{Zfullduality}) one finds 
\begin{eqnarray}
\left\langle | \phi^e |^2 \right\rangle_{\beta, M^e\!,\,\lambda^e\!,\,q^e\!,\,M^m\!,\,\lambda^m\!,\,q^m} & \equiv &
- \, \frac{1}{V} \frac{\partial }{\partial M^e} \, \ln Z(\beta, M^e\!,\lambda^e\!,q^e\!,\,M^m\!,\lambda^m\!,q^m)  
\nonumber \\
& = & 
 \left\langle | \widetilde{\phi}^m |^2 \right\rangle_{\widetilde{\beta}, \widetilde{M}^e\!,\,\widetilde{q}^{\,e}\!,\,\widetilde{\lambda}^e\!,
\,\widetilde{M}^m\!,\,\widetilde{\lambda}^m\!,\,\widetilde{q}^{\;m}}  \; ,
\label{duality_phi2}
\end{eqnarray}
with $\left\langle | \widetilde{\phi}^m |^2 \right\rangle = - 1/V \partial \ln Z / \partial M^m $.
According to the self-duality relation (\ref{duality_phi2}), the electric and magnetic field expectation 
values are converted into each other when 
changing from weak to strong coupling and simultaneously interchanging electric and 
magnetic coupling parameters. In a similar 
way self-duality relations for various observables and correlation functions can be obtained, 
where electric and magnetic fields and their 
parameters are interchanged, when switching between the weak and strong coupling 
domains of the theory. We stress once more, that the
self-duality relations we discuss here hold identically for fermionic and for bosonic matter. 

We remark again that the Boltzmann factor \eqref{boltzmann_both} which 
we use in \eqref{Zfull} is complex, such that we cannot use Monte Carlo simulations
for self-dual QED directly in the form \eqref{boltzmann_both}, \eqref{Zfull}. This can be 
overcome by switching to a worldline representation, which we discuss in Appendix \ref{app:worldline}.

\subsection{Comments on coupling dyons}
\label{sec:dyonic_theories}

In addition to purely electric and purely magnetic matter, one can also couple dyonic matter by identifying the lattice and its dual, e.g., 
by specifying a map $F: \Lambda\rightarrow \tilde\Lambda$ which sends $x\in\Lambda$ to $x+\frac{\hat s}{2}\in \tilde\Lambda$. 
We also define the map $\tilde F:\tilde \Lambda\rightarrow \Lambda$ which sends $\tilde x\in\tilde \Lambda$ to $\tilde x+\frac{\hat s}{2}\in\Lambda$. 
The maps $F$ and $\tilde F$ do not have to necessarily be shifts by the same amount, and we could have just as well defined 
$\tilde F: \tilde x \rightarrow \tilde x-\frac{\hat s}{2}$. We now work with only a single species of matter, the dyonic field $\phi_x \in \mathds{C}$,
that lives on the sites $x$ of the lattice and carries both an electric charge $q^e$ and a magnetic charge $q^m$. 
The corresponding action is obtained by replacing $\phi^e_x \rightarrow \phi_x$ in the mass and quartic terms of \eqref{Selectric} and 
by replacing the hopping terms as follows,
\be
\phi_x^{e \, *} \, e^{\, i \, q^e \, {A^e_{x,\mu}}} \, \phi^e_{x+\hat \mu} \; \rightarrow \; \phi_x \, e ^{\, i \, q^e\,{A^e_{x,\mu}} \, + \, i \, q^m \, 
{\tilde A^m_{F(x),\mu}}} \, \phi_{x+\hat \mu}\;.
\label{dyonhop}
\ee
The field $\phi^m$ from the previous section is dropped completely\footnote{We remark that we could have placed the dyonic field also on the dual lattice by 
coupling the dual of the electric field there.}. Under a duality transformation the dyon field is mapped to the dual lattice, and since in the construction here we 
identify original and dual lattice, the duality relations \eqref{Zfullduality} apply also to the dyonic case, of course with the modification that we have only a
single mass parameter $M$ and only a single quartic coupling $\lambda$. As in the previous subsection, also here the construction  
carries over to fermionic dyon matter in an obvious way.

Notice, however, that coupling dyons makes it difficult to eliminate the sign problem even if they are bosonic. In the previous subsection and Appendix 
\ref{app:worldline} the complex action problem was solved by switching to a worldline formulation for the magnetic matter, which converts the 
dependence of the magnetic matter action on the magnetic gauge field $\widetilde{A}^m$ into terms of the form 
$e^{i \, q^m \, \widetilde{A}^m_{\tilde x,\mu} \, \widetilde{k}_{\tilde x,\mu}}$ where the 
$\widetilde{k}_{\tilde x,\mu} \in \mathds{Z}$ are the flux variables that describe the dynamics 
of the magnetic matter in the worldline language (see Appendix \ref{app:worldline}). In this form the $\widetilde{A}^m$  then can be integrated out together 
with the constraint factors \eqref{constraintintegral} to generate more general constraints that couple the Villain and the flux variables. However, when dyons
are considered, the worldline expansion also generates terms of the form $e^{i \, q^e \, A^e_{x,\mu} \, {k}_{x,\mu}}$ that can not be integrated out, since the 
action for the electric gauge field $A^e$ is quadratic. Further the $\mathcal S$-duality transformation does not change the situation, 
since it merely interchanges the form of 
the electric and the magnetic gauge action, i.e., one of the two is always quadratic.  Hence dyons generically have a complex action problem. 

However, it is well known that a $\theta$-term can add or remove an electric charge from a monopole, which is the famous Witten effect \cite{Witten:1979ey}, 
and we will exploit this effect to relate some dyonic theories to their simpler cousins discussed in the previous subsection, 
where indeed a worldline representation 
solves the complex action problem. We will review the Witten effect and its relation to the complex action problem in the next section when we introduce the 
$\theta$-term for the Villain action. For the moment let us just note that when we change $\theta\rightarrow \theta \pm 2\pi$ we can absorb this shift by 
changing the dyonic charges $q^e \rightarrow q^e \mp q^m$ and $q^m \rightarrow q^m$. 
Obviously this can be used to relate theories with sign problem to theories that can be
simulated without the sign problem. This idea becomes more powerful when the shift of $\theta$ is combined with the duality transformation such that 
the group of transformations becomes $\SL(2,\mathbb Z)$. We will see that this allows us to connect some theories which have 
dyons\footnote{As we will see, such dyonic theories in general have smeared electric and magnetic charges, as the introduction of the 
$\theta$-term will force the Witten effect to endow monopoles with electric charges which are generically spread across multiple lattice links.} 
and a nonzero $\theta$-term, to theories without the complex action problem. This will be discussed in detail in Section~\ref{sec:sign-problem}.

\section{QED with a $\theta$-term} 

We generalize our formulation of self-dual U(1) lattice gauge theory and self-dual lattice QED further by adding a $\theta$-term. We 
first construct a suitable discretization of the topological charge and analyze its properties. Subsequently we add the $\theta$-term 
to the gauge field Boltzmann factor and identify the corresponding duality transformation. In yet another generalization step we
construct a self-dual Boltzmann factor at finite $\theta$ for the 
gauge fields, which then is the basis for fully self-dual U(1) lattice gauge theory and self-dual lattice QED 
both now containing also the $\theta$-term. 

Before we go through these steps, let us comment on our construction. It is well known that there are many ways to discretize a $\theta$-term. We will, 
however, insist that the $\theta$-term capturea the topological aspecta of the continuum U(1) theory. In the continuum the $\theta$-term of the U(1) gauge 
theory is
\be
S_\theta \; = \; {i\theta} \, Q \; ,
\ee
where
\be
Q \;  = \; \frac{1}{8\pi^2}\int \! d^4x \sum_{\mu < \nu \atop \rho < \sigma} F_{\mu\nu} F_{\rho\sigma}\epsilon^{\mu\nu\rho\sigma} \; ,
\ee
is the topological charge. It is well known that the $\theta$-term does not depend on the local details of the gauge field $A_\mu$, and only 
depends on the topological class of the gauge bundle. In other words, the variation of the above action with respect to small local changes of the gauge field $A_{\mu}$ is zero.

Moreover when $\theta$ is shifted $\theta\rightarrow \theta+2\pi$, the Euclidean path integral weight is unchanged in the continuum, 
because\footnote{This is only true on spin manifolds. On non-spin manifolds the $\theta$-periodicity is 
$\theta\rightarrow \theta+4\pi$ because $Q$ can be half-integer.} $Q\in \mathbb Z$. 

The above features are both ruined in the presence of a dynamical monopole\footnote{Such an operator is constructed by 
excising a contour $\mathcal C$ from the space-time manifold, and imposing that on a small sphere linking the contour $\mathcal C$ 
the flux is $\int F =2\pi q_m$, where $q_m\in \mathbb Z$ is the charge of the monopole operator.}. Indeed the Witten effect would endow the 
worldline of the monopole with electric charge $\theta/ 2\pi$, hence generating an explicit dependence on the gauge field $A_\mu$ 
along the monopole worldline, turning the monopole into a dyon. We will see a manifestation of this in our lattice construction below. 
The $\theta$-periodicity can be restored if the theory contains an entire tower of dyons of all electric charges, but generically it will 
be destroyed by the presence of dynamical magnetic matter.

To construct the desired $\theta$-term, we will therefore be guided by a theory without monopoles being gauge-invariant, independent of the 
local details of the gauge field $A_{x,\mu}$ and that has a $2\pi$-periodicity in $\theta$. To construct such a term it is sufficient to construct the 
appropriate topological charge $Q$. The most natural way to construct such a topological charge is to identify the lattice $\Lambda$ and its dual 
$\widetilde{\Lambda}$ by defining a map $\Lambda\rightarrow \widetilde{\Lambda}$, as was discussed in \cite{Sulejmanpasic:2019ytl}. Since such 
maps are not unique, the definition of the topological charge $Q$ will not be unique either, and the most general definition of the topological charge will 
not be local. All of these definitions will turn out to be equivalent when monopoles are absent, because of the topological nature of $Q$. 

We will see, however, that if such a $\theta$ term is introduced, any attempt to define an ultra-local kinetic term will fail to satisfy self-duality. 
In particular electric-magnetic duality transformation of such an ultra-local action would produce a local, but not ultra local action, and therefore 
violate the self-dual covariance. When we insist on self-duality we will be forced to consider local, but not ultra-local actions. 

It is well known that $\theta$-periodicity and self-duality generate an $\SL(2,\mathbb Z)$ duality group. Therefore, insisting on 
the exact $\SL(2,\mathbb Z)$ structure of a lattice theory will force us to abandon ultra-locality. 

\subsection{Construction of a suitable topological charge}
\label{subsec:topocharge}

Following \cite{Sulejmanpasic:2019ytl} we now introduce a topological charge for the Villain formulation but already 
generalize the discretization presented in \cite{Sulejmanpasic:2019ytl} to a form that will turn out to be useful 
for the construction of self-dual theories with a $\theta$-term. More specifically we 
introduce a whole family $Q_T[F^e]$ of lattice discretizations for the topological charge 
that are labelled by an integer-valued vector $T$,
\begin{eqnarray}
Q_T[F^e] & \equiv & \frac{1}{8 \pi^2} \sum_x \sum_{\mu < \nu \atop \rho < \sigma} 
F^e_{x,\mu \nu} \, \epsilon_{\mu \nu \rho \sigma} \, F^e_{x-\hat{\rho} - \hat{\sigma} +T, \,\rho \sigma} 
\label{QTdef}  \\
 & = & \frac{1}{8 \pi^2} \sum_x \sum_{\mu < \nu \atop \rho < \sigma} 
(dA^e + 2\pi n)_{x,\mu \nu} \, \epsilon_{\mu \nu \rho \sigma} \, (dA^e + 2\pi n)_{x-\hat{\rho} - \hat{\sigma} +T,\,\rho \sigma} \; ,
\nonumber
\end{eqnarray}
where we have introduced $T=\sum_\mu t_\mu \hat\mu$ with 
$t_\mu \in \mathds{Z}$. Note that $F^e_{x,\mu \nu}$ is again given by the combination
(\ref{Fdef}) of the exterior derivative and the Villain variable, i.e., $F^e_{x,\mu \nu} = (dA^e)_{x,\mu \nu} + 2\pi n_{x,\mu \nu}$.
We now want to show that the definition has the desired properties of the continuum topological charge, when magnetic monopoles are not present. Since monopoles are not present, we will assume that Villain variables obey the closedness constraints (\ref{Villainconstraint}). 
Thus the topological properties of  $Q_T[F^e]$ as we define it in (\ref{QTdef}) are specific to the case when monopoles are 
absent. 

For $T= 0$ the definition of $Q_0[F^e]$ is essentially a direct lattice discretization of the continuum topological charge 
$Q \propto \int d^4 x  F_{\mu \nu}(x) \epsilon_{\mu \nu \rho \sigma} F_{\rho \sigma}(x)$ with an additional shift
of the second factor $F^e_{x,\rho \sigma}$ by one lattice unit in the negative $\rho$ and $\sigma$ directions\footnote{Without this shift we would not be able to show the topological properties of $Q_T$, and the $\theta$-term in the absence of monopoles would lose the independence on 
$A_{x,\mu}^e$ and the $\theta$-periodicity.}. This, in the absence of monopoles, will have as a consequence that the 
$\theta$-term depends only on the villain variables $n_{x,\mu\nu}$ and not on gauge fields, and will also imply $\theta$ periodicity by $2\pi$. 
Following \cite{Sulejmanpasic:2019ytl}, this $\theta$-term may be interpreted as defining a map from a lattice to a dual lattice, and then making a natural product of $F^e_{x,\mu \nu}$ on a plaquette 
$(x,\mu \nu)$ with the corresponding dual field strength on the plaquette dual to $(x,\mu \nu)$. For $T \neq 0$ this
definition from \cite{Sulejmanpasic:2019ytl}
is generalized to a product of $F^e_{x,\mu \nu}$ with the dual field strength on the dual plaquette shifted by
$T$, which we implement by adding $T$ in the argument of the second factor in \eqref{QTdef}. Note that when 
$T$ becomes larger than the extent of the lattice the periodic boundary conditions will identify the shifted site 
$x +T$ with some site of the lattice, such that only a finite number of vectors $T$ correspond to 
distinct definitions of $Q_T[F^e]$.

We will now show, that $Q_T[F^e]$ is topological in nature and that $Q_T[F^e]$ gives the same result for all vectors $T$
as long as the Villain variables are closed. We proceed towards this goal in several steps where we discuss and use properties 
of $Q_T[F^e]$ that are partly proven in Appendix \ref{app:B}. 

The first property shown in Appendix \ref{app:B} is the fact that  $Q_T[F^e]$ is invariant under adding an arbitrary 
exterior derivative $(dB)_{x,\mu \nu}$ to the field strength $F^e_{x,\mu \nu}$, i.e.,
\begin{equation}
Q_T[F^e + d B] \; = \; Q_T[F^e] \; ,
\label{property1}
\end{equation}
where $B_{x,\mu}$ is an arbitrary 1-form, i.e., an arbitrary set of link-based fields. We stress again that this property only holds
when the Villain variables $n_{x,\mu \nu}$ in $F^e_{x,\mu \nu} = (dA^e)_{x,\mu \nu} + 2\pi \, n_{x,\mu \nu}$ obey the 
closedness condition (\ref{Villainconstraint}).

The property (\ref{property1}) immediately implies that $Q_T[F^e]$ is independent of the exterior derivative 
$(dA^e)_{x,\mu \nu}$ in $F^e_{x,\mu \nu} = (dA^e)_{x,\mu \nu} + 2\pi \, n_{x,\mu \nu}$. Thus the topological charge
depends only on the Villain variables $n_{x,\mu \nu}$ and we may write (compare the definition (\ref{QTdef}))
\begin{equation}
Q_T[F^e] \; = \; \frac{1}{2} \sum_x \sum_{\mu < \nu \atop \rho < \sigma} 
n_{x,\mu \nu} \, \epsilon_{\mu \nu \rho \sigma} \, n_{x-\hat{\rho} - \hat{\sigma} +T,\,\rho \sigma} \; \;
\mbox{with} \; \; (d n)_ {x,\mu \nu \rho}  = 0 \; \; \forall x,  \mu \! < \! \nu \! < \! \rho \, .
\label{Qnn}
\end{equation}  

The fact that the Villain variables obey the closedness condition (\ref{Villainconstraint}) can be used together with the 
Hodge decomposition (see Eq.~(\ref{AppA:hodge}) of Appendix \ref{app:diff_forms}) to write the Villain variables in the form
\begin{equation} 
n_{x,\mu \nu} \; = \; (d \, l)_{x,\mu \nu} \; + \; h_{x,\mu \nu} \; ,
\label{hodge}
\end{equation}
where the condition $(d\,n)_{x,\, \mu \nu \rho} = 0$ implies that we do not have a contribution
$(\partial \, c)_{x,\, \mu \nu \rho}$ in the Hodge decomposition (\ref{AppA:hodge}) of $n_{x,\mu \nu}$. In (\ref{hodge}) 
$l_{x,\mu} \in \mathds{Z}$ is an integer-valued 1-form which due to $d^2 l = 0$ does not contribute to 
$dn$. The second term in the Hodge decomposition (\ref{hodge}) are the closed integer-valued harmonic contributions 
$h_{x,\mu \nu}$ which obey $(dh)_{x,\mu\nu\rho} = 0$ and cannot be written as exterior derivatives. We 
may parameterize them in the form ($\mu < \nu$)
\begin{equation}
h_{x,\mu \nu} \; = \; \omega_{\mu \nu} \, \sum_{i=1}^{N_\rho} \sum_{j=1}^{N_\sigma} 
\delta_{x,i \hat\rho + j \hat\sigma}^{(4)} 
\; \; \; \; \; \mbox{with} \; \; \; \; \; \rho \neq \mu,\nu \, ; \, \sigma \neq \mu,\nu \, ; \, \rho \neq \sigma \; ,
\label{harmonics}
\end{equation}
where $N_\rho$ and $N_\sigma$ denote the lattice extents in the $\rho\,$- and $\sigma$-directions and 
$\delta_{x,i \hat\rho + j \hat\sigma}^{(4)}$ is the 4-dimensional Kronecker delta.
In other words $h_{x,\mu \nu}$ is the constant $\omega_{\mu \nu} \in \mathds{Z}$ for $\mu$-$\nu$ plaquettes  
$(x,\mu \nu)$ that have their root site $x$ in the $\rho$-$\sigma$ plane that is orthogonal to the $\mu$-$\nu$ plane 
and contains the origin. For all other plaquettes $h_{x,\mu \nu} = 0$. It is easy to see that the harmonics $h_{x,\mu \nu}$ 
are closed forms, i.e., they obey $(dh)_{x,\mu\nu\rho} = 0$. We stress that our choice for the parameterization of the $h_{x,\mu \nu}$ 
is not unique, since they can be deformed by adding arbitrary exterior derivatives. 

In Appendix \ref{app:B} we show the following result for the topological charge\footnote{In the notation for the final expression on  the rhs.\ of 
\eqref{Q_omega} we assume antisymmetry of the $\omega_{\mu \nu}$.} 
\begin{equation}
Q_T[2\pi n] =  Q_T[2\pi h] =  \omega_{12} \, \omega_{34}-  \omega_{13}  \omega_{24} + 
\omega_{14}  \omega_{23}  = \frac{1}{8} \epsilon_{\mu \nu \rho \sigma}  \omega_{\mu \nu} \omega_{\rho \sigma} 
 \in \mathds{Z} \;,\;\;\forall \; T \; .
\label{Q_omega}
\end{equation}  
The first identity follows already from (\ref{property1}) and shows that the topological charge depends only on the harmonic
contributions to the Villain variables. The second step is the explicit evaluation of  $Q_T[2\pi n] = Q_T[2\pi h]$ for the 
parameterization (\ref{harmonics}) of the harmonics, which, as shown in Appendix~\ref{app:B},  turns out to be independent of the vector $T$ that 
appears as a parameter in the definitions (\ref{QTdef}) and (\ref{Qnn}). This completes the proof of our statement that 
for closed Villain variables $Q_T[F^e] = Q_T[2\pi n] = Q_T[2\pi h]$ is independent of $T$ due to its topological nature. 
In addition the definition of the 
topological charge is integer-valued and can be computed uniquely from the harmonics in the Hodge decomposition 
of the Villain variables.

\subsection{The Witten effect}
\label{subsec:witten}

Before we come to performing the duality transformation in the next section let us discuss an interesting physical 
aspect of the abelian $\theta$-term, the Witten effect, which constitutes an important consistency check of our formulation. 
The Witten effect states that the $\theta$-term endows a magnetic monopole with the minimal possible magnetic charge 
$m = 2 \pi$ with an electric charge $q = \theta/2\pi$. 

We here consider a general definition of the topological charge given by the superposition of different discretizations 
$Q_T$,
\be\label{eq:Q_def}
Q \; = \; \sum_{T}\gamma_T \, Q_T[dA^e+2\pi n]\;,
\ee
where the $Q_T$ are given by \eqref{QTdef} and we require $\sum_T\gamma_T=1$.
Now consider a magnetic Wilson loop defined by
\be
\prod_{(\tilde x,\mu) \in \widetilde{C}} e^{\, i \,  \widetilde{A}^m_{\tilde x,\mu}} \; ,
\label{eq:magneticloop}
\ee
along some contour $\widetilde{C}$ on the dual lattice. Combining the Wilson loop with the closedness factor \eqref{constraintintegral}
and integrating over the $\widetilde{A}^m$ we find
\begin{eqnarray}
\hspace*{-8mm}&& \int \!\! D[\widetilde{A}^m] \, e^{-i \sum_x \! \sum_{\mu < \nu < \rho} A^m_{x,\mu \nu \rho} (dn)_{x,\mu \nu \rho}} \!\!
\prod_{(\tilde x,\mu) \in \widetilde{C}} \!\! e^{\, i \, \widetilde{A}^m_{\tilde x,\mu}} 
\\
\hspace*{-8mm} && \qquad = \; \int \!\! D[\widetilde{A}^m] \!\!
\prod_{(\tilde x,\mu) \in \widetilde{C}} \!\! e^{\, i \, \widetilde{A}^m_{\tilde x,\mu} 
[ 1 + \sum_{\nu<\rho<\sigma} \epsilon_{\mu\nu\rho\sigma}(dn)_{x+\hat\mu,\nu\rho\sigma}]} \!\!\!
\prod_{(\tilde x,\mu) \not\in \widetilde{C}} \!\!
e^{\, i \, \widetilde{A}^m_{\tilde x,\mu} \sum_{\nu<\rho<\sigma} \epsilon_{\mu\nu\rho\sigma}(dn)_{x+\hat\mu,\nu\rho\sigma} } ,
\nonumber
\end{eqnarray}
where in the second step we rewrote the $A^m$ in terms of the dual fields $\widetilde{A}^m$. Obviously this integral 
imposes the new constraint
\be
\sum_{\nu<\rho<\sigma}\epsilon_{\mu\nu\rho\sigma}(dn)_{x+\hat\mu,\nu\rho\sigma} \; = \; \begin{cases}
-1 &\text{if $(\tilde x,\mu)\in \widetilde C$} \; \; ,\\
0 &\text{if $(\tilde x,\mu) \notin \widetilde C$ \; .}
\end{cases} 
\label{modconstraint}
\ee
This constraint modifies the condition \eqref{Villainconstraint},  which in its original form ensures the complete absence of monopoles, while now 
along the contour $\widetilde C$ of the magnetic Wilson loop \eqref{eq:magneticloop} monopole charges are inserted. 

Let us now inspect $Q_T$ in the presence of the modified constraint \eqref{modconstraint} induced by the 
magnetic Wilson loop. We find (use $Q_T[dA^e] = 0$),
\begin{multline}
Q_T[dA^e+2\pi n] = \frac{1}{4\pi} \sum_x \sum_{\mu < \nu \atop \rho < \sigma} n_{x,\mu\nu}
\epsilon_{\mu\nu\rho\sigma}(dA^e)_{x+T-\hat\rho-\hat\sigma,\,\rho\sigma} \\
+\frac{1}{4\pi} \sum_x \sum_{\mu < \nu \atop \rho < \sigma} (dA^e)_{x,\mu\nu}\epsilon_{\mu\nu\rho\sigma}
n_{x+T-\hat\rho-\hat\sigma,\,\rho\sigma}+Q_T[2\pi n]\;.
\label{witten_aux}
\end{multline}
Reorganizing the sums one finds
\begin{eqnarray}
\hspace*{-8mm} && \sum_x \sum_{\mu < \nu \atop \rho < \sigma}   n_{x,\mu\nu} \, 
\epsilon_{\mu\nu\rho\sigma}(dA^e)_{x+T-\hat\rho-\hat\sigma,\,\rho\sigma}
= \sum_x \sum_{\mu < \nu \atop \rho < \sigma}  n_{x-T+\hat\rho+\hat\sigma,\mu\nu} \, \epsilon_{\mu\nu\rho\sigma}(dA^e)_{x,\,\rho\sigma} 
\label{witten_aux2}
\\
\hspace*{-8mm}  && \quad = \sum_x \sum_{\rho<\sigma} N_{x,\rho\sigma}(dA^e)_{x,\, \rho\sigma}
= \sum_x \sum_{\mu} A^e_{x,\mu} (\partial N)_{x,\mu}
= \sum_x \sum_\mu \! A_{x+T,\mu}  \sum_{\nu<\rho<\sigma} \!\! \epsilon_{\mu\nu\rho\sigma}(dn)_{x+\hat\mu,\nu\rho\sigma}\;, 
\nonumber
\end{eqnarray}
where
\be
N_{x,\, \rho\sigma}=\sum_{\mu<\nu }n_{x-T+\hat\rho+\hat\sigma,\mu\nu} \, \epsilon_{\mu\nu\rho\sigma} \; .
\ee
In the third step of \eqref{witten_aux2} we used the partial integration formula \eqref{AppA:partint}, and again some 
reordering of terms to get to the final form. In a similar way we find for the second sum in \eqref{witten_aux},
\be
 \sum_x \sum_{\mu < \nu \atop \rho < \sigma} (dA^e)_{x,\mu\nu} \, \epsilon_{\mu\nu\rho\sigma} \, 
n_{x+T-\hat\rho-\hat\sigma,\rho\sigma} \; = \; \sum_x \sum_\mu 
 A_{x-T+\hat s,\mu} \sum_{\nu<\rho<\sigma} \!\!  \epsilon_{\mu\nu\rho\sigma}(dn)_{x+\hat\mu,\nu\rho\sigma} \; ,
\label{witten_aux3}
\ee
so that inserting \eqref{witten_aux},  \eqref{witten_aux2} and \eqref{witten_aux3} in \eqref{eq:Q_def} we find
\be
Q[dA^e+2\pi n]= \frac{1}{2\pi} \sum_{x}\sum_{\mu} \sum_T 
\gamma_T \frac{A_{x+T,\mu}+A_{x-T+\hat s,\mu}}{2} \sum_{\nu<\rho<\sigma} \!\! \epsilon_{\mu\nu\rho\sigma}(dn)_{x + \hat\mu,\nu\rho\sigma} + Q[2\pi n]\;.
\ee
Since in the presence of the magnetic Wilson loop $\prod_{(\tilde x,\mu) \in \widetilde{C}} e^{\, i \,  \widetilde{A}^m_{\tilde x,\mu}}$ upon integrating over 
the $\widetilde{A}^m$ we have
$\sum_{\nu<\rho<\sigma} \epsilon_{\mu\nu\rho\sigma}(dn)_{x,\nu\rho\sigma}=-1$ whenever $(\tilde x,\mu)\in \widetilde{C}$, and since $Q$ 
comes with a weight $e^{-i{\theta}Q}$ in the partition function, a magnetic Wilson loop generates the contribution
\be
\prod_{(\tilde x,\mu)\in \widetilde{C}} e^{ \, i \frac{\theta}{2\pi} \sum_T \gamma_T \frac{A^e_{x+T,\mu}+A^e_{x-T+ \hat s,\mu}}{2}}\;,
\label{newterm}
\ee
which is the $A^e$-dependent part of $Q[dA^e+2\pi n]$ that gets added to the topological part $Q[2\pi n]$ given by the harmonic contributions
as stated in \eqref{Q_omega}. 
The interpretation of the above formula is that the monopole gets an electric charge $q^e=\frac{\theta}{2\pi}$, although smeared to the 
neighborhood of the dual link $(\tilde x,\mu)$. This is the famous Witten effect \cite{Witten:1979ey}.

When we discussed the coupling of dyonic matter in Subsection \ref{sec:dyonic_theories}, we already announced, that in this context the 
Witten effect gives rise to a new duality ${\cal T}$ which relates shifts of the $\theta$ angle by $2 \pi$ to a shift of the magnetic charge. 
The remainder of this subsection is devoted to discussing the duality ${\cal T}$. 

Similar to the derivation of the worldline expansion of magnetic matter discussed in Appendix~\ref{app:worldline} we may expand the hopping terms of the dyon 
action (compare \eqref{dyonhop}) to bring the dependence of the dyon partition sum on the two gauge fields
into the form 
\begin{equation}
\prod_{(x,\mu)} e^{ \, i \, \big[q^e \, {A}^e_{x,\mu}  \, + \, q^m \, \widetilde{A}^m_{F(x),\mu} \big] \, k_{x,\mu} } \; = \; 
\prod_{(x,\mu)} e^{ \, i \, q^e \, {A}^e_{x,\mu} \, k_{x,\mu} }  \; \prod_{(x,\mu)} e^{ \, i \, q^m \, \widetilde{A}^m_{F(x),\mu} \, k_{x,\mu} } \; ,
\label{dyonflux}
\end{equation}
where $k_{ x,\mu} \in \mathds{Z}$ are the flux variables for dyon matter (compare Appendix~\ref{app:worldline}). Obviously the second factor on the 
rhs.\ generalizes the insertion of the magnetic Wilson loop considered in \eqref{eq:magneticloop}. Upon integrating out the magnetic gauge field this 
generates the modified constraint 
\begin{equation}
\sum_{\nu<\rho<\sigma}\epsilon_{\mu\nu\rho\sigma}(dn)_{x+\hat\mu,\nu\rho\sigma} \; = \; - \, q^m \, k_{x,\mu} \quad \forall \; (x,\mu) \; ,
\label{modconstraint2}
\end{equation}
which now replaces \eqref{modconstraint}. However, the steps discussed in the previous 
paragraphs go through essentially unchanged also with the new constraint, and the additional term that is generated by the topological 
charge is a generalization of \eqref{newterm} given by
\be
\prod_{(x,\mu)}  e^{ \, i \frac{\theta}{2\pi} \, q_m \,   \sum_T \gamma_T \frac{A^e_{x+T,\mu}+A^e_{x -T+ \hat s,\mu}}{2} \, k_{x,\mu} } 
\prod_{(x,\mu)} \delta \left( \sum_{\nu<\rho<\sigma}\epsilon_{\mu\nu\rho\sigma}(dn)_{x+\hat\mu,\nu\rho\sigma} \, + \, q^m \, k_{x,\mu}\right) \; 
e^{ - \, i \, \theta \, Q[2\pi n]} \; ,
\label{newterm2}
\ee 
where we also wrote explicitly the constraint \eqref{modconstraint2}, using the already familiar product over Kronecker 
deltas. We also display the remaining topological contribution.
The first factor in \eqref{newterm} may be combined with the flux terms \eqref{dyonflux} and we replace
\begin{equation}
\prod_{(x,\mu)} e^{ \, i \, \big[q^e \, {A}^e_{x,\mu}  \, + \, q^m \, \widetilde{A}^m_{F(x),\mu} \big] \, k_{x,\mu} } \; \rightarrow \; 
\prod_{(x,\mu)}  e^{ \, i \, \big[q^e \, {A}^e_{x,\mu}  \, + \, q^m \, \sum_T \gamma_T \frac{A^e_{x+T,\mu}+A^e_{x -T+ \hat s,\mu}}{2}
\, + \, q^m \, \widetilde{A}^m_{F(x),\mu} \big] \, k_{x,\mu} } \; ,
\label{dyonflux2}
\end{equation}
where we have reinstated the terms with $\widetilde{A}^m$ that upon integration generate the constraint in \eqref{newterm2}, given that 
the topological term is kept in the form $Q[2\pi n]$.

Note that the required 1-form gauge invariance \cite{Sulejmanpasic:2019ytl}
\be\label{eq:1-form_gauge}
\begin{split}
&A^e_{x,\mu}\rightarrow A^e_{x,\mu}+2\pi k_{x,\mu} \; ,\\
&n_{x,\mu\nu}\rightarrow n_{x,\mu\nu}-(dk)_{x,\mu\nu} \; ,
\end{split}
\ee 
can be achieved by a shift of $A^m_{\tilde x,\mu}$ ,
\be\label{eq:1-form_gauge_supp}
\widetilde{A}^m_{F(x),\mu} \rightarrow \widetilde{A}^m_{F(x),\mu} -\frac{\theta}{2} 
\sum_T \gamma_T\left(k_{x+T-\hat\mu,\mu}+k_{x-T+s-\hat\mu,\mu}\right)
\; .
\ee
Indeed the remainder of the action, has phase terms
\be
i\theta Q[2\pi n]+i\sum_x \! \sum_{\mu < \nu < \rho} \epsilon^{\mu\nu\rho\sigma}(dn)_{x,\mu \nu \rho}\tilde 
A^m_{F(x)-\hat \sigma,\sigma} \; ,
\ee 
which are invariant under the transformations \eqref{eq:1-form_gauge} and \eqref{eq:1-form_gauge_supp}. Note that $Q[2\pi n]$ is now no longer necessarily an integer, because it is a sum of integers weighted by $\gamma_T$, and must be kept in the action even if $\theta\in 2\pi\mathbb Z$. In fact it is crucial that this term is not dropped, as then the gauge symmetry \eqref{eq:1-form_gauge_supp} would be ruined. 

The rhs.\ of \eqref{dyonflux2} shows that in the presence of magnetic matter, the topological term generates an additional contribution 
to the electric charge, such that the combined electric charge is given by $q^e + q^m \theta / 2 \pi$, although this charge is generally 
smeared across multiple lattice links, weighted by $\gamma_T$. In fact, no choice of the 
coefficients $\gamma_T$ in the definition \eqref{eq:Q_def} allows for an ultra-local Witten effect. 
Indeed, if even only a single $\gamma_T$ is nonzero, i.e., $\gamma_T=1$ with $\gamma_{T'}=0 \; \forall \; T'\ne T$, this will produce the spread 
of the electric charge. However, independent of the 
details of the definition of $Q$ we find that shifting  $\theta \rightarrow \theta + 2 \pi$ is equivalent to 
shifting the electric charge $q^e \rightarrow q^m + q^e$. Thus, if we 
shift $\theta \rightarrow \theta - 2 \pi$ and simultaneously $q^e \rightarrow q^m + q^e$ this becomes an invariance. We will refer to this 
invariance as ${\cal T}$ duality, as is conventional, and summarize it as follows,
\begin{equation}
\theta \, \rightarrow \, \overline{\theta} \, = \, \theta - 2 \pi \; , \; \;
{\bm q} \, \rightarrow \, \overline{\bm q} \, = \, T \, {\bm q} \; \; \mbox{with} \; \;   
T=\begin{pmatrix}
 1 & 1\\
 0&1
\end{pmatrix}\;,
\label{eq:T_matrix}
\end{equation}
where again we use the vector of charges ${\bm q} = (q^e, q^m)^t$. 
We will later see that the ${\cal S}$ duality defined in \eqref{eq:S_matrix}, i.e., the self-dual transformation, and the 
 duality ${\cal T}$ together generate the group SL$(2,\mathbb Z)$. 

In the next two subsections we will see that a generalized form of the ${\cal S}$ duality holds also for the theory with a
$\theta$-term, but that the discretization we discussed so far will not map the action into the same form under a duality transformation. 
We will then go on to construct lattice actions which enjoy an exact self-duality, at the price of abandoning the ultra-local 
structure we so far employed. That such a non-ultra-local structure may be expected is already suggested by the discussion 
of the Witten effect in this subsection. 

\subsection{$\theta$-term and duality transformation}

We now generalize the partition sum (\ref{Zfull}) of self-dual lattice QED further by adding a $\theta$-term and discuss the 
form of the duality transformation in the presence of such a term. This will not yet give rise to a self-dual theory, which will be identified 
only in the next subsection after another necessary generalization of the action.  

We may write the partition sum that now also contains the $\theta$-term in the form
\begin{equation}
Z(\beta, \theta, M^e\!,\lambda^e\!,q^e\!,\,M^m\!,\lambda^m\!,q^e) \equiv \! \int \!\! D[A^e] \! \int \!\! D[A^m] 
B_{\beta,\theta} [A^e,A^m]  
Z_{M^e\!,\,\lambda^e\!,\,q^e} [A^e] \widetilde{Z}_{M^m\!,\,\lambda^m\!,\,q^m} \big[\widetilde{A}^m\big] ,
\label{Zfulltheta}
\end{equation}
where we have generalized the Boltzmann factor (\ref{boltzmann_both}) by adding the $\theta$-term,
\begin{equation}
B_{\beta,\theta}[A^e,A^m]  \; \equiv \; \sum_{\{ n \}} \; 
e^{ \, -\frac{\beta}{2} \sum_{x} \! \sum_{\mu < \nu} \big( F^e_{x,\mu \nu} \big)^2 } \; 
e^{\, - i \, \theta \, Q_0[F^e]} \; \;
e^{\,- \, i \sum_x \! \sum_{\mu < \nu < \rho} A^m_{x,\mu \nu \rho} (dn)_{x,\mu \nu \rho}}  .
\label{boltzmann_theta1}
\end{equation}
Note that for now
we use the $T = 0$ discretization $Q_0[F^e]$ from the family $Q_T[F^e]$ of possible equivalent lattice forms of the 
topological charge we have introduced in Subsection~\ref{subsec:topocharge} in Eq.\ (\ref{QTdef}). 
We will generalize the $\theta$-term 
further in the next subsection when we construct the self-dual form of the Boltzmann factor.

Both, the gauge field action and the topological charge $Q_0[F^e]$ are quadratic in $F^e = d A^e \, + \, 2 \pi \, n$, such that we can combine them 
in a quadratic form. The Boltzmann factor thus reads 
\begin{equation}
B_{\beta,\theta}[A^e,A^m]  \; = \,  \sum_{\{ n \}}  
e^{ \, -\frac{\beta}{2} \sum_{x, \mu < \nu \atop y, \rho < \sigma } 
 F^e_{x,\mu \nu} \, M_{x,\mu \nu | y, \rho \sigma} \, F^e_{y,\rho \sigma} } \; 
e^{ \,  i  \sum_{x,\mu < \nu} (\partial A^m)_{x,\mu \nu} \, n_{x,\mu \nu} },
\label{boltzmann_theta2}
\end{equation}
where we used the partial integration formula (\ref{AppA:partint}) from Appendix \ref{app:diff_forms} to also rewrite the exponent of the last 
exponential that upon integration over $A^m$ generates the constraints, which in the absence of magnetically charged matter are the 
closedness constraints \eqref{Villainconstraint} or, when magnetic matter is coupled, the modified constraints \eqref{modconstraint}. 
The kernel $M_{x,\mu \nu | y, \rho \sigma}$ 
of the quadratic form \eqref{boltzmann_theta2} is defined as (note that in our notation $\mu < \nu$ and $\rho < \sigma$)
\begin{equation}
M_{x,\mu \nu | y, \rho \sigma} \; = \; \delta_{\mu \rho} \, \delta_{\nu \sigma} \, \delta_{x,y}^{\; (4)} 
\; + \; i \, \xi \;  \epsilon_{\mu \nu \rho \sigma} \, \delta_{x- \hat{\rho} - \hat{\sigma},y}^{\; (4)} \qquad \mbox{with}
\quad \xi \; \equiv \; \frac{\theta}{4 \pi^2 \beta} \; .
\label{kernelM}
\end{equation}
Using Fourier transformation (see Appendix \ref{app:C} for our conventions) we may diagonalize the lattice site dependence of $M$. Using 
\eqref{Afourier} we find for the Fourier transform of $M$,
\begin{equation}
\widehat{M}(p)_{\mu \nu | \rho \sigma} \; = \; \sum_{x} e^{i p\cdot (x-y)} M_{x,\mu \nu | y, \rho \sigma}=
\delta_{\mu \rho} \, \delta_{\nu \sigma} 
\; + \; i \, \xi \; \epsilon_{\mu \nu \rho \sigma} \; e^{ \, i p_\rho \, + \, i p_\sigma} \; .
\label{Mhat}
\end{equation}
It is straightforward to see that for $|\xi| < 1$ (this is a sufficient condition) we have $\det \widehat{M} \neq 0$, such that the matrix 
$\widehat{M}$ in Fourier space is invertible and thus also the real space matrix $M$. One finds (see Appendix \ref{app:C}),
\begin{eqnarray}
\widehat{M}(p)_{\mu \nu | \rho \sigma}^{\; - 1} & = & 
\frac{\delta_{\mu \rho} \, \delta_{\nu \sigma} 
\, - \, i \, \xi \, \epsilon_{\mu \nu \rho \sigma} \; e^{ \, i p_\rho \, + \, i p_\sigma}}{
1 \; + \; \xi^2 \; e^{ \, i p \cdot \hat{s}}} 
\nonumber \\
& = & 
\sum_{k = 0}^\infty (-\xi^2)^k \left[
\delta_{\mu \rho} \, \delta_{\nu \sigma} \, e^{ \, i p \cdot k \hat{s}}
\, - \, i \, \xi \, \epsilon_{\mu \nu \rho \sigma} \; e^{ \, i p_\rho \, + \, i p_\sigma \, + \, i p \cdot k \hat{s}}
\right] ,
\label{Mhatinverse}
\end{eqnarray}
where $\hat s=\hat 1+\hat 2+\hat 3+\hat 4$ and in the second step we 
have expanded the denominator using the geometric series which converges exponentially for 
$|\xi| < 1$. Using this second form and (\ref{Ainvfourier}) we can evaluate $M^{-1}_{\, x,\mu \nu | y, \rho \sigma}$ and obtain,
\begin{equation}
M^{-1}_{\, x,\mu \nu | y, \rho \sigma} \; = \; \sum_{k = 0}^\infty (-\xi^2)^k \left[
\delta_{\mu \rho} \, \delta_{\nu \sigma} \, \delta_{ x - k \hat{s},y}^{\;(4)}
\; - \; i \, \xi \, \epsilon_{\mu \nu \rho \sigma} \; \delta_{ x - \hat{\rho} - \hat{\sigma} - k \hat{s},y}^{\; (4)}
\right] \; .
\label{Minverse}
\end{equation}

We now apply the generalized Poisson resummation formula proven in 
Appendix \ref{app:D} for $N = 6V$ and find\footnote{We remark that it is easy to show that for $|\xi| < 1$ all eigenvalues of $M$ have
positive real parts, which guarantees the existence of the Gaussian integral in the Poisson resummation.}
 (use again Eq.~(\ref{constraintsexponent}) to first rewrite the second exponent in (\ref{boltzmann_theta2}))
\begin{equation}
B_{\beta,\theta}[A^e,A^m]  = C_M(\beta) \!   \sum_{\{ p \}} \!
e^{\! -\frac{\widetilde{\beta}}{2} \sum_{x, \mu < \nu \atop y, \rho < \sigma } \!
(\partial  A^m + 2\pi p)_{x,\mu \nu} \, M^{-1}_{\, x,\mu \nu | y, \rho \sigma} \, (\partial A^m + 2\pi p)_{y,\rho \sigma} }
 e^{- i \sum_{x,\mu < \nu} (dA^e)_{x,\mu\nu} \, p_{x,\mu \nu} } \! , 
\label{boltzmann_theta2_poisson}
\end{equation} 
with 
\begin{equation}
\widetilde{\beta} \; \equiv \;  \frac{1}{4 \pi^2 \beta} 
\quad , \qquad
\sum_{\{ p \}} \; \equiv \; \prod_{x, \mu < \nu} \; \sum_{p_{x,\mu \nu} \in \mathds{Z}} 
\quad , \qquad
C_M(\beta) \; \equiv \; \frac{1}{(\sqrt{2 \pi \beta})^{6V} \sqrt{ \det M }} \; .
\end{equation}
$\sum_{\{ p \}}$ denotes the sum over all configurations of 
plaquette occupation numbers $p_{x,\mu \nu} \in \mathds{Z}$. 

Note that the first exponent in (\ref{boltzmann_theta2_poisson}) with $M^{-1}_{\, x,\mu \nu | y, \rho \sigma}$ given by (\ref{Minverse}) 
gives rise to a lattice action that combines terms at arbitrary distances shifted relative to each other by 
$k \hat{s}$ with $k \in \mathds{N}_0$, 
such that this action is not ultra-local, i.e., it connects terms at arbitrary distances $k \hat{s}$. 
However, the corresponding terms are suppressed exponentially with $k$, due to the
condition $|\xi| < 1$. Such types of non-ultra local lattice actions with exponential suppression of shifted terms are widely used, 
with prominent examples being fixed point actions \cite{Hasenfratz:1998jp} or the  overlap operator 
\cite{Neuberger:1997fp}, and it is known that the exponential 
suppression of distant terms guarantees a local continuum limit \cite{Hernandez:1998et}. As already mentioned,  such lattice actions are referred to as 
local but non-ultra-local actions. 
  
Let us now complete the duality transformation. The final step of the duality transformation is to express the electric 
and the magnetic gauge fields, as well as the plaquette occupation numbers in terms of their counterparts on 
the dual lattice, i.e., we again use Eqs.~(\ref{convertdual1}) and (\ref{convertdual2}). We find
\begin{equation}
B_{\beta,\theta}[A^e,A^m]  \; = \; C_M(\beta) \!   \sum_{\{ p \}} 
e^{ \, -\frac{\widetilde{\beta}}{2} \sum_{\tilde x, \mu < \nu \atop \tilde y, \rho < \sigma } 
\widetilde{F}^m_{\tilde x,\mu \nu} \, \widetilde{M}_{\tilde x,\mu \nu | \tilde y, \rho \sigma} \, 
\widetilde{F}^m_{\tilde y,\rho \sigma} } \; 
e^{ \, - \, i  \sum_{\tilde x,\mu < \nu < \rho }  \widetilde{A}^e_{ \tilde x,\mu \nu \rho} \, (d \widetilde{p})_{\tilde x,\mu \nu \rho} },
\label{boltzmann_theta_dual}
\end{equation} 
where $\widetilde{F}^m_{\tilde x,\mu \nu} = (d\widetilde{A}^m + 2\pi \widetilde{p})_{\tilde x,\mu \nu}$.  
The kernel $\widetilde{M}_{\, \tilde{x},\mu \nu | \tilde{y}, \rho \sigma}$ is identified from 
$M^{-1}_{\, x,\mu \nu | y, \rho \sigma}$ when replacing $(\partial A^m \, + \, 2\pi p)_{x,\mu\nu}$ by the dual 
expression $\sum_{\mu^\prime < \nu^\prime} \epsilon_{\mu \nu \mu^\prime \nu^\prime}  
(d \widetilde{A}^m \, + \, 2 \pi \tilde{p})_{\tilde{x} - \hat{\mu}^\prime - \hat{\nu}^\prime,\mu^\prime \nu^\prime}$ in the exponent 
of (\ref{boltzmann_theta2_poisson}) using (\ref{convertdual2}), i.e.,
\begin{equation}
\widetilde{M}_{\, \tilde{x},\mu \nu | \tilde{y}, \rho \sigma} \; \equiv \; \sum_{\mu^\prime < \nu^\prime \atop \rho^\prime < \sigma\prime}
\epsilon_{\mu \nu \mu^\prime \nu^\prime} \, 
M^{-1}_{\, x + \hat{\mu}^\prime + \hat{\nu}^\prime,\mu^\prime \nu^\prime | 
y + \hat\rho^\prime +\hat\sigma^\prime, \rho^\prime \sigma^\prime}
\, \epsilon_{\rho^\prime \sigma^\prime \rho \sigma} \; \bigg|_{x \rightarrow \tilde{x} \atop y \rightarrow \tilde{y}  } \; .
\label{derivdual}
\end{equation}
Inserting the explicit form (\ref{Minverse}) of $M^{-1}_{\, x,\mu \nu | y, \rho \sigma}$ and summing over the indices of the epsilon
tensors that appear in (\ref{derivdual}) we obtain in a few lines of algebra 
\begin{equation}
\widetilde{M}_{\, \tilde{x},\mu \nu | \tilde{y}, \rho \sigma} \; = \; \sum_{k = 0}^\infty (-\xi^2)^k  
\left[ \delta_{\mu \rho} \, \delta_{\nu \sigma} \, \delta_{ \tilde x - k \hat{s},\tilde y}^{\;(4)}
\; - \; i \, \xi \, \epsilon_{\mu \nu \rho \sigma} \; \delta_{ \tilde x - \hat{\rho} - \hat{\sigma} - k \hat{s}, \tilde y}^{\; (4)}
\right] \; .
\label{Mtilde}
\end{equation}

It is important to note that as for the case without $\theta$-term, the Boltzmann factor in its dual form 
(\ref{boltzmann_theta_dual}) is structurally similar to the original form
(\ref{boltzmann_theta2}) we started from. The dual form lives on the dual lattice and the Villain variables $n$ were replaced by the 
dual plaquette occupation numbers $\widetilde{p}$. The electric and magnetic gauge fields interchanged their role, such that $\widetilde{A}^m$ 
now appears in the quadratic form, while $\widetilde{A}^e$ generates the constraints for the dual plaquette occupation numbers.

However, the original kernel $M_{\, x,\mu \nu | y, \rho \sigma}$ Eq.~\eqref{kernelM} and the dual kernel 
$\widetilde{M}_{\, \tilde{x},\mu \nu | \tilde{y}, \rho \sigma}$ Eq.~\eqref{Mtilde} differ. More specifically, the original 
kernel $M_{\, x,\mu \nu | y, \rho \sigma}$ is ultra-local, while the dual kernel $\widetilde{M}_{\, \tilde{x},\mu \nu | \tilde{y}, \rho \sigma}$
is a sum over terms that are shifted relative to each other by $k \hat{s}$ with $k \in \mathds{N}_0$. 
Comparing this form with our definition of $Q_T$ in Eq.~(\ref{QTdef}) we
find that for a fixed $k$ the second term in (\ref{Mtilde}) gives rise to the topological charge $Q_T [\widetilde{F}^m]$
with $T =  - k \hat s$, i.e., $Q_{- k \hat s}[\widetilde{F}^m]$, with 
$\widetilde{F}^m_{\tilde x, \mu \nu} = (d \widetilde{A}^m)_{\tilde x, \mu \nu} + 2\pi \widetilde{p}_{\tilde x, \mu \nu}$. 
Since integrating over $\widetilde{A}^e$ in (\ref{boltzmann_theta_dual}) generates the closedness constraints for 
the dual plaquette occupation numbers $\widetilde{p}_{\tilde x, \mu \nu}$, the topological charge $Q_{-k\hat s}[\widetilde{F}^m]$
obeys all properties we showed for the original definition (\ref{QTdef}). In particular it is independent of $T = - k \hat s$, such that 
$Q_{-k\hat s}[\widetilde{F}^m] = Q_0[\widetilde{F}^m] \; \forall k \in \mathds{Z}$. Thus we may sum up $k$ in the term generated by the 
second factor of (\ref{Mtilde}) such that the corresponding term in the exponent reads
\begin{equation}
- \, \frac{\widetilde{\beta}}{2} \, 8 \pi^2 \, Q_0[\widetilde{F}^m] \, (- i \xi) \sum_{k = 0}^\infty (- \xi^2)^k 
\; = \; i \,  \frac{\theta}{4 \pi^2\beta^2 + \theta^2/4 \pi^2} \, Q_0[\widetilde{F}^m] \; = \; - \,  i \, \theta^\prime \, Q_0[\widetilde{F}^m] \; ,
\end{equation}
where $ \theta^\prime \equiv - \, \theta / (4 \pi^2\beta^2 + \theta^2/4 \pi^2)$.
Thus we may rewrite the dual form (\ref{boltzmann_theta_dual}) as 
\begin{equation}
B_{\beta,\theta}[A^e,A^m]  \; = \; C_M \!   \sum_{\{ p \}} 
e^{ -\frac{\widetilde{\beta}}{2} 
\sum_{k = 0}^\infty (-\xi^2)^k 
\sum_{\tilde{x}, \mu < \nu} 
\widetilde{F}^m_{\tilde{x},\mu \nu} \, 
\widetilde{F}^m_{\tilde{x} - k \hat{s},\mu \nu} \; - \; i \theta^\prime \, Q_0[\widetilde{F}^m]} \;
e^{ \, - \, i  \sum_{\tilde x,\mu < \nu< \rho} \widetilde{A}^e_{ \tilde x,\mu \nu \rho} \,(d \widetilde{p}\,)_{\tilde x,\mu \nu \rho} },
\label{boltzmann_theta_dual2}
\end{equation} 
where we have again used the partial integration formula to re-express the exponent in the last factor. The dual 
form (\ref{boltzmann_theta_dual2}) has to be compared to the original form of the Boltzmann factor (\ref{boltzmann_theta1}). 
One sees that after interchanging the gauge fields with their dual counterparts, multiplying the trivial overall factor, and replacing 
$\beta$ by $\widetilde{\beta}$ and $\theta$ by $\theta^\prime$, the two expressions are almost identical.
The only remaining difference is that the gauge field action has the form 
\begin{equation}
\frac{\tilde{\beta}}{2} \!
\sum_{k = 0}^\infty \! (-\xi^2)^k \!\!\!
\sum_{\tilde{x}, \mu < \nu} 
\widetilde{F}^m_{\tilde{x},\mu \nu} \, 
\widetilde{F}^m_{\tilde{x} - k \hat{s},\mu \nu}
  = 
\frac{\tilde{\beta}}{2} \!
\sum_{k = 0}^\infty \! (-\xi^2)^k \! \!\!
\sum_{\tilde{x}, \mu < \nu} \!\!
(d \widetilde{A}^m \! + \! 2\pi \widetilde{p}\,)_{\tilde{x},\mu \nu} 
(d \widetilde{A}^m \! +\!  2\pi \widetilde{p}\,)_{\tilde{x} - k \hat{s},\mu \nu}  \; ,
\label{FFshifted}
\end{equation}
i.e., the gauge field action is a superposition of 
terms where the two factors of the field strength 
$\widetilde{F}^m$ are shifted relative to each other
by $k \hat{s}$ with $k \in \mathds{N}_0$. The $k = 0$ term corresponds to the ultra-local action we have started from, but the duality 
transformation has generated the non-ultra-local extension, which for $|\xi| <1$ still gives rise to a 
local continuum limit as we discussed above. We remark, however, that also the restriction $|\xi| < 1$, which corresponds to
$\beta > |\theta| / 4 \pi^2$, is an unphysical restriction for a proper self-dual lattice version of U(1) lattice gauge 
theory with a $\theta$-term we are aiming at.


\subsection{Identification of the exactly self-dual theory}
\label{sec:selfdual_theta}

Having developed the duality transformation in the presence of a $\theta$-term we are now ready to identify the fully
self-dual discretization that includes the $\theta$-term. We have seen that the duality transformation has converted the ultra-local
lattice action of the original theory into a non-ultra-local lattice action for the dual theory. The key insight is that in order to obtain 
self-duality, we must not start with an ultra-local discretization for the original theory, but with a more general ansatz that allows terms
at arbitrary distances with exponentially decreasing coefficients. We write the Boltzmann factor 
in the form (compare (\ref{boltzmann_theta2}))
\begin{equation}
B_{\beta,\theta}[A^e,A^m]  \; \equiv \,  \sum_{\{ n \}}  
e^{ \, -\frac{\beta}{2} \sum_{x, \mu < \nu \atop y, \rho < \sigma } 
F^e_{x,\mu \nu} \, K_{x,\mu \nu | y, \rho \sigma} \, F^e_{y,\rho \sigma} } \; 
e^{ \, i  \sum_{x,\mu < \nu} (\partial A^m)_{x,\mu \nu} \, n_{x,\mu \nu} },
\label{boltzmann_theta3}
\end{equation}
where $K$ is a new kernel that we will specify below. Again the electric gauge field $A^e$ describes the 
dynamics and the magnetic gauge field $A^m$ generates the constraints.

The key step towards self-duality is to identify a kernel $K$ that is form-invariant under inversion. 
This can be done directly in momentum 
space, and from inspection of \eqref{Mhat} and \eqref{Mhatinverse} one finds that an ansatz of the form 
\begin{equation}
\widehat{K}(p)_{\mu \nu | \rho \sigma} \; = \; 
\frac{\delta_{\mu \rho} \, \delta_{\nu \sigma} 
\, + \, i \, \frac{\gamma}{2} \, \epsilon_{\mu \nu \rho \sigma} \, [ e^{ \, i p_\rho \, + \, i p_\sigma} + e^{ \,- i p_\mu \, - \, i p_\nu} ]  }{
\sqrt{1 \; + \; \frac{\gamma^2}{2} \, [ 1 + \cos( p \cdot \hat{s})] }} \; ,
\label{Khat}
\end{equation} 
with 
\begin{equation}
\widehat{K}(p)_{\mu \nu | \rho \sigma}^{\, -1} \; = \; 
\frac{\delta_{\mu \rho} \, \delta_{\nu \sigma} 
\, - \, i \, \frac{\gamma}{2} \, \epsilon_{\mu \nu \rho \sigma} \, [ e^{ \, i p_\rho \, + \, i p_\sigma} + e^{ \,- i p_\mu \, - \, i p_\nu} ]  }{
\sqrt{1 \; + \; \frac{\gamma^2}{2} \, [ 1 + \cos( p \cdot \hat{s})] }} \; ,
\label{Khatinverse}
\end{equation}
defines a gauge action and a topological term where the original kernel and the inverse kernel that is used in the 
dual theory have the same momentum dependence, and thus the same structure also in real space\footnote{Note that \eqref{Khat} is not the only 
choice of a kernel that keeps the electric-magnetic duality of the theory, however, no ultra-local choice is possible. 
We remark that the choice \eqref{Khat} violates hypercubic symmetries. This can indeed be corrected, as we discuss in  
Appendix~\eqref{app:gen_top_charge}.}. Here $\gamma$ is a
real parameter that we do not need to specify here and later will relate to the parameters $\beta$ and $\theta$.
Note that we have symmetrized the part that generates the topological charge in $K$ to 
$\epsilon_{\mu \nu \rho \sigma} \, [ e^{ \, i p_\rho \, + \, i p_\sigma} + e^{ \,- i p_\mu \, - \, i p_\nu} ]$,
which is related to a combination of $Q_0[F^e]$ and $Q_{-\hat s}[F^e]$.  With this choice the normalization 
factor in the denominator is regular for all $\gamma$, i.e., the argument of the square root is always positive, such that 
we have solved the problem of an unphysical restriction of $\beta$ and $\theta$ we faced in the naive attempt in the 
previous subsection. 

It is obvious, that the non-trivial denominator $\sqrt{1 \; + \; \frac{\gamma^2}{2} \, [ 1 + \cos( p \cdot \hat{s})] }$ that appears in 
the momentum space kernels \eqref{Khat} and \eqref{Khatinverse} gives rise to non-ultra-local local real
space kernels $K$ and $K^{-1}$, which we now discuss. It is straightforward to identify the real space equivalent of 
$1 \; + \; \frac{\gamma^2}{2} \, [ 1 + \cos( p \cdot \hat{s})]$ which is the Helmholtz-type lattice operator 
\begin{equation}
H_{x,y} \; = \; \delta^{\;(4)}_{x,y} \; + \; \frac{\gamma^2}{4} \, \left[
 \delta^{\;(4)}_{x + \hat{s},y} + 2 \delta^{(4)}_{x,y} + \delta^{\; (4)}_{x - \hat{s},y} \right] 
\; .
\label{helmholtz}
\end{equation}
Thus we find in real space,
\begin{eqnarray}
K_{x,\mu \nu | z, \rho \sigma} & = & \sum_z H^{-\frac{1}{2}}_{x,y} \, 
\bigg[ \delta_{\mu \rho} \, \delta_{\nu \sigma} \, \delta_{y,z}^{\; (4)} 
\; + \; i \, \frac{\gamma}{2} \;  \epsilon_{\mu \nu \rho \sigma} \, 
\left[ \delta_{y- \hat{\rho} - \hat{\sigma},z}^{\; (4)}  + \delta_{y + \hat{s} - \hat{\rho} - \hat{\sigma},z}^{\; (4)} \right] \bigg] \; ,
\label{kernelK} \\
K_{x,\mu \nu | y, \rho \sigma}^{\; -1} & = & \sum_z H^{-\frac{1}{2}}_{x,y} \, 
\bigg[ \delta_{\mu \rho} \, \delta_{\nu \sigma} \, \delta_{y,z}^{\; (4)} 
\; - \; i \, \frac{\gamma}{2} \;  \epsilon_{\mu \nu \rho \sigma} \, 
\left[ \delta_{y- \hat{\rho} - \hat{\sigma},z}^{\; (4)}  + \delta_{y + \hat{s} - \hat{\rho} - \hat{\sigma},z}^{\; (4)} \right] \bigg] \; ,\label{kernelKinverse}
\end{eqnarray}
where in real space the inverse square root $H^{-\frac{1}{2}}$ of the Helmholtz operator may be implemented with the spectral
theorem or a series expansion. Since the denominator $\sqrt{1 \; + \; \frac{\gamma^2}{2} \, [ 1 + \cos( p \cdot \hat{s})] }$ 
in the momentum space kernels $\widehat{K}(p)$ and $\widehat{K}(p)^{\, -1}$  is regular for all values of $\gamma$, 
the inverse Fourier transforms, i.e., the real space kernels $K$ and $K^{\, -1}$ will have entries that decrease 
exponentially with increasing lattice distance, i.e., they are local (but not ultra-local of course).

It is important to note that the operator $H^{-\frac{1}{2}}_{x,y}$ is composed from simple operators $\delta_{x + n \hat s, y}$ 
for integers $n$. When applying these shifts on the terms with the epsilon tensor in \eqref{kernelK} and 
\eqref{kernelKinverse} this generates terms proportional to $Q_{-\hat sk}[F^e]$ for different values of  $k$. If there are no dynamical monopoles then we can replace all $Q_{-k\hat s}[F^e]$  by $Q_0[F^e]$. 

Thus for the terms
with the epsilon tensor in \eqref{kernelK} and \eqref{kernelKinverse} we may replace the action of  $H^{-\frac{1}{2}}_{x,y}$
simply by multiplication with $1/\sqrt{1 + \gamma^2}$. We now use this fact to simplify the topological part 
of the quadratic form of the Boltzmann factor. 

We may write the Boltzmann factor \eqref{boltzmann_theta3} in the form 
\begin{equation}
B_{\beta,\theta}[A^e,A^m]  \; \equiv \,  \sum_{\{ n \}}  
e^{ \, - \beta S_g[F^e] \; - \; i \, \theta Q_0[F^e]} \; 
e^{ \, i  \sum_{x,\mu < \nu} (\partial A^m)_{x,\mu \nu} \, n_{x,\mu \nu} },
\label{boltzmann_theta4}
\end{equation}
where the gauge field action is given by
\begin{equation}
S_g[F^e] \; = \; \frac{1}{2} \sum_{x, \mu < \nu \atop y, \rho < \sigma } 
F^e_{x,\mu \nu} \, H^{-\frac{1}{2}}_{x,\mu \nu | y, \rho \sigma} \, F^e_{y,\rho \sigma}  \; ,
\end{equation}
and using the simplification of $H^{-\frac{1}{2}}_{x,y}$ for the topological part discussed above,
we identify the topological angle $\theta$ as 
\begin{equation}
\theta  \; = \; \beta \, 4 \pi^2 \, \frac{\gamma}{\sqrt{1 + \gamma^2}} \quad \Rightarrow \quad
\gamma \; = \; \frac{\theta}{\sqrt{(4 \pi^2 \beta)^2 -\theta^2}} \; .
\label{find_gamma}
\end{equation}
Note that while we obtained this for a model without dynamical magnetic matter, we can take the above formulas 
as the definition of the $\theta$ term even in this case. 

In a similar way we may identify the bare electric charge parameter 
$e$ through the pre-factor of the $F^e_{\mu \nu} F^e_{\mu \nu}$ term
which gives rise to the relation (again replacing $H^{-\frac{1}{2}}_{x,y}$  by $1/\sqrt{1+\gamma^2}$)
\begin{equation}
\frac{1}{e^2} \; = \frac{\beta}{\sqrt{1+\gamma^2}} \; .
\label{find_e}
\end{equation}
Using \eqref{find_gamma} and \eqref{find_e} we may express the two auxiliary parameters $\beta$ and $\gamma$ 
in terms of the bare charge parameter $e$ and the topological angle $\theta$,
\begin{equation}
\beta \; = \; \frac{1}{2\pi } \sqrt{ \left( \frac{2\pi}{e^2}\right) + \left(\frac{\theta }{2 \pi} \right)^2} \quad , \qquad  \gamma \; = \; \frac{\theta \, e^2}{4 \pi^2}  \; .
\end{equation}

Repeating the steps of the duality transformation from the previous section we use the generalized Poisson resummation formula
and find the dual form of the Boltzmann factor
\begin{equation}
B_{\beta,\theta}[A^e,A^m]  \; = \; C_K(\beta)  \!   \sum_{\{ p \}} 
e^{ \, -\frac{\widetilde{\beta}}{2} \sum_{\tilde x, \mu < \nu \atop \tilde y, \rho < \sigma } 
\widetilde{F}^m_{\tilde x,\mu \nu} \, \widetilde{K}_{\tilde x,\mu \nu | \tilde y, \rho \sigma} \, 
\widetilde{F}^m_{\tilde y,\rho \sigma} } \; 
e^{ \,  i  \sum_{\tilde x,\mu < \nu} (\partial \widetilde{A}^e)_{ \tilde x,\mu \nu} \, \widetilde{p}_{\tilde x,\mu \nu} },
\label{boltzmann_theta3_dual}
\end{equation} 
where (it is straightforward to establish the result $\det K = 1$ we use here)
\begin{equation}
\tilde{\beta} \; \equiv \; \frac{1}{4\pi^2 \beta} 
\qquad \mbox{and} \qquad 
C_K(\beta) \; \equiv \; \frac{1}{(\sqrt{2 \pi  \beta})^{6V} \sqrt{\det K}} \; = \;  \frac{1}{(2 \pi  \beta)^{3V}} \; . 
\label{dualfactors}
\end{equation}
The dual kernel $\widetilde{K}_{\tilde x,\mu \nu | \tilde y \rho \sigma}$ is identified in analogy to (\ref{derivdual}) and
we find
\begin{eqnarray}
\widetilde{K}_{\, \tilde{x},\mu \nu | \tilde{y}, \rho \sigma} & \equiv & \sum_{\mu^\prime < \nu^\prime \atop \rho^\prime < \sigma\prime}
\epsilon_{\mu \nu \mu^\prime \nu^\prime} \, 
K^{-1}_{\, x + \hat{\mu}^\prime + \hat{\nu}^\prime,\mu^\prime \nu^\prime | 
y + \hat\rho^\prime +\hat\sigma^\prime, \rho^\prime \sigma^\prime}
\, \epsilon_{\rho^\prime \sigma^\prime \rho \sigma} \; \bigg|_{x \rightarrow \tilde{x} \atop y \rightarrow \tilde{y}  } \nonumber \\
& = & \sum_z H^{-\frac{1}{2}}_{x,y} \, 
\bigg[ \delta_{\mu \rho} \, \delta_{\nu \sigma} \, \delta_{y,z}^{\; (4)} 
\; - \; i \, \frac{\gamma}{2} \;  \epsilon_{\mu \nu \rho \sigma} \, 
\left[ \delta_{y- \hat{\rho} - \hat{\sigma},z}^{\; (4)}  + \delta_{y + \hat{s} - \hat{\rho} - \hat{\sigma},z}^{\; (4)} \right] \bigg] \; .  
\label{Ktilde}
\end{eqnarray}
Obviously, up to the opposite sign of the imaginary part, 
also the dual kernel $\widetilde{K}_{\tilde x,\mu \nu | \tilde y, \rho \sigma}$ has the structure of the original
kernel $K_{x,\mu \nu | y, \rho \sigma}$ in Eq.~\eqref{kernelK},
and comparing (\ref{boltzmann_theta3}) and (\ref{boltzmann_theta3_dual}) we find that we have indeed 
constructed a self-dual Boltzmann factor that also includes the $\theta$-term. 
Writing the dual form of the kernel as 
\begin{equation}
B_{\beta,\theta}[A^e,A^m]  \; = \; C_K(\beta)  \!   \sum_{\{ p \}} 
e^{ \, - \widetilde{\beta} S_g[\widetilde{F}^m] \; - \; i \,  \widetilde{\theta} \, Q_0[\widetilde{F}^e] } \; 
e^{ \,  i  \sum_{\tilde x,\mu < \nu} (\partial \widetilde{A}^e)_{ \tilde x,\mu \nu} \, \widetilde{p}_{\tilde x,\mu \nu} } \;  ,
\label{B_selfdual_final}
\end{equation}
we here identify the dual topological angle as
\begin{equation}
\widetilde{\theta} \; = \; - \, 4 \pi^2 \widetilde{\beta} \frac{\gamma}{\sqrt{1+\gamma^2}} \; .
\end{equation}

We now may summarize the
self-duality relation of the Boltzmann factor with $\theta$-term as
\begin{equation}
\beta^{\, \frac{3V}{2}} \, B_{\beta,\theta}[A^e,A^m]  \; = \; \widetilde{\beta}^{\, \frac{3V}{2}} \,  
\widetilde{B}_{\widetilde{\beta},\widetilde{\theta}} \big[\widetilde{A}^m,\widetilde{A}^e\big]  \; ,
\label{Boltzmann_theta_duality_final}
\end{equation}
where we have again distributed the pre-factors symmetrically and use 
\begin{equation}
\widetilde{\beta} \; = \; \beta \, f
\quad, \qquad  
\widetilde{\theta} \; = \; - \, \theta \, f 
\qquad \mbox{where} \qquad 
f \; \equiv \; \frac{1}{(2 \pi \beta)^2} 
\; .
\label{couplingsdual}
\end{equation}
Alternatively we may write the transformation of the couplings completely in terms of the more physical parameters 
$e$ and $\theta$ (use \eqref{find_gamma} and \eqref{find_e})  
\begin{equation}
\frac{1}{\widetilde{e}^{\;2}} \; = \; \frac{1}{e^2}  \, f \quad, \qquad   
\widetilde{\theta} \; = \; - \, \theta \, f 
\qquad \mbox{with} \qquad 
f \; \equiv \; \frac{1}{\left(\frac{2 \pi}{e^2}\right)^2 + \left(\frac{\theta}{2 \pi}\right)^2 } 
\; .
\label{couplingsdual_phys}
\end{equation}
These equations constitute the generalization of the duality relation Eq.~(\ref{boltzmann_both_duality}) to the case of a 
Boltzmann factor that also includes the topological term. We remark at this point that the self-duality of QED with a $\theta$-term
can be extended to an even more general definition of the topological charge, that also fully implements all lattice symmetries. 
For the corresponding discussion see Appendix~\ref{app:gen_top_charge}.

As before we check that repeating the duality transformation provides the identity map. This property follows from
$\; \widetilde{\!\!\widetilde{A}}^{\atop e} = - A^e$, $\; \widetilde{\!\!\widetilde{A}}^{\atop m} = - A^m$ and the trivial identities
\begin{equation}
\widetilde{\!\widetilde{\beta}} \; = \; \beta \quad , \quad \widetilde{\!\widetilde{\theta}} \; = \; \theta \qquad \mbox{and} \qquad
C_K(\beta) \, C_K(\widetilde{\beta}) \; = \; 1 \; .
\end{equation}

We may now use the Boltzmann factor \eqref{B_selfdual_final} in the partition sum (\ref{Zfulltheta}) and based on 
\eqref{Boltzmann_theta_duality_final} obtain the self-duality relation for full QED with a $\theta$-term,
\begin{eqnarray}
\hspace*{-10mm} && \beta^{\frac{3V}{2}} \, Z(\beta, \theta, M^e\!,\lambda^e\!,q^e\!,\,M^m\!,\lambda^m\!,q^m) \; = \; 
\widetilde{\beta}^{\frac{3V}{2}}  \,
Z\big(\widetilde{\beta}, \widetilde{\theta}, \widetilde{M}^e\!,\widetilde{\lambda}^e\!,
\widetilde{q}^{\,e}\!,\widetilde{M}^m\!,\widetilde{\lambda}^m\!,\widetilde{q}^{\;m}\big) \; ,
\label{Zfulldualitytheta}  \\
\hspace*{-10mm}  && \qquad \mbox{with} \quad \;
\widetilde{\beta} \, = \, \beta \, f
\;, \; 
\widetilde{\theta} \, = \, - \, \theta \, f 
\;, \; 
f \, = \, \frac{1}{4 \pi^2 \beta^2} 
\;, 
\nonumber \\
\hspace*{-10mm}  && \hspace{18mm}
\widetilde{M}^e \, = \, M^m  \; , \;
\widetilde{\lambda}^e  \, = \, \lambda^m \; , \;
\widetilde{M}^m  \, = \, M^e \; , \; 
\widetilde{\lambda}^m \, = \, \lambda^e \; , \; 
\nonumber \\
\hspace*{-10mm}  && \hspace{18mm}
\widetilde{\bf q} \, = \, S \, {\bf q} \; \; , \; \;   
S=\begin{pmatrix}
 0 & 1\\
 -1&0
\end{pmatrix}\;,
\nonumber
\end{eqnarray}
where again we use the vector of charges ${\bf q} = (q^e, q^m)^t$. 

As before we can generate self-duality relations for observables by evaluating derivatives 
of $\ln Z$ with respect to the couplings. When considering
observables for the matter fields, which are obtained from derivatives with respect to 
$M^e\!,\,\lambda^e\!,\,M^m$ or $\lambda^m$, the
self-duality relations generalize in a straightforward way. For example the 
relation (\ref{duality_phi2}) now reads, 
\begin{equation}
\left\langle | \phi^e |^2 \right\rangle_{\beta, \theta, M^e\!,\,\lambda^e\!,\,q^{\,e}\!,\,M^m\!,\,\lambda^m\!,\,q^{\,m}} \; = \; 
 \left\langle | \phi^m |^2 \right\rangle_{\widetilde{\beta}, \widetilde{\theta}, \widetilde{M}^e\!,\,\widetilde{\lambda}^e\!,
\widetilde{q}^{\,e}\!,\,\widetilde{M}^m\!,\,\widetilde{\lambda}^m\!,\,\widetilde{q}^{\,m}} \; .
\label{duality_phi2_theta}
\end{equation}
However, self-duality relations for observables that contain the gauge fields, i.e., observables that are generated by derivatives with respect to 
$\beta$ or $\theta$ require a little more work, since via $\gamma$ as given in (\ref{find_gamma}) the two parameters mix, and additional terms 
appear in the self-duality relations. We derive two such self-duality relations which for notational convenience we discuss without matter fields, 
i.e., we derive them for pure gauge theory with a $\theta$-term (adding matter fields is trivial). 

Applying a derivative with respect to $\beta$ on the Boltzmann factor in the form of Eq.~\eqref{B_selfdual_final} generates the insertion 
of the action $S_g$, such that we find (the matter couplings were omitted as arguments in all expressions  
since we discuss the pure gauge case),

\begin{equation}
- \frac{\partial \ln Z(\beta, \theta)}{\partial \beta}   \, = \,  \big\langle  S_g \big\rangle_{\!\beta, \theta} \, + \,   
\beta \left\langle \! \frac{\partial S_g }{\partial \beta} \! \right\rangle_{\!\!\!\beta, \theta}   \!\!\! = \,
\big\langle  S_g \big\rangle_{\!\beta, \theta}  \, + \, 
\beta  \frac{\partial \gamma }{\partial \beta} \big\langle S_g \big\rangle_{\!\beta, \theta}   
\,  = \,
\big\langle  S_g \big\rangle_{\!\beta, \theta}  \, - \, 
\frac{\Gamma \, \theta}{\beta} \big\langle S_g \big\rangle_{\!\beta, \theta} .
\label{deriv_dualrel1}
\end{equation}
The second terms on the right hand sides come from the $\beta$-dependence of the action $S_g$ via the 
$\beta$-dependence of $\gamma$ and we use the notation $S_g^{\prime} = d \, S_g / d \gamma$.
It is straightforward to compute the derivative $\frac{\partial \gamma }{\partial \beta}$ that appears after the second step from the 
explicit expression (\ref{find_gamma}), and we write the result in the form 
\begin{equation}
\frac{\partial \gamma }{\partial \beta} \; = \; - \frac{ \Gamma \, \theta}{\beta} \qquad \mbox{with} \qquad 
\Gamma \; \equiv \; \frac{ \beta^3 (4\pi^2)^2}{\big( (4 \pi^2 \beta)^2 - \theta^2\big)^{3/2}}
\; = \; \frac{ \widetilde{\beta}^3 (4\pi^2)^2}{\big( (4 \pi^2  \widetilde{\beta})^2 -  \widetilde{\theta}^2\big)^{3/2}} \; \equiv \; 
\widetilde{\Gamma} \; ,
\end{equation}
where we have factored out the combination $\Gamma$ which is invariant under the duality transformation as follows immediately 
from the transformation properties (\ref{couplingsdual}). Using the self-duality relation (\ref{Zfulldualitytheta}) for the partition sum 
we can apply the derivative with respect to $\beta$ also on the dual form,
\begin{eqnarray}
- \frac{\partial \ln Z(\beta, \theta)}{\partial \beta}  & = &  - 
\frac{\partial}{\partial \beta} \ln \left( (2\pi \beta)^{-3V} Z(\widetilde{\beta}, \widetilde{\theta}) \right) \; = \; \frac{3 V}{\beta}
- \frac{d \widetilde{\beta}}{ d \beta} \frac{\partial \ln Z(\widetilde{\beta}, \widetilde{\theta})}{\partial \widetilde{\beta}}
\nonumber \\
& = & \frac{3 V}{\beta} - \frac{1}{4 \pi^2 \beta^2} \left[ \big\langle  S_g \big\rangle_{\widetilde{\beta}, \widetilde{\theta}}  \, - \,
\frac{\widetilde{\Gamma} \, \widetilde{\theta} }{\widetilde{\beta}} 
\big\langle S_g^{\prime} \big\rangle_{\widetilde{\beta}, \widetilde{\theta}} \right] \; .
\label{deriv_dualrel2}
\end{eqnarray} 
Setting equal the right hand sides of (\ref{deriv_dualrel1}) and (\ref{deriv_dualrel2}) one obtains after a few steps of 
trivial reordering of terms (use also $\widetilde{\beta} = 1/4\pi^2 \beta$ and $\widetilde{\Gamma} = \Gamma$) 
the final form of the self-duality relation 
\begin{equation}
\beta \, \big\langle  S_g \big\rangle_{\!\beta, \theta}  \; - \; 
\Gamma \, \theta \, \big\langle S_g^{\prime } \big\rangle_{\!\beta, \theta}  \; = \; 3V \; - \; 
\widetilde{\beta} \,\big\langle  S_g \big\rangle_{\widetilde{\beta}, \widetilde{\theta}}  \; + \;
\Gamma \, \widetilde{\theta} \, \big\langle S_g^{\prime} \big\rangle_{\widetilde{\beta}, \widetilde{\theta}} \; .
\label{S_theta_duality}
\end{equation}
It is easy to check that for $\theta = 0$ this self-duality relation reduces to the 
self-duality relation (\ref{sumruleF2}) which we derived for pure
gauge theory without $\theta$-term (use $\theta = 0 \Rightarrow \gamma = 0$, 
the fact that $S_g$ at $\gamma = 0$
reduces to the ultra-local action we used initially, and  
$\langle F^2 \rangle =  \langle S_g \!\!\mid_{\gamma = 0} \rangle / 3V$). The generalization 
to the non-ultralocal action that is needed for $\theta \neq 0$ then generates the additional terms with $S_g^{\prime}$.

In exactly the same way we may also study derivatives $i \partial / \partial \theta$ that generate expectation values of
the topological charge $Q_0$, and following the same steps as above find the corresponding self-duality relation 
\begin{equation}
\beta \, \big\langle  Q_0 \big\rangle_{\!\beta, \theta}  \; - \; 
i \, \Gamma \, \beta \, \big\langle S_g^{\prime} \big\rangle_{\!\beta, \theta}  \; = \; 
- \; \widetilde{\beta} \,\big\langle  Q_0 \big\rangle_{\widetilde{\beta}, \widetilde{\theta}}  \; + \;
i \, \Gamma \, \widetilde{\beta} \, \big\langle S_g^{\prime} \big\rangle_{\widetilde{\beta}, \widetilde{\theta}} \; .
\label{Q_theta_duality}
\end{equation}
This self-duality relation vanishes for $\theta = 0$ (as expected), since $\langle  Q_0 \rangle \big|_{\theta = 0} = 0$ and
due to $\theta = 0 \Rightarrow \gamma = 0$ also the second term disappears because of $S_g^{\prime} \big|_{\gamma = 0} = 0$. 


\subsection{The $\SL(2,\mathbb Z)$ structure of dyonic lattice QED with a $\theta$-term}
\label{sec:sign-problem}

In the course of this paper we have identified two transformations of our self-dual lattice version of QED with a $\theta$-term. The first 
one is the duality transformation ${\cal S}$ itself, which we here write in terms of the bare charge parameter $e$, the topological angle $\theta$ 
and the vector of electric and magnetic charges $\bm q \equiv (q^e,q^m)^t$ (compare \eqref{couplingsdual_phys}),
\begin{eqnarray}
&& \frac{1}{e^2} \; \rightarrow \; \frac{1}{\widetilde{e}^{\;2}} \; = \;  \frac{1}{e^2}  \, f
\quad, \qquad  
\theta \; \rightarrow \widetilde{\theta} \; = \; - \, \theta \, f 
\qquad \mbox{where} \qquad 
f \; \equiv \; \frac{1}{\left(\frac{2 \pi}{e^2}\right)^2 + \left(\frac{\theta}{2 \pi}\right)^2 } \; ,
\nonumber \\
&&
\bm q \; \rightarrow \; \widetilde{\bm q} \; = \; S \, \bm q  \qquad \mbox{with} \qquad  
S \; =\; \begin{pmatrix}
0 & 1\\
-1 &0
\end{pmatrix} \; .
\label{S_trafo}
\end{eqnarray}

The second transformation we identified is ${\cal T}$ which shifts $\theta$ by $2 \pi$ that was discussed in 
Subsection~\ref{subsec:witten} in the context of the Witten effect. It acts only on $\theta$ and the charge vector 
$\bm q$ and is defined as follows (see \eqref{eq:T_matrix})
\begin{eqnarray}
&& \theta \; \rightarrow \; \overline{\theta} \; = \; \theta - 2 \pi \; , 
\nonumber \\
&&
\bm q \; \rightarrow \; \overline{\bm q} \; = \; T \, \bm q  \qquad \mbox{with} \qquad  
T \; = \; \begin{pmatrix}
 1 & 1\\
 0&1
\end{pmatrix}\; .
\label{T_trafo}
\end{eqnarray}

We begin the discussion of the overall duality structure by noting that the matrices $S$ and $T$ 
are the generators\footnote{See for example \cite{conrad} for an elementary introduction to the group $\SL(2,\mathds{Z})$.} of the group 
$\SL(2,\mathbb Z)$, which is the group of $2 \times 2$ matrices $M$ with integer-valued elements and determinant 1, i.e., 
\begin{equation}
M \; = \; \begin{pmatrix}
 a & b\\
 c &d
\end{pmatrix} \quad \mbox{with} \quad a,b,c,d \, \in \, \mathds{Z} \quad \mbox{and} \quad ad - bc \, = \, 1 \; .
\label{M_matrix}
\end{equation}
And since they are generators, combinations of the matrices $S$ and $T$ implement arbitrary $\SL(2,\mathbb Z)$ 
transformations on the charge vector $\bm q$. 

Furthermore, also the action of the dualities ${\cal S}$ and ${\cal T}$ on $1/e^2$ and $\theta$ 
can be identified with the group $\SL(2,\mathbb Z)$. 
In order to see that, we combine the gauge coupling $e$ and the topological angle $\theta$ into a joint complex-valued
coupling $\tau$, the so-called modular parameter of U(1) gauge theory (see, e.g., \cite{Cardy:1981fd,Shapere:1988zv,Deligne:1999qp}), 
defined as
\begin{equation}\label{eq:tau}
\tau \; \equiv \; i \, \frac{2\pi}{e^2}-\frac{\theta}{2\pi} \; .
\end{equation}
It is straightforward to see from \eqref{S_trafo} and \eqref{T_trafo} 
that our two generators ${\cal S}$ and ${\cal T}$ act as
\begin{equation}
{\cal S}: \; \tau \, \rightarrow \, \widetilde{\tau} \; = \; -\frac{1}{\tau} \quad \mbox{and} \quad 
{\cal T}: \; \tau \, \rightarrow \, \overline{\tau} \; = \; \tau \, + \, 1 \; . 
\label{tautrafo}
\end{equation}
The action of a general $\SL(2,\mathbb Z)$ transformation on a complex number $\tau$ is defined 
as\footnote{Strictly speaking this implements the projected group 
$\PSL(2,\mathbb Z)$, which is $\SL(2,\mathbb Z)$ with matrices $M$ and $-M$ identified.}
\be
\tau \; \rightarrow \;  M\, \tau \; \equiv \; \frac{a \tau +b}{c\tau+d} \; ,
\label{tautrafo_full}
\ee
where $a,b,c,d$ are the entries of an $\SL(2,\mathbb Z)$ matrix $M$ as given in \eqref{M_matrix}. 
It is straightforward to see that choosing the generators 
$M = S$ and 
$M = T$ gives rise to the transformations $\tau \rightarrow - 1/\tau$ and $\tau \rightarrow \tau + 1$, i.e., the action of 
our symmetries ${\cal S}$ and  ${\cal T}$  as stated in \eqref{tautrafo}. Thus our transformations ${\cal S}$ and  ${\cal T}$ 
generate the full set of group transformations \eqref{tautrafo_full}. We conclude that combining our two transformations 
${\cal S}$ and ${\cal T}$ as stated in \eqref{S_trafo} and \eqref{T_trafo} gives rise to a full $\SL(2,\mathbb Z)$
invariance of our self-dual lattice formulation of QCD with a $\theta$-term.  

We now ignore the other couplings of the dyonic matter fields, such as
$M$ and $\lambda$, and only focus on the couplings 
$1/e^2$ and $\theta$ combined into the complex coupling $\tau$, as well as the charge vector $\bm q$. Only those
couplings are now listed as arguments of the partition function. 
Furthermore it is convenient to redefine the partition function as  $\overline Z(\tau,\bm q) \equiv \beta^{\frac{3V}{2}}Z(\tau,\bm q)$. 
Then the ${\cal S}$ and ${\cal T}$ symmetry relations are written as
\begin{equation}
\overline{Z}(\tau, \bm q) \; = \; \overline{Z}(S \tau, S \bm q) \qquad \mbox{and} \qquad 
\overline{Z}(\tau, \bm q) \; = \; \overline{Z}(T \tau, T \bm q) \; ,
\end{equation}
and according to the discussion above these two transformations generate the full $\SL(2,\mathbb Z)$ invariance given by
\begin{eqnarray}
\hspace*{-20mm} && \overline{Z}(\tau, \bm q) \; = \; \overline{Z}(M \tau, M \bm q) \qquad \mbox{with} \quad 
\label{SL2Z_final}
\\
\hspace*{-20mm} && 
M \; = \; \begin{pmatrix}
 a & b\\
 c &d
\end{pmatrix} \; , \; \quad a,b,c,d \, \in \, \mathds{Z} \; , \;  ad - bc \, = \, 1 \; , \;  M\, \tau \; = \; \frac{a \tau +b}{c\tau+d}\; .
\nonumber
\end{eqnarray} 
These equations describe the symmetry content of dyonic self-dual QED with a $\theta$-term.

\begin{figure}[p] 
   \centering
   \includegraphics[width=0.98\textwidth]{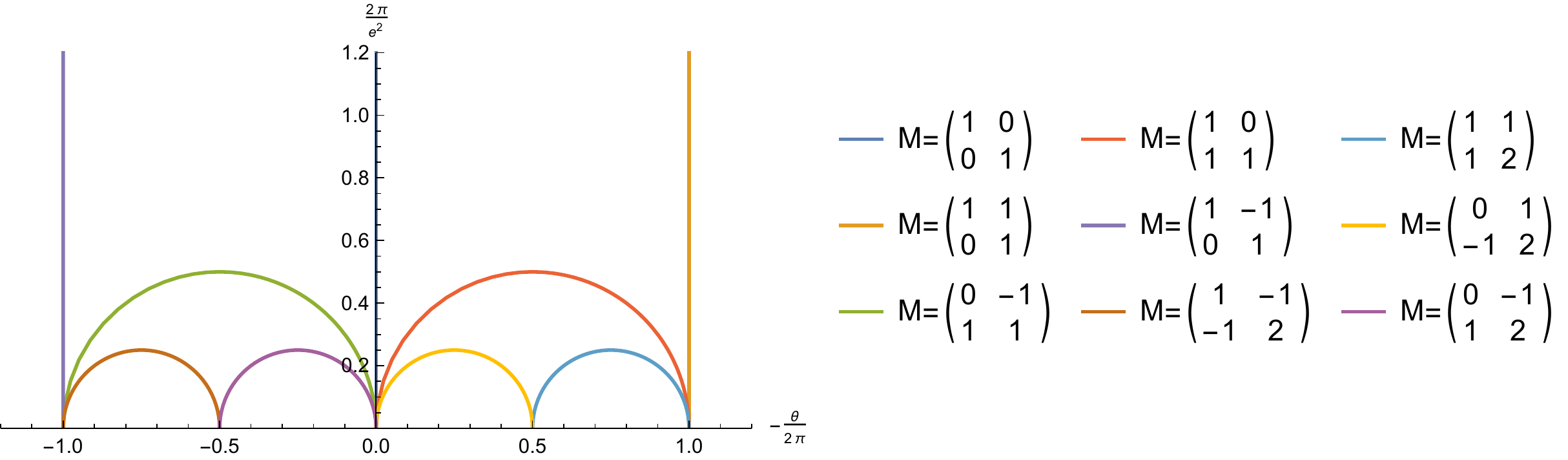} 
   \caption{Mapping of theories in the complex $\tau$ plane that are related by different elements of $\SL(2,\mathbb Z)$.  On the 
   vertical axis we plot Im $\!\tau =  2\pi / e^2$ and on the horizontal axis Re $\!\tau = - \theta / 2\pi$. Theories that 
   have $\theta = 0$ and thus can be directly simulated are located on the positive vertical axis (marked with a thick black line). Different 
   elements of $\SL(2,\mathbb Z)$ map this family of theories to equivalent theories with non-zero $\theta$ which are represented 
   by the semi-circles (see the legend for which element $M$ of $\SL(2,\mathbb Z)$ was used).}  
   \label{fig:SL2Z}
\vskip20mm
   \centering
   \includegraphics[width=0.6\textwidth]{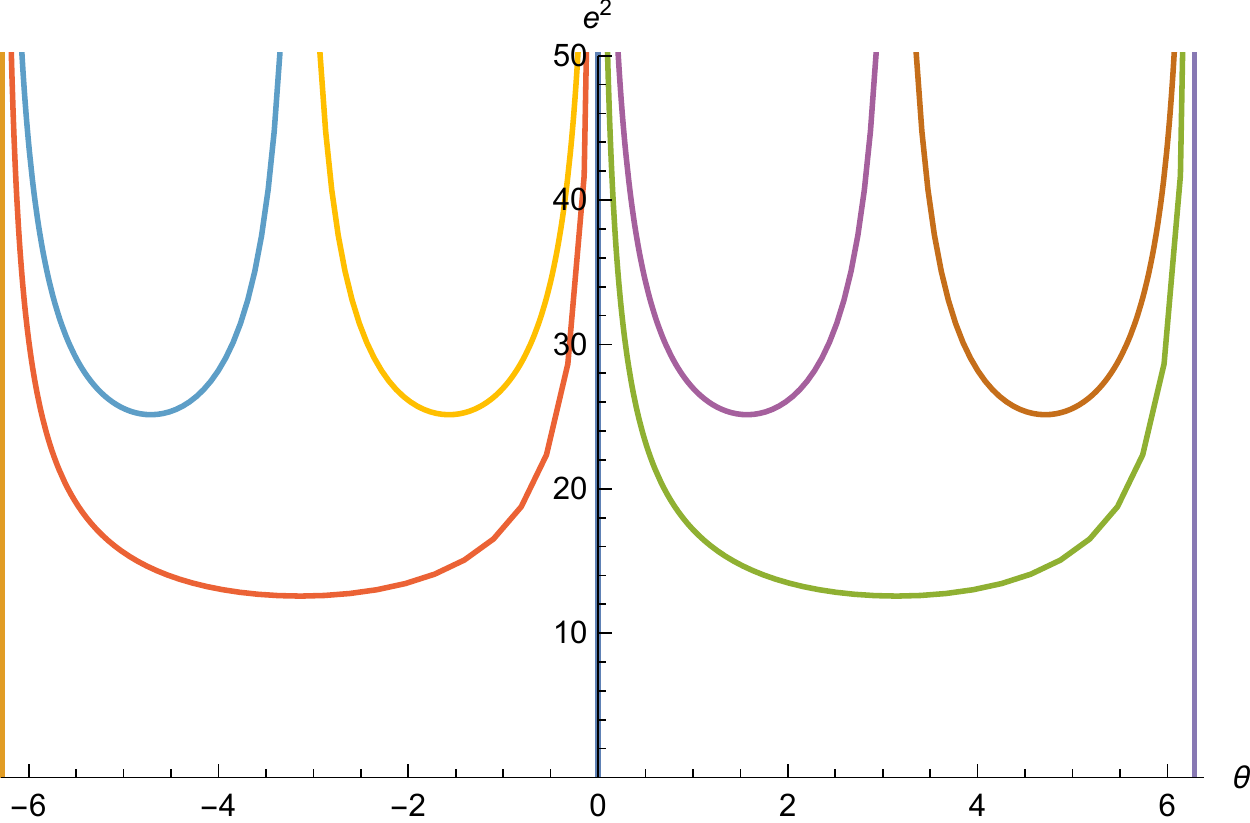} 
   \caption{The same as Fig.~\ref{fig:SL2Z}, but now plotted in the $(\theta,e^2)$ plane.}  
   \label{fig:SL2Z_e_theta}
\end{figure}

As we have seen in Sec.~\ref{sec:dyonic_theories}, dyonic theories generically have a sign problem. In addition, also theories with a $\theta$-term have a 
sign problem. However, the $\SL(2,\mathbb Z)$ structure we identified above allows us to map theories which can be recast in terms of wordlines without the 
sign-problem (i.e., with the purely electric and/or purely magnetic matter, at least one of which is bosonic and considered at $\theta=0$) onto other 
theories that naively do have a sign problem. This map can be rather nontrivial. To illustrate the idea, let us discuss an example.
We consider a matrix $M\in \SL(2,\mathbb Z)$ given by
\be
M=\begin{pmatrix}
1 & 0\\
1 & 1
\end{pmatrix}\;.
\ee
We have seen that a theory described by $(\tau,\bm q)$ is equivalent to a theory with parameters $(M\tau,M\bm q)$. 
Furthermore we can  simulate the family of theories with $\tau= i y, y\in\mathbb R_+$ and, say a matter field with charge vector $\bm q=(n,0)^t$ with 
$n\in \mathbb Z$. This family of theories is dual to theories with $\bar\tau=M\tau= \frac{i y}{i y+1}$ and $\bar{\bm q}=M\bm q=(n,n)^t$. 
Identifying $\text{Re}(\bar \tau)=\frac{\bar\theta}{2\pi}$ and $\text{Im}(\bar \tau)=\frac{2\pi}{\bar{e}^2}$, we find that the dual theories have parameters 
and matter charges charges given by  ($y \in \mathds{R}_+$)
\be
\bar{e}^2=2\pi\frac{1+y^2}{y}\;,\qquad \bar\theta=2\pi\frac{y^2}{1+y^2}\;,\qquad {\text{ and    }}\qquad \bar{\bm q} = \begin{pmatrix}
n\\
n
\end{pmatrix} \; .
\ee
Obviously this is a family of theories that has a complex action problem, which, using our $\SL(2,\mathbb Z)$ mapping we may study via simulating the 
corresponding sign free set of theories. 
The above family of theories is shown as an orange semi-circle in the $\tau$ plane in Fig.~\ref{fig:SL2Z}, or equivalently, in the $(\theta,e^2)$ plane
in Fig.~\ref{fig:SL2Z_e_theta} (again the orange curve). Note that in particular the self-dual point of the original theory 
$y=1$ is mapped to
 $e^2=4\pi$ and $\bar \theta = \pi$ of the dual theory. 

Figures \ref{fig:SL2Z} and \ref{fig:SL2Z_e_theta} show several $\SL(2,\mathbb Z)$ families of theories that are dual to the family of theories 
which obeys $\text{Im}(\tau)=0$, i.e., the $\theta=0$ theories, that can be simulated as long as the matter field is not dyonic, i.e., the matter fields are 
purely electric or purely magnetic. The legend shows the matrices $M$ that need to be applied to the matter fields in the exactly same manner as in the 
example above. Note that all the interesting theories which are related to the sign-problem free theory are at strong electric coupling $\bar e^2\ge 4\pi$, as is 
evident from Fig.~\ref{fig:SL2Z_e_theta}.


\section{Concluding remarks}

In this work we discussed the construction of U(1) gauge theories in 4d on the lattice with $\theta$-terms. 
We showed that modified Villain actions allow for a construction which admits the inclusion of $\theta$-terms in a natural way. 
However, a naive ultra-local $\theta$-term ruins the self-dual nature of the theory, mapping the ultra-local theory to a merely local one. 
We showed that by generalizing the modified Villain actions to include non-ultra-local terms, we can restore the exactly self-dual 
nature of the model. The construction is reminiscent of the inability to regulate a lattice theory in an ultra-local way, while preserving the 
axial symmetry in 4d lattice gauge theories. Indeed, the Nielsen-Ninomiya theorem \cite{Nielsen:1980rz,Nielsen:1981xu,Nielsen:1981hk} 
prohibits a lattice regularization for which the matrix $\gamma^5$ anti-commutes with the Dirac operator, which is essential for the axial symmetry. 
However, using the Ginsparg-Wilson relation \cite{Ginsparg:1981bj} and its solution by Neuberger \cite{Neuberger:1997fp}, L\"uscher showed 
\cite{Luscher:1998pqa} that an axial symmetry can be defined. But such constructions, while local, are not ultra-local. In fact 4d U(1) 
gauge theories, both free and interacting, have anomalies \cite{Kravec:2013pua,Gaiotto:2014kfa,Komargodski:2017dmc,Komargodski:2017smk} 
and it is not a priory a big surprise that for maintaining the correct symmetry and anomaly structure on the lattice one must resort to a local, 
but not ultra-local form of the action. Yet many examples exist of lattice discretization of theories with anomalies, 
which maintain their ultra-local form. One example is a $\theta=\pi$ $2d$ abelian gauge theory, whose lattice action was discussed in 
\cite{Gattringer:2018dlw,Goschl:2018uma,Sulejmanpasic:2019ytl,Anosova:2019quw,Sulejmanpasic:2020lyq,Sulejmanpasic:2020ubo}, 
and may be chosen ultra-local, even though such models may have `t Hooft anomalies involving their internal symmetries. Such models are also 
connected to half-integral spin-chains, which furnish a regulator of the Hilbert space.

It is well known, that the free abelian gauge theory has an $\SL(2,\mathbb Z)$ structure in the continuum, generated by self-duality and shifts 
of the $\theta$-angle by $2\pi$. As Witten showed in \cite{Witten:1995gf}, the free U(1) gauge theory partition function transforms as a 
modular form, with a modular weight determined by the Euler characteristic $\chi$ and the signature of the spin-for manifold $\sigma$. In 
\cite{Honda:2020txe} this was interpreted as a `t Hooft anomaly (see Section 4.2.2 of \cite{Honda:2020txe}) at special points of the modular 
parameter $\tau$, where the theory enjoys a $\mathbb Z_6$ symmetry, which is a subgroup of $\SL(2,\mathbb Z)$. As explained by Honda and 
Tanizaki in \cite{Honda:2020txe}, this $\mathbb Z_6$ symmetry has a mixed anomaly with gravity. On the other hand, the mixed axial-gravitational 
anomaly of fermions is also well known (see, e.g., \cite{Delbourgo:1972xb,Eguchi:1976db,Alvarez-Gaume:1983ihn}). Furthermore, the role of gravitational 
anomalies in lattice discretization was noted in \cite{Hellerman:2021fla}, where it was proven (under some mild assumptions) that there exists 
no lattice discretization\footnote{The proof was for a lattice discretization of Hilbert space, where time is viewed as continuous.} in two dimensions 
unless its pure gravitational anomaly vanishes. One may then naturally wonder if the inability to write ultra-local actions is related 
to the involvement of mixed-gravitational anomalies.

\vskip6mm
\noindent
{\bf Acknowledgments:} 
We thank Simeon Hellerman, Nabil Iqbal, Masataka Watanabe for discussions related to this work. We also thank Simon Hands and Simon Catterall 
who initially encouraged us to write this work for a collection they are editing. TS is supported by the Royal Society University Research Fellowship. 

\vskip8mm

\noindent
{\LARGE \bf Appendices}

\appendix

\section{Notation and results for lattice differential forms}\label{app:diff_forms}
In this appendix we summarize the notation and some basic results for differential forms on the lattice, which
we use in our paper. For a more general presentation see, e.g., \cite{wallace,Sulejmanpasic:2019ytl}. In the main part of the paper we work 
in $d = 4$ dimensions, but this appendix is kept general with arbitrary $d$.

We consider a $d$-dimensional hypercubic lattice $\Lambda$ 
and set the lattice spacing to $a = 1$. The spatial extents of the lattice  
are denoted as $N_\mu, \, \mu = 1,2 \, ... \, d$ and we use periodic boundary conditions. Sites are denoted as $x$. Links
are denoted as $(x,\mu)$, plaquettes as $(x, \mu \nu)$ with $\mu < \nu$, 3-cubes as $(x, \mu \nu \rho)$ with 
$\mu < \nu < \rho$ et cetera. 

We can identify these elements with $r$-cells: Sites are 0-cells.  A link $(x,\mu)$ is a 1-cell and contains the 0-cells 
(sites) $x$ and $x+\hat \mu$ where by $\hat \mu$ we denote the unit vector in direction $\mu$. 
A plaquette $(x,\mu \nu)$ is a 2-cell and contains the sites $x$, $x+\hat{\mu}$, 
$x+\hat{\nu}$ and $x+\hat{\mu} + \hat{\nu}$, as well as the links $(x,\mu)$, $(x + \hat{\mu},\nu)$, $(x + \hat{\nu},\mu)$
and $(x,\nu)$. In an equivalent way 3-cells are the 3-cubes that contain 8 sites, 12 links and 6 plaquettes and similarly for 
higher $r$-cells. The site $x$ in the definition of an $r$-cell we will sometimes refer to as \emph{root site} or
\emph{root}.  

We assign integers or real numbers to the $r$-cells and refer to these as $r$-forms. We denote $r$-forms as 
$f_{x,\,\mu_1 \mu_2 \, ... \, \mu_r} \in \mathds{R}$ or $\mathds{Z}$ with $\mu_1 < \mu_2 < \, ... \, < \mu_r$, i.e., 
the $r$-forms are labelled by the root site and the ordered indices $\mu_j, \, j = 1,2, \, ... \, r$ that label the 
corresponding $r$-cell. It is convenient to not only label the $r$-forms with the ordered indices 
$\mu_1 < \mu_2 < \, ... \, < \mu_r$ but allow for arbitrary ordering with the convention that 
$f_{x,\,\mu_1 \, ... \, \mu_j \, ... \, \mu_k \, ... \, \mu_r} = - f_{x,\,\mu_1 \, ... \, \mu_k \, ... \, \mu_j \, ... \, \mu_r}$,
i.e., $r$-forms are antisymmetric in their indices. This also implies that an $r$-form vanishes when two or more of the 
indices are equal. One may use this generalized labelling with non-ordered indices also for the cells and the sign of the
permutation of the indices defines an orientation for the cells.

We define two discrete differential operators that act on $r$-forms, the \emph{exterior derivative} 
$d$ that takes an $r$-form into an $(r\!+\!1)$-form with the 
definition that $d$ acting on a $d$-form gives zero. The other differential operator is the 
\emph{boundary operator} $\partial$
that takes an $r$-form into an $(r\!-\!1)$-form. When acting on a $0$-form $\partial$ gives zero. The two differential 
operators are defined as
\begin{eqnarray}
(d\,f)_{x,\,\mu_1 \mu_2 \, ... \, \mu_r} & \equiv & 
\sum_{j = 1}^r (-1)^{j+1} \, \Big[ \,f_{x+\hat{\mu}_j, \,\mu_1 \, ... \, {\mu}^{\!\!\!^o}_j  \, ... \, \mu_r} \, - \, 
f_{x, \,\mu_1 \, ... \, {\mu}^{\!\!\!^o}_j \, ... \, \mu_r} \,\Big] \; ,
\nonumber \\
(\partial f)_{x,\,\mu_1 \mu_2 \, ... \, \mu_r} & \equiv & \sum_{\nu = 1}^d \, \Big[ \, f_{x, \, \mu_1 \, ... \, \mu_r \, \nu} \, - \, 
f_{x - \hat{\nu},\, \mu_1 \, ... \, \mu_r \, \nu} \, \Big] \; ,
\label{AppA:diffdef}
\end{eqnarray}
where in the first equation ${\mu}^{\!\!\!^o}_j$ indicates that $\mu_j$ is dropped from the list of 
indices. We remark that due to the 
antisymmetry of the $r$-forms with respect to interchange of their indices, in the second line of (\ref{AppA:diffdef})
only those $(r\!+\!1)$-forms contribute to the sum where $\nu$ is different from all indices $\mu_1, \mu_2 \, ... \, \mu_r$.   
The ordering of these indices and $\nu$ determines the sign of the contribution. 

Both differential operators are nilpotent, i.e., $d^{\,2} = 0$ and $\partial^{\,2} = 0$. Furthermore one may show the 
following partial integration formula, where $f$ is an $r$-form and $g$ an $(r\!-\!1)$-form,
\begin{equation}
\sum_{x \atop \mu_1 <  \, ... \, < \,\mu_r} \!\!\!\!\!
f_{\, x, \, \mu_1 \, ... \, \mu_r} \; (d\,g)_{\,x,\,\mu_1 \, ... \, \mu_r} \quad = \quad (-1)^r  \!\!\!\!\!\!\!
\sum_{x \atop \mu_1 <  \, ... \, <  \, \mu_{r-1}} \!\!\!\!\!
(\partial f)_{\,x, \, \mu_1 \, ... \, \mu_{r-1}} \; g_{\,x,\,\mu_1 \, ... \, \mu_{r-1}} \; .
\label{AppA:partint}
\end{equation}
The \emph{Hodge decomposition} states that an arbitrary $r$-form $f$ can be written as the sum of the boundary 
operator $\partial$ acting on an $(r\!+\!1)$-form $p$, the exterior derivative operator $d$ acting on an $(r\!-\!1)$-form $q$
and a \emph{harmonic} or \emph{defect} $r$-form $h$ that obeys $d \, h = 0$ and $\partial \, h = 0$,
\begin{equation}
f_{\, x, \, \mu_1 \, ... \, \mu_r} \; = \; (\partial \, p)_{\,x,\,\mu_1 \, ... \, \mu_r} \; + \; 
(d\,q)_{\,x,\,\mu_1 \, ... \, \mu_r} \; + \; h_{\, x, \, \mu_1 \, ... \, \mu_r} \; .
\label{AppA:hodge}
\end{equation}

The \emph{dual lattice} $\tilde\Lambda$ is defined by a one-to-one identification of the $r$-cells of the 
original lattice with the $(d\!-\!r)$-cells of the dual lattice. The sites $\tilde{x}$ of $\tilde\Lambda$ are
at the centers of the $d$-cells of $\Lambda$, i.e., $\tilde{x} = x + \frac{1}{2} ( \hat{1} + \hat{2} + ... + \hat{d}\,)$. 
There is a natural identification of the $r$-forms $f_{x,\,\mu_1 \, ... \, \mu_r}$ on $\Lambda$ with the $(d\!-\!r)$-forms on 
$\tilde\Lambda$. We denote these as $\tilde{f}_{\tilde{x},\,\nu_{r+1} \, ... \, \nu_{d}}$ but stress that the numerical value 
is of course the same as the $r$-form $f_{x,\,\mu_1 \, ... \, \mu_r}$ we identify it with. The identification is given by
\begin{equation}
f_{x,\,\mu_1 \, ... \, \mu_r} \; = \; \sum_{\nu_{r+1} < \,\nu_{r+2} < \, ... \, < \,\nu_{d}}
\epsilon_{\mu_1 \, ... \, \mu_r \, \nu_{r+1} \, ... \, \nu_{d}} \; 
\tilde{f}_{\tilde{x} - \hat{\nu}_{r+1} - \hat{\nu}_{r+2} \, ... \, - \hat{\nu}_{d} , \, \nu_{r+1} \, ... \, \nu_{d}} \; .
\label{AppA:dualforms}
\end{equation} 
Finally, when switching between the original lattice $\Lambda$ and the dual lattice $\tilde \Lambda$ the exterior
derivative $d$ and the boundary operator are converted into each other,
\begin{equation}
(\partial \, f)_{x,\,\mu_1 \, ... \, \mu_r} \; = \; \sum_{\nu_{r+1} < \,\nu_{r+2} < \, ... \, < \,\nu_{d}}
\epsilon_{\mu_1 \, ... \, \mu_r \, \nu_{r+1} \, ... \, \nu_{d}} \; 
( d\, \tilde{f})_{\tilde{x} - \hat{\nu}_{r+1} - \hat{\nu}_{r+2} \, ... \, - \hat{\nu}_{d} , \,\nu_{r+1} \, ... \, \nu_{d}} \; .
\label{AppA:diffconvert}
\end{equation}


\section{Auxiliary results for the topological charge}\label{app:B}

In this appendix we collect some results for the topological charge $Q_T[F]$ which can be obtained with 
elementary although lengthy algebra. As a first result we show (\ref{property1}), i.e., the fact that the topological 
charge $Q_T[F]$ is invariant when adding the exterior derivative $(dB)_{x,\mu\nu}$ of an arbitrary 1-form $B_{x,\mu}$ 
to the field strength $F_{x,\mu\nu} = (dA^e)_{x,\mu\nu} + 2 \pi \, n_{x,\mu\nu}$, as long as the Villain variables 
$n_{x,\mu\nu}$ obey the closedness condition (\ref{Villainconstraint}), i.e., $(dn)_{x,\mu \nu \rho} = 0$.
The expression $Q_T[F \!+\! dB]$ can be trivially decomposed into three contributions, 
\begin{eqnarray}
Q_T[F \!+\! dB] & \!\!\! =  \!\!\!  &  \frac{1}{8 \pi^2} \sum_x \sum_{\mu < \nu \atop \rho < \sigma} 
(F + dB)_{x,\mu \nu} \, \epsilon_{\mu \nu \rho \sigma} \, (F + dB)_{x-\hat{\rho} - \hat{\sigma} + T,\rho \sigma} 
\; = \; Q_T[F] \, + \,Q_T[(dB)]  
\nonumber   \\
&  \!\!\!  +  \!\!\!  &  \frac{1}{8 \pi^2} \sum_x \sum_{\mu < \nu \atop \rho < \sigma} 
\Big[ F_{x,\mu\nu} (dB)_{x-\hat{\rho} - \hat{\sigma} +T,\rho \sigma} \, + \, 
F_{x-\hat{\rho} - \hat{\sigma} +T,\rho \sigma}(dB)_{x,\mu\nu} \, \Big]  \, \epsilon_{\mu \nu \rho \sigma} \; .
\label{AppendixB_e1}
\end{eqnarray}
For proving the claimed independence of $dB$ we need to show that the second and third terms on the right hand side 
vanish. We begin with the third term: After inserting the explicit definition of the exterior derivatives $(dB)_{x,\mu\nu}$ and 
reshuffling the summation indices the third term can be written in the form (a factor $1/8 \pi^2$ was dropped) 
\begin{eqnarray}
\hspace*{-8mm} && \sum_x \sum_{\mu < \nu \atop \rho < \sigma} \epsilon_{\mu \nu \rho \sigma} 
\Big[ B_{x,{\rho}} \; ( F_{x + \hat{\rho} + \hat{\sigma} -T,\mu \nu} - F_{x + \hat{\rho} -T,\mu \nu} ) \; - \; B_{x,{\sigma}} \; ( F_{x + \hat{\rho} + \hat{\sigma} -T,\mu \nu} - F_{x + \hat{\sigma} -T,\mu \nu} ) \; +
\nonumber \\
\hspace*{-8mm}  && B_{x,{\mu}} \; ( F_{x - \hat{\rho} - \hat{\sigma} +T,\rho \sigma} - 
F_{x - \hat{\rho} - \hat{\sigma} - \hat{\nu} +T,\rho \sigma} ) \; - \; 
B_{x,{\nu}} \; ( F_{x - \hat{\rho} - \hat{\sigma} +T,\rho \sigma} - 
F_{x - \hat{\rho} - \hat{\sigma} - \hat{\mu} +T,\rho \sigma} \; ) \Big]  \; .
\label{AppB_aux1}
\end{eqnarray}
For a fixed site $x$ we now may identify all the contributions that multiply $B_{x,\mu}$ 
for a fixed value of $\mu$. For example, the terms that
multiply $B_{x,1}$ are given by
\begin{eqnarray}
\hspace*{-8mm} 
\hspace*{-10mm} && F_{x +T  - \hat{3} - \hat{4}, 34}  - F_{x +T -\hat{2} - \hat{3} - \hat{4}, 34}  - 
F_{x +T -\hat{2} - \hat{4}, 24}  + 
F_{x +T -\hat{2} - \hat{3} - \hat{4}, 24} + F_{x +T -\hat{2} - \hat{3}, 23} 
\nonumber \\
\hspace*{-10mm} &&
 -   \, F_{x +T -\hat{2} - \hat{3} - \hat{4}, 23}  + F_{x -T + \hat{1} + \hat{4}, 23}  - F_{x -T + \hat{1}, 23}  - 
 F_{x -T  + \hat{1} + \hat{3}, 24}  
\label{AppB_aux2}
 \\
\hspace*{-10mm} &&
 +  \, F_{x -T  + \hat{1}, 24}  + F_{x -T  + \hat{1} + \hat{2}, 34} - F_{x - T  + \hat{1}, 34} 
\,  = \, (dF)_{x + T  -\hat{2} - \hat{3} - \hat{4}, 234} \, - \, (dF)_{x - T + \hat{1}, 234}  \; = \; 0 ,
 \nonumber
\end{eqnarray}
where in the last line we have combined the contributions into the exterior derivatives 
$(dF)_{x, \mu\nu\rho} = (d^{2}A)_{x, \mu\nu\rho} + 2\pi (dn)_{x, \mu\nu\rho} = 0$, which vanish due to $d^2 = 0$ and the 
closedness condition $(dn)_{x, \mu\nu\rho} = 0$ we implemented for the Villain variables. 
The same steps can be repeated for the factors 
that multiply the other $B_{x,\mu}$ thus establishing that the third term on the rhs.\ of (\ref{AppendixB_e1}) vanishes.

Let us now explore the second term on the right hand side of (\ref{AppendixB_e1}), i.e., the term $Q[(dB)]$. Dropping again a factor 
of $1/8\pi^2$ the term reads
\begin{eqnarray}
\hspace*{-8mm}  && \sum_x \sum_{\mu < \nu \atop \rho < \sigma} (dB)_{x,\mu\nu} \, \epsilon_{\mu \nu \rho \sigma} \, 
(dB)_{x - \hat{\rho} - \hat{\sigma} + T,\rho \sigma} \; = \; 
\sum_x \sum_{\mu < \nu \atop \rho < \sigma} (dB)_{x + \hat{\rho} + \hat{\sigma} - T,\mu\nu} \, 
\epsilon_{\mu \nu \rho \sigma} \, (dB)_{x,\rho \sigma} \; = 
\nonumber \\ 
\hspace*{-8mm}  && \sum_x \sum_{\mu < \nu \atop \rho < \sigma} \! \epsilon_{\mu \nu \rho \sigma} 
\Big[ B_{x + \hat{\rho}, \sigma} - B_{x, \sigma} - B_{x + \hat{\sigma}, \rho} + B_{x, \rho} \Big] \, 
(dB)_{x + \hat{\rho} + \hat{\sigma} - T,\mu\nu} \; = 
 \nonumber \\
\hspace*{-8mm}  && \sum_x \sum_{\mu < \nu \atop \rho < \sigma}  \! \epsilon_{\mu \nu \rho \sigma} 
\Big[ B_{x, \rho} \Big( (dB)_{x + \hat{\rho} + \hat{\sigma} - T,\mu\nu} - 
(dB)_{x + \hat{\rho} - T,\mu\nu} \Big) 
 \nonumber \\
\hspace*{-8mm}  && \hspace{40mm}
- \, B_{x, \sigma}
 \Big( (dB)_{x + \hat{\rho} + \hat{\sigma} - T,\mu\nu} - (dB)_{x + \hat{\sigma} - T,\mu\nu} \Big) \Big],
\end{eqnarray}
where we performed trivial shifts of site indices, explicitly wrote out one of the exterior derivatives, and organized the terms such 
that we identify the factors multiplying $B_{x,\rho}$ and $B_{x,\sigma}$. As with the third term we now collect all factors that 
multiply $B_{x,\mu}$ for a fixed $\mu$. For the case of $B_{x,1}$ this factor is given by 
\begin{eqnarray}
&& (dB)_{x - T + \hat{1} + \hat{4}, 23} \; - \; (dB)_{x - T + \hat{1}, 23} \; - \; 
(dB)_{x - T + \hat{1} + \hat{3}, 24} \;  + \; (dB)_{x + T + \hat{1}, 24} \qquad \; \\
&& \hspace*{20mm} + \; (dB)_{x - T + \hat{1} + \hat{2}, 34} \; - \; (dB)_{x - T + \hat{1}, 34}  \; = \; 
(d^{2}B)_{x - T + \hat{1}, 234} \; = \; 0 \; . \qquad 
\nonumber
\end{eqnarray}
In the second line we have identified the terms with $(d^{2}B)_{x - T + \hat{1}, 234}$, which vanishes due to the 
nilpotency of the exterior derivative. In a similar way one can treat the terms that multiply $B_{x,\mu}$ for all values of $\mu$
and show that also the second term on the rhs.\ of (\ref{AppendixB_e1}) vanishes. Thus we have proven (\ref{property1}), i.e.,
$Q_T[F \!+\! dB] = Q_T[F]$ for configurations of the Villain variables that obey the closedness condition $(dn)_{x, \mu\nu\rho} = 0$.

The second result we show in this appendix is the fact that when using the Hodge decomposition (\ref{hodge}) for the Villain variables 
the result for the topological charge $Q_T[F]$ at arbitrary $T$ is given by (\ref{Q_omega}) when the harmonic contribution is parameterized 
as in (\ref{harmonics}). Obviously
\begin{equation}
Q_T[F] =  Q_T[2\pi n] = Q_T[2\pi dl + 2\pi h]  = Q_T[2\pi h]  = \frac{1}{2} \sum_x \sum_{\mu < \nu \atop \rho < \sigma} 
\epsilon_{\mu\nu\rho\sigma} h_{x, \mu\nu} h_{x - \hat{\rho} - \hat{\sigma} + T, \rho\sigma} \; ,
\end{equation}
where in the first step we used (\ref{Qnn}), then inserted the Hodge decomposition 
(\ref{hodge}) and subsequently used (\ref{property1}). 
In the final step $Q_T[2\pi h]$ was written explicitly. Inserting the parametrization  
(\ref{harmonics}) for the harmonic contribution $h$ we find
\begin{eqnarray}
Q_{T}[2\pi h] & = & \frac{1}{2} \sum_x \sum_{\mu < \nu \atop \rho < \sigma} \epsilon_{\mu\nu\rho\sigma} \, \omega_{\mu\nu} \, 
\omega_{\rho\sigma} \, \sum_{i=1}^{N_\rho} \sum_{j=1}^{N_\sigma} \, \delta_{x, \, i \hat{\rho} + j \hat{\sigma}} \sum_{n=1}^{N_\mu} 
\sum_{l=1}^{N_\nu} \, \delta_{x - \hat{\rho} - \hat{\sigma} + T, \, n \hat{\mu} + l \hat{\nu}} 
\\
& = & \frac{1}{2} \sum_{\mu < \nu \atop \rho < \sigma} \epsilon_{\mu\nu\rho\sigma} \, \omega_{\mu\nu} \, \omega_{\rho\sigma} \, 
\sum_{n=1}^{N_\mu} \sum_{l=1}^{N_\nu} \sum_{i=1}^{N_\rho} \sum_{j=1}^{N_\sigma}  \, 
\delta_{ i \hat{\rho} + j \hat{\sigma} - \hat{\rho} - \hat{\sigma} + T, n \hat \mu + l \hat \nu} 
\nonumber \\
& = &\frac{1}{2} \sum_{\mu < \nu \atop \rho < \sigma} \epsilon_{\mu\nu\rho\sigma} \, \omega_{\mu\nu} \, \omega_{\rho\sigma} \, 
\sum_{n=1}^{N_\mu} \sum_{l=1}^{N_\nu} \sum_{i=1}^{N_\rho} \sum_{j=1}^{N_\sigma} \, 
\delta_{0, (n-t_\mu) \hat \mu + (l - t_\nu) \hat \nu + (1-i) \hat{\rho} + (1-j) \hat{\sigma} } \;  .
\nonumber 
\end{eqnarray}
In the first step we summed over $x$ to remove the first Kronecker delta. In the second step we used  
that $T = t_\mu\hat{\mu} + t_\nu\hat{\nu} + t_\rho\hat{\rho} + t_\sigma\hat{\sigma}$ for mutually distinct $\mu, \nu, \rho, \sigma$.
Then it is obvious that 
only a single term remains where the Kronecker deltas give 1, such that the  topological charge reduces to
\begin{equation}
Q_T[2\pi h] \;  = \; \frac{1}{2} \sum_{\mu < \nu \atop \rho < \sigma} \epsilon_{\mu\nu\rho\sigma} \, \omega_{\mu\nu} \, \omega_{\rho\sigma} 
\; = \; \omega_{12}\omega_{34} - \omega_{13}\omega_{24} + \omega_{14}\omega_{23} \; \in \; \mathds{Z} \; ,
\end{equation}
which is the result  (\ref{Q_omega}).


\section{Properties of the kernels $M$ and $K$ from Fourier transformation}\label{app:C}

In Sections 4.2 and 4.3 we combine the gauge field action and the $\theta$-term into quadratic forms with kernels 
$M_{x,\mu \nu | y, \rho \sigma}$ defined in Eq.~(\ref{kernelM}) and similarly in Section 4.3 with a new kernel 
$K_{x,\mu \nu | y, \rho \sigma}$ defined in Eq.~(\ref{kernelK}). In order to diagonalize the space-time dependence
of these kernels, which we here generically denote as $A_{x,\mu \nu | y, \rho \sigma}$,  we use Fourier transformation, 
i.e., a similarity transformation with the unitary matrices 
$U_{p,x} \equiv V^{-1/2} e^{-i p \cdot x}$, where the momenta are given by 
$p_\mu = 2 \pi n_\mu /N_\mu \; , \; n_\mu = 0,1,2 \; ... \; N_\mu -1$
(we use periodic boundary conditions). We find 
\begin{equation}
\sum_{x,y} U_{p,x}^\star \, A_{\, x,\mu \nu | y, \rho \sigma} \, U_{q,y} \; = \;
\delta^{\; (4)}_{p,q} \; \widehat{A}(p)_{\mu \nu | \rho \sigma} \; , 
\label{Afourier}
\end{equation}
such that 
\begin{equation}
A_{\, x,\mu \nu | y, \rho \sigma}^{\; -1} \; = \; 
\frac{1}{V} \sum_{p} e^{ip(y-x)} \, \widehat{A}(p)_{\mu \nu | \rho \sigma}^{\, -1} \; .
\label{Ainvfourier}
\end{equation}
The Fourier transform $\widehat{A}(p)_{\mu \nu | \rho \sigma}$ is a $6 \times 6$ matrix (we order the 
indices $\mu < \nu$, $\rho < \sigma$). Both, the Fourier transforms of $M$ and of $K$ have a similar structure
given by\footnote{For $K$ we actually use a symmetrized version of the term with the $\epsilon$-tensor, but the
inversion strategy is identical to the one outlined here.} (again we use the generic notation $\widehat{A}$ for both kernels),
\begin{equation}
\widehat{A}(p)_{\mu \nu | \rho \sigma} \; = \; f(p)
[ \delta_{\mu \rho} \, \delta_{\nu \sigma} 
\, + \, i \, a \, \epsilon_{\mu \nu \rho \sigma} \; e^{ \, i p_\rho \, + \, i p_\sigma} ] \; ,
\end{equation}
where $f(p)$ is some function of the momenta and $a \in \mathds{R}$ some real-valued parameter. It is straightforward to show
that 
\begin{equation}
\sum_{\rho < \sigma} [ \delta_{\mu \rho} \, \delta_{\nu \sigma} 
\, + \, i \, a \, \epsilon_{\mu \nu \rho \sigma} \; e^{ \, i p_\rho \, + \, i p_\sigma} ] [ \delta_{\rho \tau} \, \delta_{\sigma \omega} 
\, - \, i \, a \, \epsilon_{\rho \sigma \tau \omega} \; e^{ \, i p_\tau \, + \, i p_\omega} ] \; = \; 
 \delta_{\mu \tau} \, \delta_{\nu \omega} (1 + a^2 e^{i p \cdot \hat{s}}) \; ,
\end{equation}
where again we use $\hat{s} \equiv \hat1 + \hat2 + \hat3 + \hat4$. 
As a consequence, the inverse momentum space kernel is given by 
\begin{equation}
\widehat{A}(p)_{\mu \nu | \rho \sigma}^{\; - 1} \; = \; \frac{
\delta_{\mu \rho} \, \delta_{\nu \sigma} 
\, - \, i \, a \, \epsilon_{\mu \nu \rho \sigma} \; e^{ \, i p_\rho \, + \, i p_\sigma} }{f(p) ( 1 + a^2 e^{i p \cdot \hat{s}})} \; ,
\end{equation}
and with \eqref{Ainvfourier} one finds the inverse kernel in real space. 


\section{Generalized Poisson resummation}\label{app:D}

In this appendix we present a short derivation of the generalized Poisson resummation formula. We consider 
a function $b(x_1,x_2,\, ... \; x_N)$ of $N$ real variables $x_j$ that has the form
\begin{equation}
b(x_1,x_2,\, ... \; x_N) \; \equiv \; \left( \, \prod_{j=1}^N \sum_{n_j \in \mathds{Z}} \right) 
e^{\, - \frac{\beta}{2} \sum_{j,k = 1}^N (x_j + 2\pi n_j) \, M_{jk} \, (x_k + 2\pi n_k)}
\; e^{\, - i \sum_{j=1}^N \zeta_j ( x_j + 2\pi n_j)} \; .
\label{poisson1}
\end{equation}
$M_{jk}$ is an invertible matrix that has eigenvalues with positive real parts and $\zeta_j$ are some
real parameters. The generalized Poisson resummation formula states that $b(x_1,x_2 \,, ... \; x_N)$ can be expressed
as 
\begin{equation}
b(x_1,x_2 \, ... \; x_N) \; = \; \left(\frac{1}{\sqrt{2\pi \beta}}\right)^{\! \!N} \!\!\! \frac{1}{\sqrt{\det M}}
\left( \, \prod_{j=1}^N \sum_{p_j \in \mathds{Z}} \right)  
e^{\, - \frac{1}{2\beta} \sum_{j,k = 1}^N (\zeta_j + p_j) \, M^{-1}_{jk} \, (\zeta_k + p_k)} \;
e^{\,  i \sum_{j=1}^N p_j \, x_j } .
\label{poisson2}
\end{equation}
To prove (\ref{poisson2}) we first note that $b(x_1,x_2 \, ... \; x_N)$ given in (\ref{poisson1}) is $2\pi$-periodic
in each of its arguments. This implies that it has the Fourier representation
\begin{equation}
b(x_1,x_2 \, ... \; x_N) \; = \; \left( \, \prod_{j=1}^N \sum_{p_j \in \mathds{Z}} \right)  \;
\widehat{b}\,(p_1,p_2 \, ... \; p_N) \; e^{\,  i \sum_{j=1}^N p_j \, x_j } \; ,
\label{poisson3}
\end{equation}
with the Fourier transforms $\widehat{b}\,(p_1,p_2 \, ... \; p_N)$ given by
\begin{equation}
\widehat{b}\,(p_1,p_2 \, ... \; p_N) \; = \; \left( \, \prod_{j=1}^N \int_{-\pi}^\pi \frac{dx_j}{2\pi} \right)
b(x_1,x_2 \, ... \; x_N) \, e^{\, - i \sum_{j=1}^N p_j \, x_j } \; .
\end{equation}
Inserting the explicit form (\ref{poisson1}) we find
\begin{eqnarray}
\widehat{b}\,(p_1,p_2\, ... \, p_N) & \!\!\!\! = \!\!\!\!\! & \left( \prod_{j=1}^N \sum_{n_j \in \mathds{Z}} 
\int_{-\pi}^\pi \frac{dx_j}{2\pi} \! \right) 
e^{ - \frac{\beta}{2} \sum_{j,k = 1}^N (x_j + 2\pi n_j)  M_{jk}  (x_k + 2\pi n_k)}
 e^{- i \sum_{j=1}^N (\zeta_j + p_j)( x_j + 2\pi n_j)} 
\nonumber \\
& \!\!\!\! = \!\!\!\!\!  & \left( \, \prod_{j=1}^N  
\int_{-\infty}^\infty \frac{dy_j}{2\pi} \right) 
e^{\, - \frac{\beta}{2} \sum_{j,k = 1}^N y_j \, M_{jk} \, y_k}
\; e^{\, - i \sum_{j=1}^N (\zeta_j + p_j) y_j} 
\nonumber \\
&  \!\!\!\! = \!\!\!\!\! &
\left(\frac{1}{\sqrt{2\pi \beta}}\right)^{\! \!N} \!\!\! \frac{1}{\sqrt{\det M}}
\; \,  e^{\, - \frac{1}{2\beta} \sum_{j,k = 1}^N (\zeta_j + p_j) \, M^{-1}_{jk} \, (\zeta_k + p_k)} \; ,
\end{eqnarray}
where in the first step we have inserted factors $e^{-i 2 \pi p_j n_j} = 1$ (note that the $p_j$ are integer) and in the 
second step have switched to new integration variables $y_j = x_j + 2\pi n_j$. In the last step the $N$-dimensional
Gaussian integral was solved. Inserting this result for $\widehat{b}\,(p_1,p_2\, ... \; p_N)$ in (\ref{poisson3}) 
completes the proof of (\ref{poisson2}).


\section{The dual worldline formulation}\label{app:worldline}

In the Section 3 we have generalized the self-dual U(1) gauge theory from Section 2 to include electric and magnetic matter
and showed that the construction is self-dual. However, the form of self-dual lattice QED discussed in Section 3 is not directly 
suitable for a numerical simulation, since the gauge field Boltzmann factor (\ref{boltzmann_both}) obviously 
is complex and does not give rise to a real and positive
weight factor that may be used in a Monte Carlo simulation. In this apppendix we now show that for bosonic matter this complex action 
problem can be overcome by switching to a worldline formulation for the magnetic 
matter\footnote{We remark that also for fermionic matter one 
may switch to a worldline formulation that solves the complex action problem 
from the gauge field Boltzmann factor. However, for fermions
the worldline configurations come with signs that are due to the Pauli principle and 
the spinor properties of the fermions. Thus the worldline 
formulation of fermions generates its own challenges that for some 
examples could be overcome with other techniques such as resummation 
or density of states techniques (see, e.g., \cite{Gattringer:2016kco}).}. First numerical results for self-dual QED based on this 
representation were presented in \cite{Anosova:2021akr}. 

In order to prepare the Boltzmann factor (\ref{boltzmann_both}) for the worldline formulation we rewrite the second exponent in 
(\ref{boltzmann_both}) by switching to the dual lattice using (\ref{convertdual1}) and the identity 
\begin{equation}
(dn)_{x,\mu\nu\rho} \; = \; - \sum_{\sigma} \epsilon_{\mu\nu\rho\sigma} \, (\, \partial \, \widetilde{n}\,)_{\tilde{x}-\hat{\sigma},\sigma} \; ,
\end{equation}
which is a direct consequence of (\ref{AppA:diffconvert}). Thus the gauge field Boltzmann factor assumes the form 
\begin{equation}
B_\beta[A^e,A^m]  \; = \; \sum_{\{ n \}} \; e^{ \, -\frac{\beta}{2} \sum_{x} \! \sum_{\mu < \nu} \big( F^e_{x,\mu \nu} \big)^2 } 
\; \prod_{\tilde{x},\mu} e^{\,  i  \widetilde{A}^m_{\tilde{x},\mu} (\partial \widetilde{n})_{\tilde{x},\mu}}  \, ,
\label{boltzmann_both_dual}
\end{equation}
where we have converted the sum in the second exponent of the Boltzmann factor into a product over all links of the dual lattice.

The second step is to use the well known worldline representation of the charged scalar in a background U(1) gauge field 
(see, e.g., \cite{Mercado:2013yta,Mercado:2013ola}). It is straightforward to convert this worldline representation to the dual lattice 
where the magnetic matter partition sum (\ref{Zmagnetic}) is defined. The worldline representation then reads (compare the appendix
of \cite{Sulejmanpasic:2019ytl} for the notation used here)
\begin{equation}
\widetilde{Z}_{M^m\!,\,\lambda^m\!,\,q^m} \big[\widetilde{A}^m\big] \; \equiv \; 
\sum_{\{ \widetilde{k} \}} \, W_{M^m\!,\,\lambda^m} \big[ \, \widetilde{k} \,\big] \; \left[
\prod_{\tilde{x}} \delta \Big( \big(\partial \widetilde{k}\,\big)_{\tilde{x}} \Big) \right] \left[
\prod_{\tilde{x},\mu} e^{\, i \, q^m \, \widetilde{A}^m_{\tilde{x},\mu} \, \widetilde{k}_{\tilde{x},\mu} } \right] \; .
\label{Zm_worldline}
\end{equation}
The partition function is a sum over configurations of the dual flux variables $\widetilde{k}_{\tilde{x},\mu} \in \mathds{Z}$ assigned to the 
links $(\tilde{x},\mu)$ of the dual lattice, where 
\begin{equation}
\sum_{\{ \widetilde{k} \}} \; \equiv \; \prod_{\tilde{x},\mu} \; \sum_{\widetilde{k}_{\tilde{x},\mu} \in \mathds{Z}} \; .
\label{fluxsum}
\end{equation}
The flux variables are subject to vanishing divergence constraints 
\begin{equation}
\big(\partial \widetilde{k}\,\big)_{\tilde{x}} \; \equiv \; \sum_{\mu = 1}^d 
\Big[ \widetilde{k}_{\tilde x,\mu} - \widetilde{k}_{\tilde x-\hat \mu,\mu} \Big] \; = \; 0 \; \; \; \; \forall \tilde x \; ,
\label{divergence0}
\end{equation}
which in (\ref{Zm_worldline}) are implemented with the product of Kronecker deltas. These constraints  enforce flux conservation 
at each site $\tilde x$ of the dual lattice, such that the $\widetilde{k}_{\tilde{x},\mu}$ form closed loops
of flux on the dual lattice. At every link $(\tilde{x},\mu)$ of the dual lattice the dual magnetic gauge field $\widetilde{A}^m_{\tilde{x},\mu}$
couples in the form  $e^{\, i \, q^m\, \widetilde{A}^m_{\tilde{x},\mu} \, \widetilde{k}_{\tilde{x},\mu} }$, which gives rise to the second product in 
\eqref{Zm_worldline}. 

The configurations of the dual flux variables $\widetilde{k}_{\tilde{x},\mu}$ come with real and positive weight factors 
$W_{M^m\!,\,\lambda^m} \big[ \, \widetilde{k} \,\big]$, that are themselves 
sums  (defined analogously to (\ref{fluxsum})) over configurations $\sum_{\{ \widetilde{a} \}}$ of auxiliary variables 
$\widetilde{a}_{\tilde{x},\mu} \in \mathds{N}_0$. The weights are given by
\begin{eqnarray}
\hspace*{-8mm}&& W_{M^m\!,\,\lambda^m} \big[ \, \widetilde{k} \,\big]  \; \equiv \; \sum_{\{ \widetilde{a} \}}
 \left[\prod_{\tilde x,\mu} \frac{1}{\left(|\widetilde{k}_{\tilde x,\mu}|\!+\!\widetilde{a}_{\tilde x,\mu}\right)! \ 
 \widetilde{a}_{\tilde x,\mu}!} \right]
\left[\prod_{\tilde x} \! I_{M^m\!,\,\lambda^m}\left(f_{\tilde x}\right)\right] \qquad \mbox{with}
\label{W_m} \\ 
\hspace*{-8mm} && I_{M^m\!,\,\lambda^m}\!\left(f_{\tilde x}\right)  \,  \equiv   \!\int_{0}^{\infty} \!\!\!\!\!\! dr\, r^{f_{\tilde x}+1} 
e^{-M^m r^2 -\lambda^m r^4} \quad , \quad
f_{\tilde x}  \,  \equiv  \! \ \sum_{\mu}
\Big[ |\widetilde{k}_{\tilde x,\mu}| \!+\! |\widetilde{k}_{\tilde x-\hat{\mu},\mu}| + 2\left( \widetilde{a}_{\tilde x,\mu} \!+\! 
\widetilde{a}_{\tilde x-\hat{\mu},\mu} \right) \Big] .
\nonumber
\end{eqnarray}
The $f_{\tilde x}$ are non-negative integer-valued combinations of the flux variables $\widetilde{k}_{\tilde{x},\mu}$ and the 
auxiliary variables $\widetilde{a}_{\tilde{x},\mu}$. For a numerical simulation the integrals $I_{M^m\!,\,\lambda^m}\!\left(f_{\tilde x}\right)$ may 
be pre-computed numerically and stored for sufficiently many values of $f_{\tilde x}$. The updates of the auxiliary variables 
can be implemented with standard techniques (see, e.g., \cite{Gattringer:2012df,Gattringer:2012ap} for details). 

With the gauge field Boltzmann factor in the form (\ref{boltzmann_both_dual}) and the dependence of the partition sum 
$\widetilde{Z}_{M^m\!,\,\lambda^m\!,\,q^m}$ on the dual magnetic gauge field $\widetilde{A}^m_{\tilde{x},\mu}$ given by the 
last factor in (\ref{Zm_worldline}) we can now completely integrate out the dual magnetic gauge field. The corresponding 
integral reads (compare (\ref{Zfull}) and use $\int \! D[{A}^m] = \int \! D[\widetilde{A}^m]$ ) 
\begin{eqnarray}
\int \!\! D[\widetilde{A}^m] \!
\left[\prod_{\tilde{x},\mu} e^{\, i  \widetilde{A}^m_{\tilde{x},\mu} (\partial n)_{\tilde{x},\mu}} \right] \!\!
\left[\prod_{\tilde{x},\mu} \, e^{\, i \, q^m \, \widetilde{A}^m_{\tilde{x},\mu} \, \widetilde{k}_{\tilde{x},\mu} } \right]  & = & 
\prod_{\tilde{x},\mu} \int_{-\pi}^\pi \!\! \frac{ d \widetilde{A}^m_{\tilde{x},\mu} }{2\pi} 
e^{\,  i  \widetilde{A}^m_{\tilde{x},\mu} \big[ q^m \, \widetilde{k}_{\tilde{x},\mu} + (\partial \widetilde{n})_{\tilde{x},\mu} \big] }
\nonumber \\ 
 & = & \prod_{\tilde{x},\mu} \delta \Big( q^m \, \widetilde{k}_{\tilde{x},\mu}  +  ( \partial \widetilde{n})_{\tilde{x},\mu} \Big) \; .
\label{kidentity}
\end{eqnarray}
Integrating out the dual magnetic gauge fields has generated link-based constraints that relate the flux variables
and the dual Villain variables via 
\begin{equation}
q^m \, \widetilde{k}_{\tilde{x},\mu} \; =  \; - \, ( \partial \widetilde{n})_{\tilde{x},\mu}
\qquad \forall (\tilde x, \mu) \; .
\label{ksolution}
\end{equation}
Note that the constraints (\ref{ksolution}) are consistent with the vanishing divergence constraints $\partial \widetilde{k} = 0$ from 
(\ref{divergence0}), due to $\partial^2 = 0$ (see Appendix \ref{app:diff_forms}).
 
Thus we may summarize the final form of self-dual scalar lattice QED with a worldline representation for the magnetic matter: 
\begin{eqnarray}
\hspace*{-12mm} && Z(\beta, M^e\!,\,\lambda^e\!,\,q^e\!,\,M^m\!,\,\lambda^m\!,\,q^m) \, =  \int \!\! D[A^e]  \sum_{\{ n \}} 
\sum_{\{ \widetilde{k} \}} \sum_{\{ \widetilde{a} \}}
\left[\prod_{\tilde{x},\mu} \delta \Big( q^m \, \widetilde{k}_{\tilde{x},\mu}  +  ( \partial \widetilde{n})_{\tilde{x},\mu} \Big) \right]
\label{Zfull_worldline1}
\\
\hspace*{-12mm}  && \qquad \qquad 
e^{ \, -\frac{\beta}{2} \sum_{x} \! \sum_{\mu < \nu} \big( F^e _{x,\mu \nu} \big)^2 } \; Z_{M^e\!,\, \lambda^e\!,\,q^e} [A^e] \;
\left[\prod_{\tilde x,\mu} \frac{1}{\left(| \widetilde{k}_{\tilde x,\mu}|\!+\!\widetilde{a}_{\tilde x,\mu}\right)! \ \widetilde{a}_{\tilde x,\mu}!} \right]
\left[\prod_{\tilde x} \! I_{M^m\!,\,\lambda^m}\left(f_{\tilde x}\right)\right] \; ,
\nonumber
\end{eqnarray}
with $F^e _{x,\mu \nu} = (d A^e \, + \, 2 \pi \, n)_{x,\mu \nu}$ and
\begin{equation}
f_{\tilde x}  \,  =  \! \ \sum_{\mu}
\Big[ | \widetilde{k}_{\tilde x,\mu}| \!+\! | \widetilde{k}_{\tilde x-\hat{\mu},\mu}| + 2\left( \widetilde{a}_{\tilde x,\mu} \!+\! 
\widetilde{a}_{\tilde x-\hat{\mu},\mu} \right) \Big] .
\label{Zfull_worldline2}
\end{equation}
Obviously all weight factors in (\ref{Zfull_worldline1}) are real and positive, such that this form now is accessible to numerical 
Monte Carlo simulations. Note that for $q^m = \pm 1$ one may use the constraints \eqref{ksolution} to completely eliminate 
the flux variables $\widetilde{k}_{\tilde x,\mu}$ such that in that case 
the Villain variables are not subject to any constraints, which in some aspects makes a 
numerical simulation of  \eqref{Zfull_worldline1} simpler than the simulation of the pure gauge 
theory (\ref{Zdef1}), where configurations of the 
Villain variables need to obey the closedness constraint (\ref{Villainconstraint}). 

We conclude the discussion of the worldline form by expressing the expectation value $\left\langle | \phi^m |^2 \right\rangle$ that
appears in the duality relation (\ref{duality_phi2}) in terms of the worldline variables. The expectation value is obtained from
a derivative of $\ln Z$ with respect to $M^m$, and this derivative can of course also be applied to $Z$ in the form (\ref{Zfull_worldline1}), 
(\ref{Zfull_worldline2}). A few lines of algebra give (use $-\partial/\partial M^m I_{M^m\!,\,\lambda^m}\left(f_{\tilde x}\right) =
 I_{M^m\!,\,\lambda^m}\left(f_{\tilde x} + 2\right)$),
\begin{eqnarray}
\left\langle | \phi^m |^2 \right\rangle_{\! \beta, M^e\!,\,\lambda^e\!,\,q^e\!,\,M^m\!,\,\lambda^m\!,\,q^m}  & = & 
- \, \frac{1}{V} \frac{\partial }{\partial M^m} \, \ln Z (\beta, M^e\!,\,q^e\!,\,\lambda^e\!,\,M^m\!,\,\lambda^m\!,\,q^m) 
\nonumber \\
& = &
\frac{1}{V} \left\langle \sum_{\tilde x} \frac{  I_{M^m\!,\,\lambda^m}\left(f_{\tilde x} + 2 \right)}{ I_{M^m\!,\,\lambda^m}\left(f_{\tilde x}\right)}
\right \rangle_{\! \beta, M^e\!,\,\lambda^e\!,\,q^e\!,\,M^m\!,\,\lambda^m\!,\,q^m} .
\end{eqnarray}


\section{Self-duality with a generalized topological charge}\label{app:gen_top_charge}

In this appendix we briefly discuss how the construction of the self-dual theory with a $\theta$-term in 
Subsection~\ref{sec:selfdual_theta}
can be generalized 
further, such that the topological charge fully implements all lattice symmetries. 

The general form of the self-dual kernel we consider (this generalizes the kernel defined in  \eqref{Khat}) is in 
momentum space written as
\be\label{eq:Kgen}
\widehat K_{\mu\nu|\rho\sigma}(p) \; = \; \frac{1}{\widehat G(p)}\left(\delta_{\mu\rho}\delta_{\nu\sigma}-i \sum_T \gamma_T \,\epsilon_{\mu\nu\rho\sigma} \, e^{ip_\rho+ip_\sigma+iT\cdot p}\right) \; ,
\ee
where $T = \sum_{\mu}t_\mu \hat\mu$ with $t_\mu\in\mathbb Z$, and $\gamma_T$ are some coefficients. Now note that
(compare also Appendix \ref{app:C})
\begin{eqnarray}
&& \left(\delta_{\mu\rho}\delta_{\nu\sigma} \, + \, i \sum_T \gamma_T \,  \epsilon_{\mu\nu\rho\sigma} \, 
e^{\, ip_\rho \, + \, ip_\sigma \, - \, iT\cdot p}\right)\left(\delta_{\rho \mu'} \delta_{\sigma\nu'} \, - \, 
i \sum_T \gamma_T \, 
\epsilon_{\rho\sigma\mu'\nu'} \, e^{\, ip_\mu' \, + \, ip_\nu' \, - \, iT\cdot p}\right)
\nonumber \\
&& \hspace{20mm}
= \; \delta_{\mu\mu'}\delta_{\nu\nu'}+\sum_{T,T'} \gamma_T \, \gamma_{T'} \, 
\epsilon_{\mu\nu\rho\sigma} \, \epsilon_{\rho\sigma\mu'\nu'} \, e^{\, ip_\rho \, + \, ip_\sigma \, + \, ip_{\mu}
\, + \, ip_{\nu} \, - \, i T\cdot p-iT'\cdot p}
\nonumber \\
&& \hspace{20mm} = \; 
\delta_{\mu\mu'} \delta_{\nu\nu'}\left(1+\sum_{T,T'}\gamma_T\gamma_{T'} 
e^{\, i \hat s\cdot p \, - \, i T\cdot p \, - \, iT'\cdot p}\right)\; ,
\end{eqnarray}
where again $\hat s= \hat 1+\hat 2+\hat 3+\hat 4$. So by choosing
\be
\widehat G(p)= \sqrt{1+\sum_{T,T'}\gamma_T\gamma_{T'}e^{ \, i(\hat s-T-T')\cdot p}} \; ,
\ee
we find that $\hat K$ is form-covariant under inversion, i.e., 
\begin{align}
&\widehat K_{\mu\nu|\rho\sigma}(p) \; = \; 
\frac{\delta_{\mu\rho}\delta_{\nu\sigma}+i \sum_T \gamma_T \, \epsilon_{\mu\nu\rho\sigma} \, 
e^{ip_\rho+ip_\sigma+iT}}{\sqrt{1+\sum_{T,T'}\gamma_T\gamma_{T'}e^{i(\hat s-T-T')\cdot p}}}\; , \\
&\widehat K_{\mu\nu|\rho\sigma}^{-1}(p)  \; = \; 
\frac{\delta_{\mu\rho}\delta_{\nu\sigma}-i \sum_T \gamma_T \, \epsilon_{\mu\nu\rho\sigma} \, 
e^{ip_\rho+ip_\sigma+iT}}{\sqrt{1+\sum_{T,T'}\gamma_T\gamma_{T'}e^{i(\hat s-T-T')\cdot p}}}\;.
\end{align}
Now a convenient choice of the coefficients $\gamma_T$ is such that $\gamma_{-T+\hat s}=\gamma_{T}$, 
which implies that the argument in the square root of $\widehat G (p)$ is real. This argument is given by
\be
1+\sum_{T,T'}\gamma_T\gamma_{T'}e^{i(\hat s-T-T')\cdot p} \; = \; 1+ \left|\sum_{T} \gamma_T e^{-i T\cdot p}\right|^2\; .
\ee
In real space this corresponds to the operator 
\be
H_{x,y}=\sum_{T,T'}\gamma_T \, \gamma_{T'} \, \delta^{\, (4)}_{x+T,y+T'} \; ,
\ee
which generalizes the Helmholtz lattice operator introduced in \eqref{helmholtz}. 

We point out that the above result does not depend on the vector $\hat s$ which singles out a direction on the lattice, 
such that now, with a suitable choice of the coefficients $\gamma_T$ the lattice symmetries can be implemented in 
$H$. The full $\theta$-term, however, does not seem to be manifestly invariant due to the appearance of the vectors
$\hat \sigma$ and $\hat\rho$ in \eqref{eq:Kgen}. In a free theory, however, the $\theta$-term is topological, 
and hence lattice symmetries are exact in this case too.

The action density is now given by
\be
\sum_{y}\sum_{\mu<\nu}\sum_{\rho<\sigma}F_{x,\mu\nu}(H^{-1/2})_{x,z} \left(\delta_{y,z}+i \sum_T\gamma_T \,\epsilon_{\mu\nu\rho\sigma} \, \delta_{y,z-\hat\rho-\hat\sigma+T}\right) F_{z,\rho\sigma} \; .
\ee
As announced, the theory can be made fully invariant under all lattice symmetries if, in addition to the conditions 
$\gamma_{-T+\hat s}=\gamma_{T}$ from above, we implement the 
following relations among the coefficients\footnote{This is easiest to see as follows: The generic $\theta$-term is defined as $Q_W=\sum_p F_{p}F_{\star W(p)}$, where $W$ is an arbitrary translation operator from the lattice to the dual lattice, 
$p$ is a plaquette of the lattice and $\star$ is the Hodge-star analogue mapping of the lattice to the dual lattice 
(compare \cite{Sulejmanpasic:2019ytl}) .  
The lattice rotation $R$ commutes with the $\star$ operator, but not with $W$. We now consider 
$R: F_p \rightarrow F_{R(p)}$, but then $Q_W\rightarrow \sum_p F_{R(p)}F_{R(\star WR^{-1}R(p))}=Q_{RWR^{-1}}$. So if we now 
define a general topological charge as $\sum_W \kappa_W Q_W$, with coefficients $\kappa_W$ that respect  
$\kappa_{R W R^{-1}}=\kappa_{W}$, 
the $\theta$-term will obey 
the lattice symmetries. Writing explicitly 
\[\sum_{W}\kappa_W Q_W \; = \; \sum_{x}\sum_{\mu<\nu}\sum_{\rho<\sigma}\sum_{\bm W} \kappa_{\bm W} F_{x,\mu\nu}
\epsilon_{\mu\nu\rho\sigma}F_{x+\frac{\hat s}{2}+\bm W-\hat \rho-\hat\sigma}\;,
\] 
where $\bm W$ is now a vector which corresponds to the map $W$. Now, setting a lattice vector, $T=\bm W+\frac{\hat s}{2}$, 
this form can be written as 
\[
\sum_{\bm W}\kappa_{\bm W} Q_{\bm W} \; = \; \sum_{x}\sum_{\mu<\nu}\sum_{\rho<\sigma}\sum_{T} \kappa_{T} F_{x,\mu\nu}\epsilon_{\mu\nu\rho\sigma}F_{x+T-\hat \rho-\hat\sigma} \; ,
\]
where $\kappa_{T-\frac{\hat s}{2}}= \kappa_T$, and the condition $\kappa_{R(\bm W)}= \kappa_{\bm W}$ translates to 
$\gamma_{T}=\gamma_{R\left(T-\frac{\hat s}{2}\right)+\frac{\hat s}{2}}$ as stated in \eqref{gamma_trafo}.},
\be
\gamma_{T}=\gamma_{R(T-\frac{\hat s}{2})+\frac{\hat s}{2}}\;,
\label{gamma_trafo}
\ee
where $R$ is an arbitrary lattice rotation.  

\newpage

\bibliographystyle{utphys}
\bibliography{bibliography}

\end{document}